\documentclass[12pt,a4paper,twoside,openright]{report}
\usepackage{Thesis_template_arx}

\begin{document}
\includepdf[pages=1, scale=0.93]{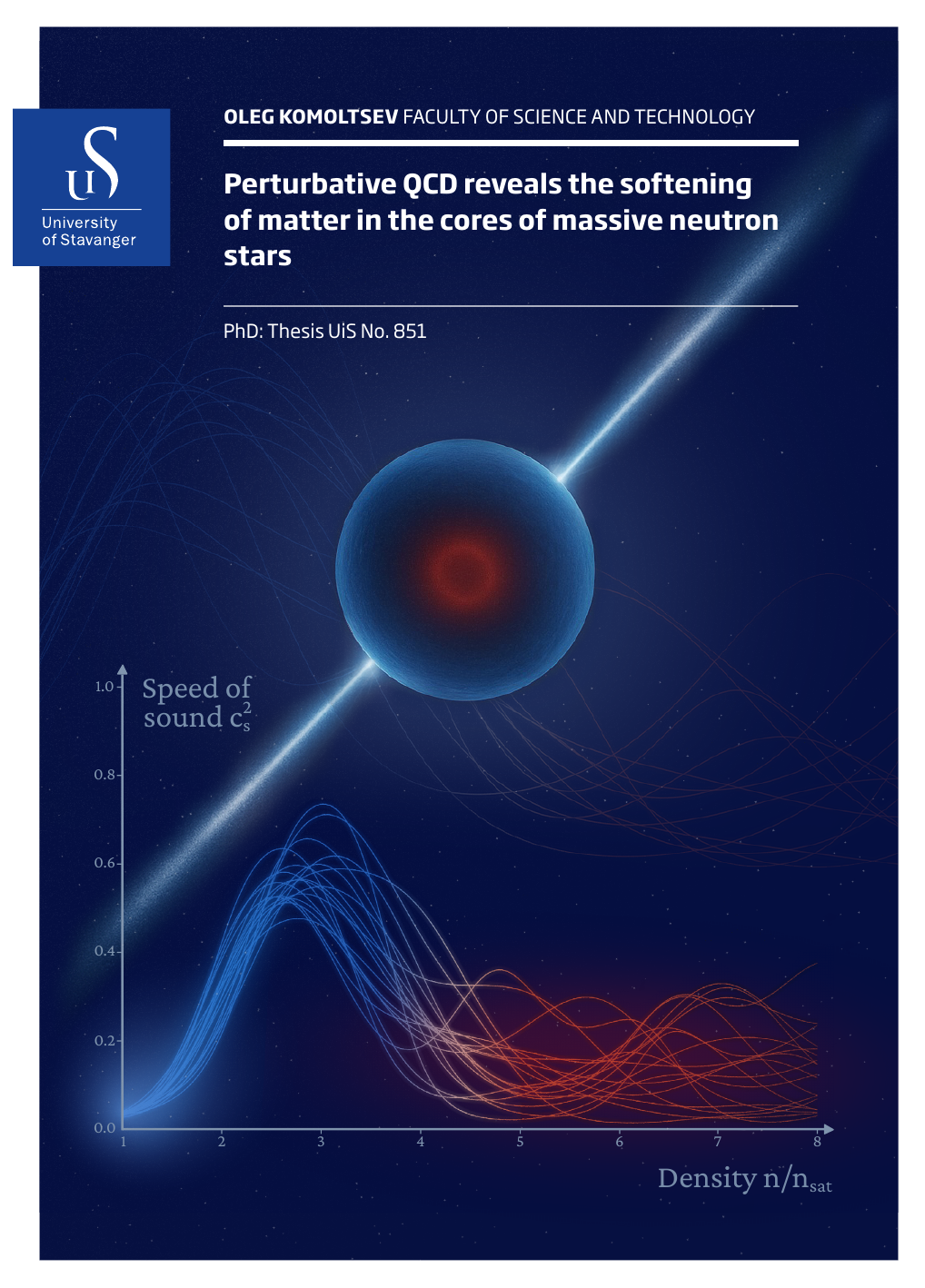}
\setcounter{page}{0}
\title{Perturbative QCD reveals the softening of matter in the cores of massive neutron stars}
\author{Oleg Komoltsev}
\year{2025}
\faculty{Faculty of Science and Technology}
\department{Department of Mathematics and Physics}

\insertTitlePage


\ISBN{978-82-8439-357-5}
\ISSN{1890-1387}
\thesisnum{851}

\insertColophon

\fontdimen2\font=4pt 

%
\addcontentsline{toc}{chapter}{Preface}
\mainfont{\chapter*{Preface}

\tolerance=1
\emergencystretch=\maxdimen
\hyphenpenalty=10000
\hbadness=10000

This thesis is submitted in partial fulfillment of the requirements for the degree of Philosophiae Doctor (PhD) at the University of Stavanger, Faculty of Science and Technology, Norway. The research has been carried out at the University of Stavanger from April 2021 to March 2025.

I would like to express my deepest gratitude to my supervisor, Aleksi Kurkela, who has dedicated an extraordinary amount of time to my development. Over the past four years, I have received his undivided attention and unwavering support with every challenge I faced. I am profoundly grateful for his guidance in the scientific world, and for numerous conversations that kept my enthusiasm for physics at its peak.

I also extend my sincere appreciation to my close scientific collaborator, Tyler Gorda, for his invaluable help with all aspects of my PhD research. My gratitude further goes to my co-supervisor, Alex Nielsen, as well as my colleagues and fellow PhD students at the University of Stavanger for providing a fantastic research environment with an incredibly friendly atmosphere. I am also grateful to Aleksi Vuorinen and Joonas Nättilä for their valuable insights and for a rewarding collaboration. Additionally, I am grateful for the many engaging scientific conversations. In particular, I thank Kenji Fukushima, Aleksas Mazeliauskas, Rahul Somasundaram, Ingo Tews, J\'{e}r\^{o}me Margueron, Hauke Koehn, Luciano Rezzolla, Krishna Rajagopal and Sanjay Reddy, for their inspiring discussions.

A special thanks goes to my parents and my brother, Ilya, without whom this scientific journey would not have been possible. Their love and support are what made this thesis possible! 
 
\vspace{15pt}

Oleg\\
Stavanger, February 2025}

\clearpage

%
\addcontentsline{toc}{chapter}{Abstract}
\mainfont{\chapter*{Abstract}

\tolerance=1
\emergencystretch=\maxdimen
\hyphenpenalty=10000
\hbadness=10000

The cores of neutron stars (NSs) contain the densest matter in the universe. Rapid advancements in neutron-star observations allow unprecedented empirical access to cold, ultra-dense Quantum Chromodynamics (QCD) matter. The combination of these observations with theoretical calculations has revealed previously inaccessible features of the equation of state (EoS) and the QCD phase diagram.

In this thesis, I demonstrate how perturbative-QCD calculations at asymptotically high baryon densities provide robust constraints on the EoS at neutron-star densities. The method for constraint propagation is based solely on thermodynamical causality, stability, and consistency of the EoS. 
By constructing a large ensemble of EoSs using Gaussian processes regression and incorporating it into a Bayesian inference of EoS, I demonstrate that the novel pQCD constraints go beyond those obtained from current astrophysical observations alone, forcing the EoS to soften at the maximum densities of stable neutron stars. 

This softening of the EoS can be interpreted as an indication of approximate conformal symmetry restoration, a sign of a first-order phase transition (FOPT), or potentially both. I show that the conformal symmetry restoration is consistent with the hypothesis of quark matter cores inside the most massive NSs. Although current astrophysical data and theoretical inputs cannot definitively distinguish between the two scenarios, they slightly favor the occurrence of a phase transition of some kind — whether a crossover to quark matter or a destabilizing FOPT — in the cores of the most massive neutron stars.

}

\clearpage

%
\addcontentsline{toc}{chapter}{List of papers}
\mainfont{\chapter*{List of papers}

\begin{enumerate}[label=\Roman*]

\item{How Perturbative
QCD Constrains the Equation of State at Neutron Star Densities. }{O. Komoltsev and A. Kurkela.}{ Phys. Rev. Lett. 128, 202701 (2022), arXiv:2111.05350 [nucl-th] \cite{Komoltsev:2021jzg}}

\item{
Ab-initio QCD Calculations Impact the Inference of the Neutron-star-matter Equation of State. }{T. Gorda, O. Komoltsev, A. Kurkela. }{Astrophys.J. 950 (2023) 2, 107, arXiv:2204.11877 [nucl-th] \cite{Gorda:2022jvk}}

\item{Bayesian uncertainty quantification of perturbative QCD input to the neutron-star equation of state. }{T. Gorda, O. Komoltsev, A. Kurkela, A. Mazeliauskas. }{JHEP 06 (2023) 002, arXiv:2303.02175 [hep-ph] \cite{Gorda:2023usm}}

\item{Strongly interacting matter exhibits deconfined behavior in massive neutron stars. }{E. Annala, T. Gorda, J. Hirvonen, O. Komoltsev,
A. Kurkela, J. Nättilä, and A. Vuorinen. }{Nature Commun. 14, 8451 (2023), \\ arXiv:2303.11356 [astro-ph.HE] \cite{Annala:2023cwx}}

\item{Equation of state at neutron-star densities and beyond from perturbative QCD. }{O. Komoltsev, R. Somasundaram, T. Gorda, A. Kurkela, J. Margueron, I. Tews. }{Phys.Rev.D 109 (2024) 9, 094030, arXiv:2312.14127 [nucl-th] \cite{Komoltsev:2023zor}}

\item{First-order phase transitions in the cores of neutron stars.\\}{O.Komoltsev. }{Phys.Rev.D 110 (2024) 7, L071502, \\ arXiv:2404.05637 [nucl-th] \cite{Komoltsev:2024lcr}}

\end{enumerate}}

\clearpage

\insertTOC


\setPageStyles


%
\mainfont{\chapter{Introduction}

\section{Neutron stars}
\label{sec:ns}

The concept of dense stars, even denser than white dwarfs, was originally proposed by Lev Landau in 1931 (published in 1932 \cite{Landau:1932uwv}). According to Chandrasekhar’s prediction \cite{1931ApJ....74...81C}, white dwarfs have a maximum mass limit beyond which the pressure of relativistic degenerate electrons is insufficient to counteract gravity. Landau speculated about an even denser form of star that could exist beyond this limit. Described by Lev Davidovich as “one gigantic nucleus,” these dense stars were theorized well before the experimental discovery of neutron stars and, remarkably, even before the discovery of the neutron itself. Following the discovery of the neutron in 1932, Walter Baade and Fritz Zwicky made the first explicit prediction of neutron stars \cite{Baade:1934wuu} as an attempt to explain the energy released during supernova explosions.

Landau’s pioneering work inspired Oppenheimer and his student Volkoff to incorporate general relativity into their analysis of dense stellar objects. Collaborating with Tolman, who had formulated the general relativistic equations for static spherically symmetric fluids, they numerically solved these equations for a non-interacting fluid of neutrons \cite{Oppenheimer:1939ne, Tolman:1939jz}. This collaboration led to the development of the Tolman-Oppenheimer-Volkoff (TOV) equations, as presented in next section \cref{eq:TOV}. These equations remain fundamental in modern astrophysics and is utilized extensively throughout this thesis.

Theoretical efforts from this period onward aimed at modeling the behavior of ultra-dense matter under extreme conditions, such as advancements made by Harrison and Wheeler, who extended the analysis by incorporating a mixture of nuclei (modeled via the liquid drop model), electrons, and a free neutron gas \cite{1965gtgc.book.....H}. Cameron further expanded on this approach by including nuclear interactions \cite{1959ApJ...130..884C} described using the Skyrme model \cite{SKYRME1958615}. He also highlighted the potential presence of hyperons — baryons containing strange quarks — at such extreme densities, with further developments contributed by Salpeter, Ambartsumyan, and Saakyan in \cite{SALPETER1960393, 1960SvA.....4..187A,1962SvA.....5..601A}. The subsequent theoretical works introduced the possibility of neutron superfluidity \cite{MIGDAL1959655, osti_4635281}, meson condensates (such as pions and kaons) \cite{WALECKA1974491} and explored potential phase transitions to quark matter at even higher densities \cite{Ivanenko1965,Ivanenko1969,SHURYAK198071}. The modern theoretical understanding of the behavior of dense matter are summarized in the following \cref{sec:intro_eos}. For a historical overview, see \cite{1983bhwd.book.....S,Schaffner-Bielich_2020,Yakovlev:2012rd}.

Despite these early theoretical predictions, it was not until 1967 that the first observational evidence of neutron stars was obtained. Jocelyn Bell-Burnell, then a PhD student of Antony Hewish, detected periodic radio signals with millisecond-to-second timescales and extraordinary precision, with period stability on the order of $10^{-10}$ to $10^{-21}$ \cite{Hewish:1968bj}. The initial hypothesis of “LGM-1” (Little Green Men) was ruled out after the discovery of additional sources with similar characteristics from different sky locations. The rapid periods and short emission timescales could only be explained by rapidly rotating neutron stars, dubbed pulsars. The theoretical work of Pacini \cite{PACINI1967} and Gold \cite{GOLD1968} proposed that the periodicity of these signals was due to rotating neutron stars with strong magnetic fields.

The first pulsar in a binary system with another neutron star, PSR1913+16, was discovered by Joseph Taylor and Russell Hulse in 1975 \cite{1975ApJ...195L..51H}. Their observations provided the first indirect evidence that the system was losing energy through gravitational wave (GW) emission.

Beyond testing general relativity, binary pulsars offer a surprisingly rich opportunity to study the behavior of cold, ultra-dense matter. Radio observations of pulsars in binary systems provide the most precise measurements of neutron star masses to date. In such systems with a visible companion, such as a white dwarf or a main sequence star, radial velocities can be measured due to the regular pulsing of the neutron star and the visibility of the companion. The ratio of their radial velocities provides the mass ratio between the two objects. If the companion’s mass is determined through methods like spectroscopy (e.g., as demonstrated in \cite{Antoniadis:2013pzd} for white dwarfs), the pulsar’s mass can be accurately inferred.

In cases where the binary companion is not directly observable, additional general relativistic corrections can be used to constrain the pulsar's mass. One such method is the Shapiro time delay \cite{PhysRevLett.13.789}, which measures the delay in the radio pulses caused by the gravitational field of the companion. The effect is maximized when the orbital plane’s inclination is close to 90$^{\circ}$, meaning it lies along the line of sight. In this alignment, the pulses pass through the companion, undergo time delays that depend on the companion’s mass, and remain detectable by radio telescopes. This allows for precise mass measurements \cite{Demorest:2010bx, Cromartie:2019kug, Fonseca:2016tux, Fonseca:2021wxt}. One of the recent breakthroughs in the astrophysical observation of pulsars is the discovery of two-solar-mass neutron stars \cite{Demorest:2010bx}, first measured using the Shapiro time delay. As will be discussed later, this discovery imposes some of the strongest constraints on the properties of cold, ultra-dense matter.

On the other side of the spectrum, different sources X-ray emission can provide simultaneous constraints on the mass and radius of neutron stars. One of such source is a hot spot on the pulsar’s surface. The rotation of a neutron star introduces periodic variations in the observed X-ray intensity, which can be modeled to infer properties of the hotspots, such as their size, temperature, and location on the star’s surface. The neutron star’s intense gravitational field causes relativistic effects, including light bending, which change the periodicity and intensity of the observed signals. Modeling these effects requires accounting for local radiation beaming due to the bulk motion of material on the rotationally deformed surface, as well as ray propagation through the star’s curved spacetime. These relativistic phenomena depend on the star’s compactness (the ratio of its mass to radius), providing simultaneous constraints on both the mass and radius of the star. Observationally, these studies are conducted using X-ray observatories like the Neutron Star Interior Composition Explorer (NICER), a soft X-ray telescope installed on the International Space Station in 2017, and XMM-Newton, a space observatory launched by the European Space Agency in 1999. The analysis of such data produces two-dimensional mass-radius probability distributions for neutron stars, as demonstrated in the studies of pulsars such as PSR J0030+0451 \cite{Miller:2019cac,Riley:2019yda} and PSR J0740+6620 \cite{Fonseca:2021wxt,Miller:2021qha,Riley:2021pdl}.

Another source of X-ray emission is thermonuclear X-ray bursts that frequently appears in low-mass X-ray binaries (LMXBs). By observing these bursts and modeling their cooling processes, it is possible to constrain the size of the emitting region. The most precise constraint to date comes from the neutron star in the binary system 4U J1702-429 \cite{Nattila:2017wtj}, achieved through direct atmosphere model fits to the time-evolving energy spectra of these bursts. Additionally, two other binary systems, 4U 1724-307 \cite{Nattila:2015jra} and SAX J1810.8$-$260 \cite{Nattila:2015jra}, have been studied using the cooling-tail method for mass-radius constraints. Another approach to constraining the NS radius involves spectral fitting in quiescent LMXBs, referring to periods of minimal accretion in these systems. By analyzing the X-ray flux and surface temperature of a NS with a reliable distance measurement, the radius of the emitting region can be constrained \cite{Steiner:2017vmg,Shaw:2018wxh}. The data is collected using ROSAT, Chandra, and XMM-Newton, which are space-based telescopes designed to observe X-ray emissions from cosmic sources. 

\begin{figure}[ht!]
    \centering
\includegraphics[width=1\textwidth]{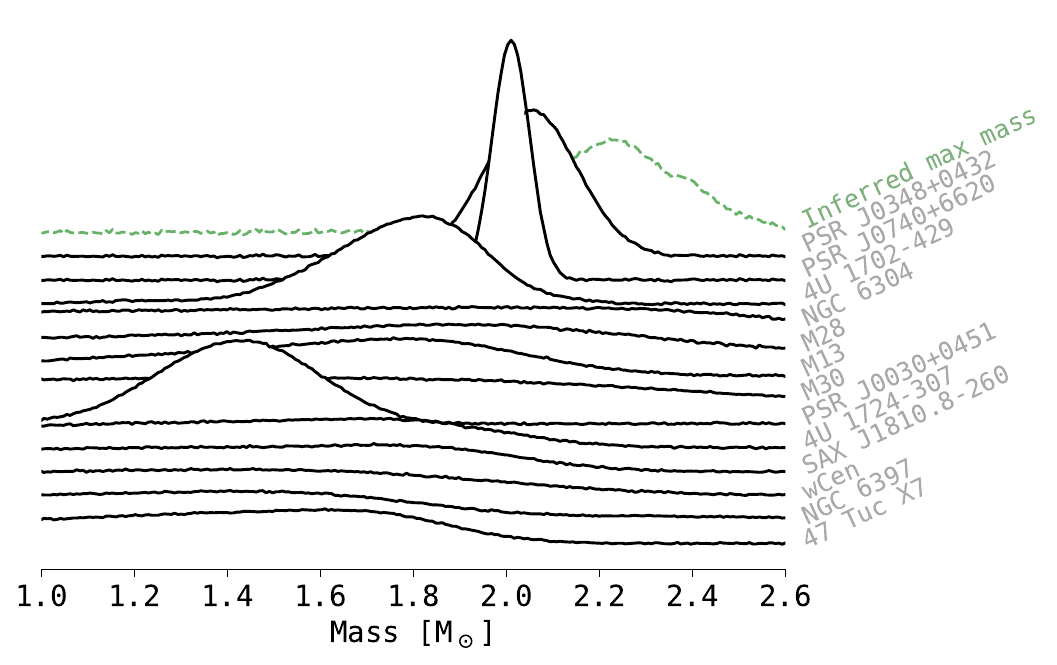}
    \caption{An artistic representation of the posterior distributions of observed masses from various X-ray and radio measurements, which is utilized in this thesis. Additionally, the inferred maximum mass of NSs is shown in green. The details of the inference are provided in the subsequent sections.}
    \label{fig:mass_measurments}
\end{figure}

Lastly, neutron stars can be studied through the rapidly evolving field of gravitational wave and multimessenger astronomy. Coalescing binary neutron stars generate a quadrupole moment, which, according to general relativity, produces ripples in spacetime known as gravitational waves. Modeling GW requires understanding the tidal deformability (TD) of compact objects—i.e., how such objects respond to the gravitational field of a nearby massive body, as in the case of a binary neutron star (BNS) merger. Determining the binary TD from observed mergers provides additional information about the neutron-star matter, as detailed in \cref{eq:Love}. GW data is collected using the ground-based detectors operated by the LIGO and Virgo Scientific Collaborations. 

Gravitational waves can be accompanied by an electromagnetic counterpart, hence the term “multimessenger astronomy”. Astrophysical modeling of BNS mergers suggests that these events exhibit distinct signatures across the electromagnetic spectrum \cite{Metzger_2012,10.1093/mnras/stt037}, including short gamma-ray bursts (sGRBs) \cite{Blinnikov:2018boq,Paczynski:1986px,Paczynski:1991aq,Narayan:1992iy,Eichler1989} and longer-lived afterglows. The timing and spectral properties of sGRBs and kilonovae (gamma-ray bursts caused by the radioactive decay of heavy r-process nuclei) can be used to constrain the behavior of matter in extreme conditions.

Such multimessenger event was the first simultaneous observation of a gravitational wave signal from a binary NS merger, GW170817 \cite{PhysRevLett.119.161101, LIGOScientific:2018cki,LIGOScientific:2018hze}, and its electromagnetic counterpart with a lag of $\lesssim$ 2 s, GRB 170817A \cite{LIGOScientific:2017ync}. The GW signal provided novel constraints on binary TD. Additional constraints arise from the electromagnetic counterpart’s properties. Astrophysical modeling of jet generation and launching suggests that the merger remnant collapsed into a black hole \cite{Margalit:2017dij,Rezzolla:2017aly,Ruiz:2017due,Shibata:2017xdx,Shibata:2019ctb}. Furthermore, the spectral properties of sRGB imply that the remnant either underwent a prompt collapse into a black hole or formed a supramassive/hypermassive neutron star that subsequently collapsed into a black hole shortly afterward.

\section{TOV equation}
\label{sec:intro_tov}
The connection between the described experimental data and the behavior of cold, ultra-dense matter is established through the TOV equation. This equation is a solution of Einstein’s equations for a spherically symmetric, static body, with the approximation that an isolated star can be modeled as a perfect fluid:
\begin{align}
\label{eq:TOV}
 \frac{dP}{dr} &= -\frac{Gm}{r^2} \e \left( 1 + \frac{P}{\e c^2} \right) \left( 1 + \frac{4\pi r^3 P}{mc^2} \right) \left( 1 - \frac{2Gm}{rc^2} \right)^{-1} \nonumber \\
 \frac{dm}{dr} &= 4\pi r^2 \e , 
\end{align}
where $r$ represents the radial coordinate, and $m(r)$ is the total mass within the radius $r$.  The initial condition for these equations are 
\begin{equation}
    P(r = 0) = P_{\mathrm{central}}, \quad \quad m(r = 0) = 0.
\end{equation}
The only input required to solve TOV equation is the equation of state (EoS). For the TOV equation EoS is expressed as the pressure as a function of energy density, $p(\e)$. The next section provides an overview of the EoS.

The solution of the TOV equation provides a sequence of neutron star masses ($M$) and radii ($R$) as a function of central pressure or density. The stability condition for neutron stars is determined by the sign of $dM/dP_{\mathrm{central}}$ \change{\cite{1988ApJ...325..722F}}:
\begin{itemize}
    \item Stable branch: $dM/dP_{\mathrm{central}} > 0$.
    \item Unstable branch: $dM/dP_{\mathrm{central}} < 0$ (mass decreases with increasing density, leading to gravitational collapse into a black hole).
\end{itemize}
The point where $dM/dP_{\mathrm{central}} = 0$ marks the maximum stable mass $M_{\mathrm{TOV}}$.

Next, I examine how GW data can be used to extract information about neutron-star matter. The early inspiral phase of two coalescing neutron stars is affected by the internal structure of NSs. To linear order, this impact can be characterized by a single parameter — the tidal deformability $\lambda$, which is defined as the ratio of the induced quadrupole moment of the star to the tidal field, i.e., how much the star deforms due to the external gravitational field of the companion.

Following the approach of \cite{Hinderer_2008,Han:2018mtj}, the relevant equations are derived using linearized metric perturbations and are presented in appendix \cref{eq:Love,eq:Love2}. By inputting an EoS, these first-order differential equations can be solved numerically alongside the TOV equation to predict the dimensionless tidal deformability $\Lambda=\lambda/M^5$ as a function of the central density. The dimensionless tidal deformability of a star is used in \cref{sec:bayesian} to calculate the binary tidal deformability $\tilde\Lambda$ (see \cref{eq:TD}), which can be compared to the observations, such as those provided by the LIGO/Virgo collaborations. 

The system of equations can also include an additional equation to calculate the total baryonic number:
\begin{align}
\label{eq:total_baryonic}
    \frac{dN}{dr}=4\pi r^2 n \left[ 1 - \frac{2m(r)}{r} \right]^{-1},
\end{align}
where n is baryon number density. This is essential for utilizing constraints derived from multi-messenger observations of events like GW170817, as described in the \cref{sec:bayesian}.

Solving the system of \cref{eq:TOV,eq:Love2,eq:total_baryonic} establishes a one-to-one correspondence between a neutron star’s mass, radius, and tidal deformability with the equation of state:
\begin{align}
    M(R) \Leftrightarrow p(\e),\quad \Lambda(R) \Leftrightarrow p(\e).
\end{align}
Consequently, neutron stars serve as natural laboratories for exploring matter under the most extreme conditions.

\section{Equation of state}
\label{sec:intro_eos} 

The theory of strong interaction, known as Quantum Chromodynamics (QCD), is a quantum field theory that describes the interactions between quarks, mediated by gluons. The QCD phase diagram, shown in \cref{fig:pd_qcd}, represents temperature $T$ as a function of baryon chemical potential $\mu$. At relatively low temperatures and low densities (corresponds to the small chemical potential), matter exists in the form of hadrons — bound states of quarks, such as protons and neutrons. In contrast, at extreme temperatures or densities, quarks become deconfined, transitioning from hadronic matter to a quark matter.
\begin{figure}[ht!]
    \centering
\includegraphics[width=0.9\textwidth, trim=20 20 0 0, clip]{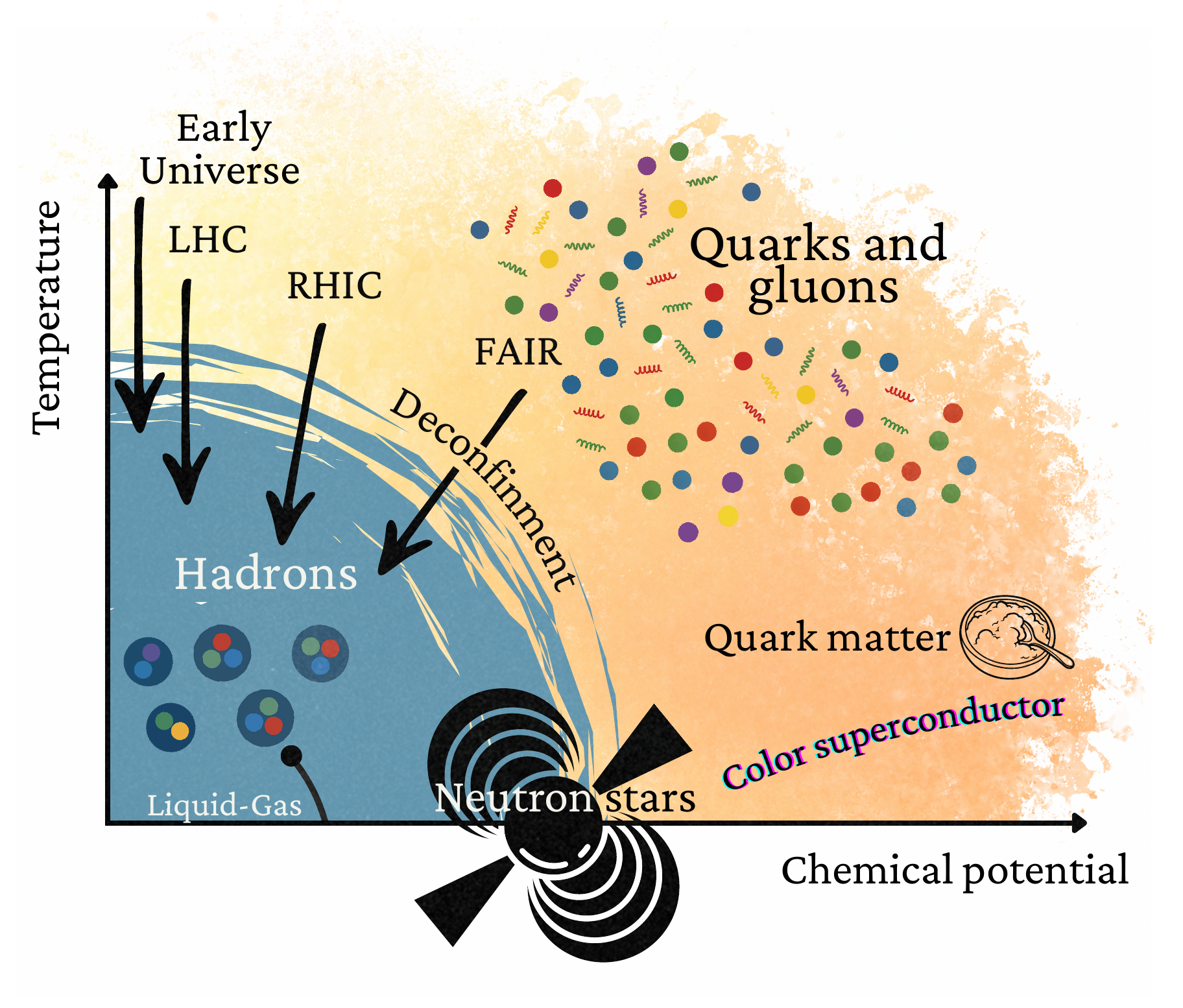}
 \caption{An artistic representation of the phase diagram of QCD.}
    \label{fig:pd_qcd}
\end{figure}

As illustrated in the figure, heavy-ion collision experiments (such as RHIC, FAIR, and LHC) primarily probe the high-temperature, low-density region of the phase diagram. In contrast, neutron stars are generally considered cold because their density far exceeds their temperature. This is particularly true for old isolated NSs, as the temperature can rise significantly during supernovae or BNS mergers. Neutron stars are currently the only observational probe of cold dense matter, with typical central densities falling within the intermediate regime between hadronic and quark matter.

In this thesis, the focus is on the zero-temperature EoSs in $\beta$-equilibrium, meaning that weak processes are balanced and have reached equilibrium. Neutron stars are transparent to neutrinos, and at zero temperature, flavor equilibration is achieved through two Urca processes (named by George Gamow and Mário Schenberg while visiting a casino called Cassino da Urca \cite{PhysRev.59.539}):
\begin{equation}
n \rightarrow p + e^- + \bar{\nu}_e, \quad
p + e^- \rightarrow n + \nu_e.
\end{equation}
These reactions establish the condition for $\beta$-equilibrium in nuclear matter:
\begin{equation}
\mu_n = \mu_p + \mu_e,
\end{equation}
where $\mu_n, \mu_p,$ and $\mu_e$ are the chemical potentials of neutrons, protons, and electrons, respectively. The local charge neutrality of NSs requires equal densities of protons and electrons, i.e., $n_{p} = n_{e}$. 

In terms of quark degrees of freedom, the condition for $\beta$-equilibrium can be written as $\mu_d=\mu_u+\mu_e$. When strange quarks are present, their contribution to charge neutrality can fully balance the quark charges. In the case of massless three-flavor quark matter (for justification, see later), the conditions for $\beta$-equilibrium, along with charge neutrality, can be expressed as:
\begin{equation}
\mu_u = \mu_d = \mu_s = \mu/3 = \mu_q,
\end{equation}
where the subscripts $u, d, s$ correspond to up, down, and strange quarks. The baryon chemical potential and quark chemical potential are denoted as $\mu$ and $\mu_q$, respectively.

Next, I summarize the current theoretical understanding of the EoS from \textit{ab initio} calculations. In principle, the behavior of strongly interacting matter is governed by the Lagrangian of QCD. In practice, calculations can only be performed with certain approximations. Several first-principles approaches derive the EoS from the QCD Lagrangian, including perturbative methods and lattice QCD. Additionally, effective field theory (EFT) provides a useful framework for systematically approximating the EoS.

Performing numerical calculations in lattice QCD for cold, ultradense matter is particularly challenging due to the sign problem (see e.g., \cite{NAGATA2022103991}). Lattice QCD relies on statistical methods, such as Monte Carlo sampling, to evaluate the partition function in the Euclidean path integral formalism. However, at finite chemical potential $\mu$, the standard Monte Carlo techniques for lattice QCD simulations does not work, as the Dirac determinant becomes complex.

\begin{figure}[ht!]
    \centering
\includegraphics[width=0.9\textwidth]{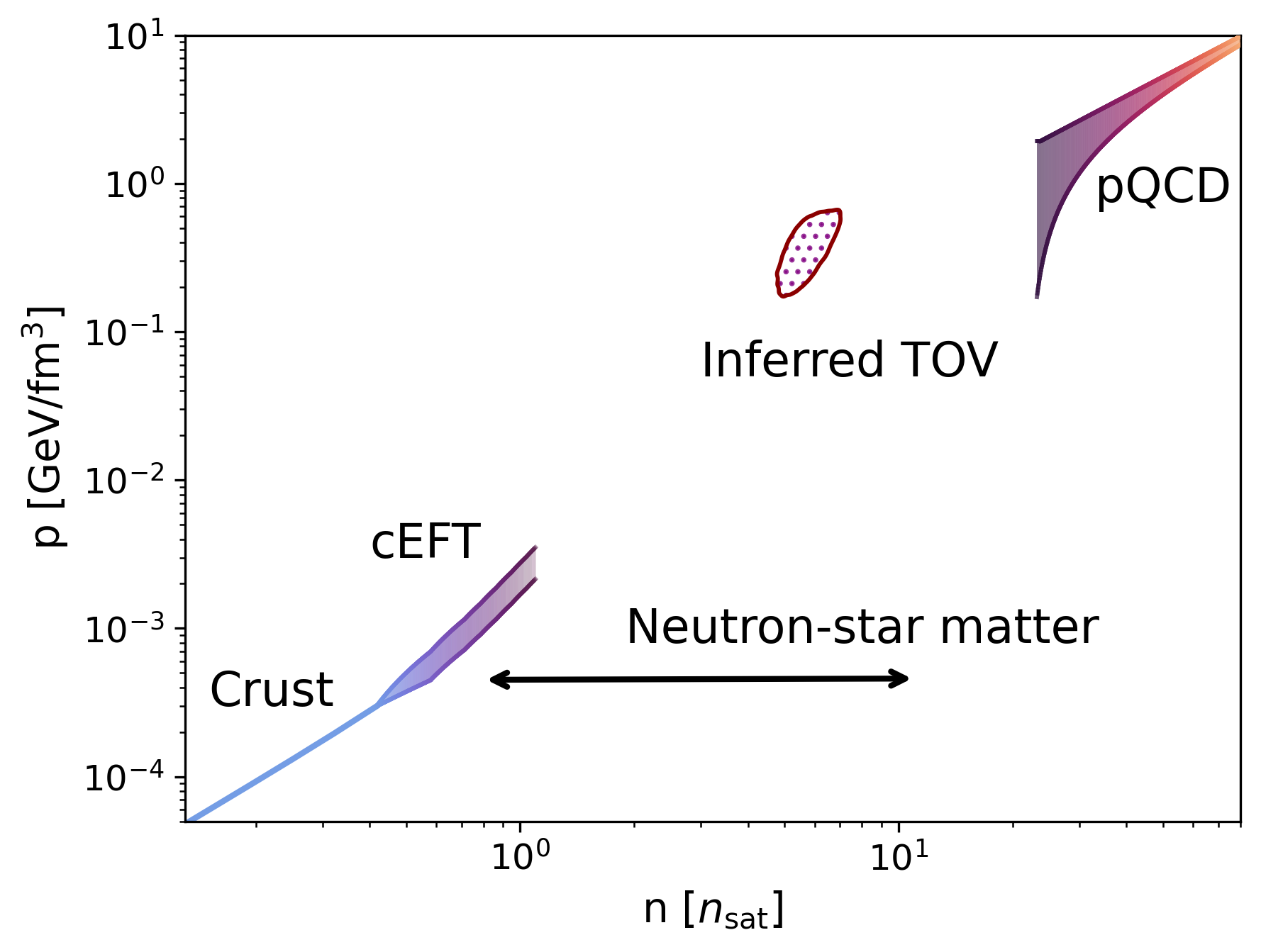}
    \caption{The summary of current theoretical inputs to the EoS of cold dense matter. It includes the crust EoS from the BPS model \cite{1971ApJ...170..299B}, cEFT calculations \cite{Hebeler:2013nza}, and the pQCD limit \cite{Kurkela:2009gj,Gorda:2021znl}. The inferred TOV region corresponds to maximal pressures and energy densities reached in NSs, with the details of the inference provided in the subsequent sections.}
    \label{fig:theory_inputs}
\end{figure}

However, EFT and perturbative methods remain accurate within their respective limits of applicability, as illustrated in \cref{fig:theory_inputs}.  The low-density regime, corresponding to hadronic matter, is constrained by Chiral Effective Field Theory (cEFT), while perturbative Quantum Chromodynamics (pQCD) calculations at high densities inform us about the behavior of quark matter. The details of both calculations are presented below. 

At small densities shown in the figure, $n\lesssim 0.5\ns$, where $\ns = 0.16/\text{fm}^3$ is the nuclear saturation density, EoS of the outer crust of a neutron star follows the BPS model \cite{1971ApJ...170..299B}, named after its authors: Baym, Pethick, and Sutherland. This calculation accounts for measured nuclear masses, the electron degeneracy pressure from a relativistic Fermi gas, and the Coulomb lattice structure of atomic nuclei.

Around nuclear saturation density, quarks are confined within hadrons, such as baryons and mesons. Nuclear interactions can be effectively described using these degrees of freedom rather than quarks and gluons, employing a Lagrangian consistent with the approximate chiral symmetry of QCD. The development of cEFT has started by Weinberg’s pioneering work \cite{1990PhLB..251..288W,1991NuPhB.363....3W} and has provided a systematic framework for expanding nuclear forces at low momenta. In this approach, nucleons interact through pion exchanges and short-range interactions, with parameters constrained by two- and few-body observables \cite{2009RvMP...81.1773E}. State-of-the-art cEFT calculations provides constraints on the EoS up to $[1.1-2]\ns$ \cite{Hebeler:2013nza,Drischler:2017wtt,Drischler:2020hwi,Tews:2018kmu, PhysRevLett.130.072701}. The cEFT band in \cref{fig:theory_inputs} corresponds to the “soft” and “stiff” EoS from \cite{Hebeler:2013nza}.

Perturbative QCD calculations are possible only at asymptotically high densities, where the QCD coupling $\alpha_s$ is small. This occurs when the baryon chemical potential $\mu$ significantly exceeds the QCD energy scale, i.e., $\mu \gg \Lambda_{\text{QCD}}$. \change{Note that the perturbative expansion is an asymptotic series, not convergent.} The expansion of the pressure in terms of $\alpha_s$ can be expressed as: 
\begin{align}
\label{eq:expansion}
p = p_{\text{FD}} + \alpha_s p_1 + \alpha_s^2 p_2 + \alpha_s^3 p_3+...
\end{align}
where $p_{\text{FD}}$ is the pressure of a free Fermi gas of quarks
while the other terms are interaction corrections. Starting from $p_2$, the dependence on $\log(\alpha_s)$ appears in the coefficients.

The relevant results of pQCD calculations, shown in \cref{fig:theory_inputs}, are constrained to zero-temperature matter composed of three flavors of massless quarks in $\beta$-equilibrium. This is well justified, as the chemical potential where the perturbative expansion is valid is much larger than the up, down, and strange quark masses, while the relevant region, as discussed later, remains below the charm threshold.

The details of the calculation and the current state-of-the-art result for the QCD grand canonical potential, which is computed at partial next-to-next-to-next-to-leading order (N$^3$LO$^*$), are presented in \cite{Kurkela:2009gj,Gorda:2021znl}. An asterisk in N$^{3}$LO$^{*}$ indicates that it is not a fully computed order. While N$^{2}$LO is fully computed, at N$^{3}$LO, only the soft contribution\footnote{\change{The contribution arises from the interactions among long-wavelength, dynamically screened gluonic fields \cite{Gorda:2021znl}.}} to the pressure is known and included.

The resulting pQCD pressure depends on the chemical potential $\mu$ and the renormalization scale $\bar{\Lambda}$, which is related to a dimensionless parameter:
\begin{equation}
\label{eq:X}
X = \frac{3\bar{\Lambda}}{2\mu}.
\end{equation}

A conventional approach for estimating theoretical uncertainties involves varying the renormalization scale by a factor of 2, resulting in $X \in [1/2,2]$. The dependence on the logarithm of $\bar{\Lambda}$ naturally arises with $2\mu/3$, motivating the choice of the central scale $X = 1$ to ensure the cancellation of logarithmic dependence in the perturbative expansion presented in \cref{eq:expansion}.

The pQCD results are most relevant for neutron star physics at the lowest density where perturbative uncertainties remain under control. A conventional choice for the lowest chemical potential is set at $\mu \geq \mu_{\mathrm{pQCD}} = 2.6$ GeV as in \cite{Kurkela:2014vha}, corresponding to number densities $n_{\mathrm{pQCD}} \gtrsim 40n_\mathrm{sat}$. This choice ensures a consistent uncertainty estimation, roughly matching the relative uncertainties of cEFT at $1.1\ns$. Both the estimation of theoretical uncertainties and the reference density at which the pQCD pressure is used are explored in detail in \cref{sec:uncertainty}. 

At these densities, Cooper pairs form due to attractive QCD interaction between quark pairs. This leads to a color-superconducting phase \cite{PhysRevLett.81.53,RevModPhys.80.1455,BARROIS1977390,BAILIN1984325,ALFORD1998247}, where the EoS receives a nonperturbative contribution of order $\mathcal{O}(\Delta^2 \mu^2)$. Here, $\Delta$ represents the color superconducting gap, an energy gap that forms at the Fermi surface of quarks. In this thesis, these contributions are neglected because their effect is suppressed relative to the leading-order pressure, which scales as $\mathcal{O}(\mu^4)$. Various models estimate the gap at densities that are not asymptotically large to be in the range of 50–150 MeV \cite{PhysRevLett.81.53,ALFORD1998247,RevModPhys.80.1455,ALFORD1999443,doi:10.1142/9789812810458_0043,Baym_2018,PhysRevD.105.036003,PhysRevLett.125.142502,BERGES1999215,PhysRevD.60.016004}.

First-principle theoretical calculations at intermediate densities between cEFT and pQCD limit are unavailable. Therefore, to explore neutron star physics and predict the mass-radius relation, it is necessary to model the EoS in the density range above the cEFT and up to the TOV density — the maximum density of a stable NS. One approach is to construct phenomenological models, which incorporate specific assumptions about the underlying physics. While numerous such models exist, their predictions vary significantly depending on the assumptions and model parameters (e.g., see the CompOSE database of NS models \cite{Typel:2013rza}).

An alternative method, which is utilized throughout this thesis and detailed in \cref{sec:bayesian}, is the model-agnostic generation of EoSs \cite{Kurkela:2014vha, PhysRevLett.120.261103, PhysRevLett.120.172703,PhysRevC.98.045804,Landry:2020vaw,Capano2020,Annala:2019puf,Dietrich:2020efo,PhysRevLett.126.061101,Raaijmakers_2021,Altiparmak_2022,Koehn:2024set,Huth2022,Hebeler:2013nza,PhysRevD.110.034035,Raaijmakers:2019dks,Miller:2019nzo,Lim:2022fap,Miller:2021qha,Landry:2018prl}. A large variety of different EoSs is generated to probe the physics of neutron star cores. The inference of a realistic EoS then involves constraining the generated EoSs with current astrophysical and theoretical inputs, excluding EoS that are incompatible with the data. The focus of this thesis is the inference of the EoS of neutron-star matter, with the main objective summarized in the next section.

\section{Thesis objective}

It is essential to incorporate all possible inputs when inferring the EoS to study the physics of neutron stars. However, prior to my PhD, QCD input was largely overlooked, except for a few groups that attempted to interpolate across two orders of magnitude in energy density and pressure between the cEFT and pQCD limits \cite{Kurkela:2014vha,Annala:2019puf,PhysRevLett.120.261103,Altiparmak_2022,PhysRevLett.120.172703,PhysRevX.12.011058}. The majority of studies instead anchored different interpolation functions to the low-density limit only, as neutron stars collapse at densities much lower than $40\ns$, where pQCD calculations become reliable. It was unclear whether pQCD provides nontrivial information about the EoS of cold dense matter.

The difference between works that interpolate all the way up to the pQCD limit and those that do not was evident (e.g., see Fig. 3 of \cite{PhysRevC.107.025801}). The key difference is in the softening of the EoS, characterized by a reduction in the speed of sound at the highest densities reached in neutron stars. This observation required further study and became the basis question that initially motivated this research.

\myboxx{white!10}{
The objective of this thesis is to explore how pQCD calculations impact the inference of the EoS of neutron-star matter.}

My collaborators and I developed a framework that utilizes thermodynamic relations to impose robust global constraints on the EoS. This allows us to propagate pQCD constraints from asymptotically high densities to the densities reached in NSs. We explicitly demonstrated the impact of the novel QCD input on the inference of the EoS of cold ultradense matter. Our findings suggest that the QCD input is crucial for understanding the physics of the cores of the most massive neutron stars.

The thesis consists of three main chapters. The first chapter is an introduction, which you are currently finishing. The second chapter explores the role of high-density calculations in EoS inference, addressing why pQCD provides nontrivial constraints for the EoS and demonstrating the impact of the QCD input. The final chapter examines what the QCD input can reveal about neutron star cores, including the potential existence of quark matter and the possibility of a first-order phase transition. 
While it remains a fundamental open question whether the phase transition occurs within the density range reached in the cores of the most massive neutron stars, I quantify the probability and find strong evidence supporting such phase change.
}


%
\mainfont{\chapter{The utility of pQCD for neutron stars} 
\label{chpt:QCD_and_NS}

In this chapter, I explore how perturbative QCD can be used to impose robust constraints on the EoS of neutron stars. First, an analytical method is derived in \cref{sec:analytic} to propagate constraints from asymptotically high densities to lower densities \cite{Komoltsev:2021jzg}. In \cref{sec:bayesian}, this method is used in Bayesian inference of the EoS of neutron-star matter, explicitly showing the impact of the QCD input \cite{Gorda:2022jvk}. The following two sections address potential caveats of this new input. \Cref{sec:uncertainty} focuses on analyzing the impact of theoretical uncertainties in QCD calculations on Bayesian inference  \cite{Gorda:2023usm}, arising from missing higher-order terms, the unphysical renormalization scale parameter $X$, and the choice of the reference density where pQCD calculations are used. Finally, \cref{sec:termination} examines how the choice of termination density, up to which the EoS is modeled, affects Bayesian inference \cite{Komoltsev:2023zor}.

\section{Analytical derivation}
\label{sec:analytic}

In this section, I demonstrate how assuming full thermodynamic potential for the low- and high-density limits, which provides the triplets of values $\{\mu_{\rm low}, n_{\rm low}, p_{\rm low}\}$ and $\{\mu_{\rm high}, n_{\rm high}, p_{\rm high}\}$, introduces global constraints on the EoS between these two points. The subscripts "low" and "high" correspond to the cEFT and pQCD limits but are used to emphasize the generality of the construction. This construction explicitly shows how information arising from pQCD calculations can propagate to neutron-star densities.

These constraints are analytic and independent of any specific interpolation function. They arise from the fundamental requirement that the EoS remains stable, causal, and consistent between the low- and high-density limits. 

These requirements and the resulting constraints on the $\mu-n$ plane are discussed in \cref{subsec:Scc}. In \cref{subsec:munp,subsec:map_to_ep}, these constraints are first extended to the three-dimensional $\mu-n-p$ space and then mapped to the $\e-p$ plane. Finally, in \cref{subsec:simpler_check} it is shown how to apply a simpler yet equivalent check against the derived constraints for the modeled EoS, which is known up to some termination density.

\subsection{Stability, causality and consistency}
\label{subsec:Scc}
While the only input required for the TOV equation for the hydrodynamic description of neutron-star matter is the EoS in the form of the pressure as a function of the energy density $p(\e)$, the complete information about the EoS is available through the thermodynamic potential. At zero temperature and finite chemical potential, in $\beta$-equilibrium, the grand canonical potential is given by $\Omega(\mu)=-p(\mu)$. Knowing the full thermodynamic potential allows access to various thermodynamic quantities, such as the pressure $p$, the chemical potential $\mu$, the number density $n=\partial_{\mu}p(\mu)$ and the energy density $\e$ calculated using \begin{align}
\label{eq:Euler}
    \e=-p+\mu n.
\end{align}

The easiest way to derive constraints is to start with the number density $n$ as a function of the chemical potential $\mu$, as presented in \cref{fig:mu_n}. The triplet $\{\mu_{\rm high},n_{\rm high},p_{\rm high}\}$ is provided by pQCD calculation for the central scale $X=1$ and $\mu_{\rm high}=\mu_{\rm QCD}=2.6$ GeV. It is represented by the purple line in the upper right corner of \cref{fig:mu_n}. A systematic discussion of uncertainty estimation related to the choice of scale $X$ and $\mu_{\rm QCD}$ is provided in \cref{sec:uncertainty}. The low-density limit $\{\mu_{\rm low},n_{\rm low},p_{\rm low}\}$, obtained using cEFT at $1.1\ns$ (corresponding to $\mu_{\rm low}\approx0.978$ GeV), is represented by the dark blue line in the bottom left corner of \cref{fig:mu_n} (corresponding to ”stiff” EoS from \cite{Hebeler:2013nza}). 

In principle, the EoS in $\beta$-equilibrated matter at zero temperature is a single line on this plane, which is unknown. Between these two limits, the only available theoretical information is that the EoS must be thermodynamically stable, causal, and able to connect the two endpoints. Consequently, not all possible interpolations between these limits will result in a valid and consistent EoS. These requirements can be summarized as follows. 
\begin{itemize}
    \item \textbf{Stability.} The grand canonical potential must be a concave function of the chemical potential.  Consequently, the number density should be a monotonically increasing single-valued function, $\partial_{\mu}n=\partial^2_{\mu}(-\Omega(\mu))\ge0$. Therefore, any lines in \cref{fig:mu_n} must be monotonic function to represent a valid EoS.
    \item \textbf{Causality.} Speed of sound cannot be bigger than the speed of light: 
    \begin{align}
    \label{eq:cs2}
        c^{-2}_s=\frac{\mu}{n}\frac{\partial n}{\partial\mu}\ge1.
    \end{align}
    This imposes a minimal slope on the number density $\partial_{\mu}n(\mu)\ge n/\mu$ for a fixed point on the $\mu-n$ plane. 
    \item \textbf{Consistency.} The EoS must simultaneous connect $n,\mu$ and $p$ of the two limits ($\e$ follows according to \cref{eq:Euler}). Since the pressure is given by $p=\int n(\mu) d\mu $, this requirement fixes the area under the curve $n(\mu)$ between two limits:
    \begin{align}
        \Delta p=p_{\h}-p_{\low}=\int^{\mu_{\h}}_{\mu_{\low}} n(\mu) d\mu 
    \end{align}
\end{itemize}

\begin{figure}[ht!]
    \centering
\includegraphics[width=0.75\textwidth]{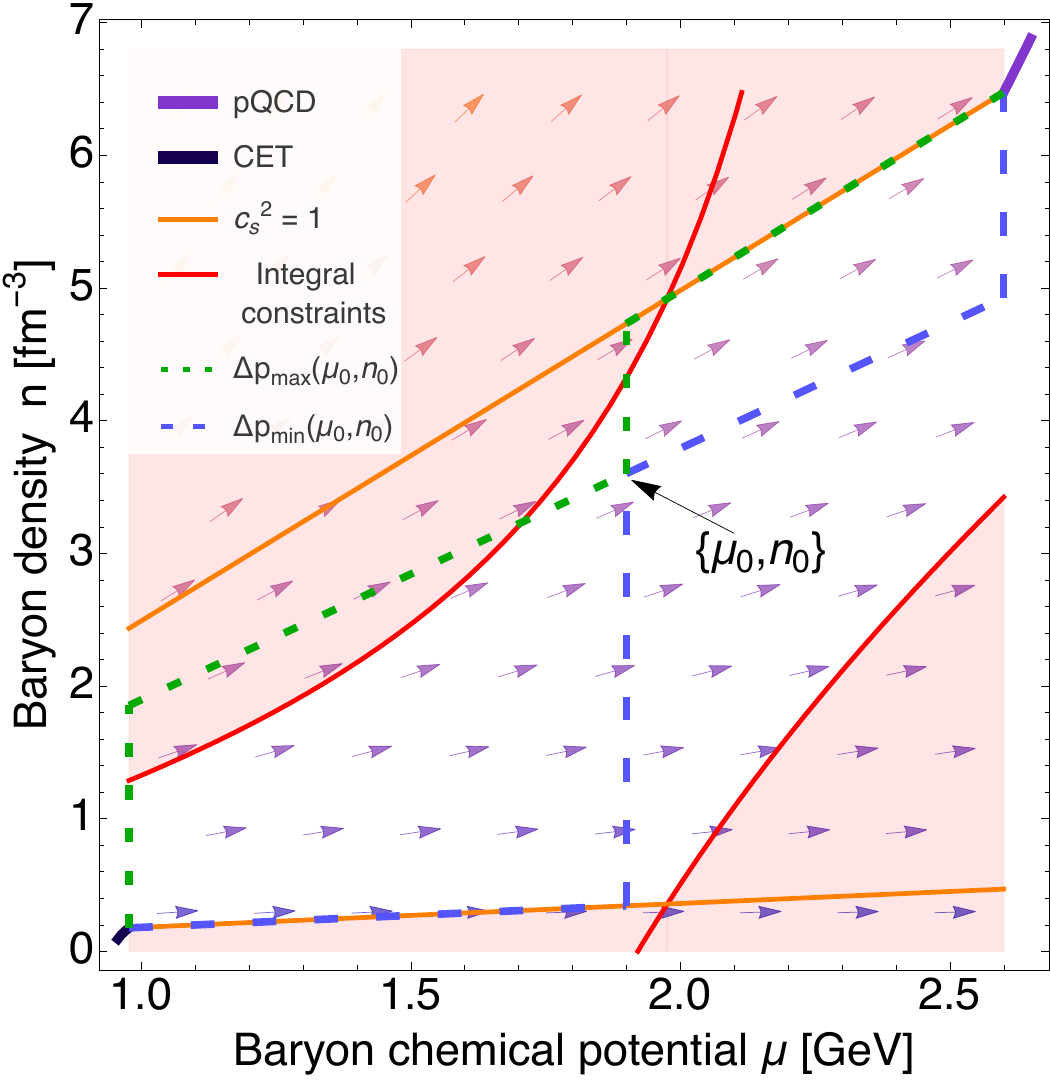}
    \caption{Baryon number density as a function of baryon chemical potential. Arrows indicate the minimal allowed slope dictated by causality, $\partial_{\mu}n(\mu) \geq n/\mu$. The red regions are excluded due to the simultaneous requirements of stability, causality, and consistency. The constructions of $\Delta p_{\rm min/max}$ for an arbitrary point ${\mu_0, n_0}$ are defined in \cref{eq:n1,eq:n2}}
    \label{fig:mu_n}
\end{figure}

Global constraints can be derived only through the interplay between different requirements. First, consider only stability and causality. The minimal slope $\partial_{\mu}n(\mu) \ge n/\mu$ can be visualized as a vector field on the $\mu-n$ plane, depicted by the arrows in \cref{fig:mu_n}. Each arrow represents the slope corresponding to a constant $c_s^2=1$. A causal EoS cannot have a slope smaller than an arrow for any given point on the plane. Solving \cref{eq:cs2} with the initial condition $\{\mu_\low, n_\low\}$ and $c^2_s=1$ results in a straight line $n(\mu)=n_\low \mu/\mu_\low$, shown as the bottom orange line in \cref{fig:mu_n}. This line indicates the excluded area (shown in red) that any stable and causal EoS cannot reach if started from the low-density limit. Similarly, the upper excluded region can be derived by starting from $\{\mu_\h, n_\h\}$ and following the arrows backwards. This results in $n(\mu)=n_\h \mu/\mu_\h$, shown as the upper orange line. These bounds were established by Rhoades and Ruffini in \cite{PhysRevLett.32.324}, and are typically represented as a rhomboid on $p-\e$ plane, corresponding to a first-order phase transition (FOPT) with a discontinuous jump in energy density $\e$ and a segment where $c_s^2 = \frac{dp}{d\epsilon} = 1$.

To utilize the consistency requirement, it is necessary to determine the absolute maximum and minimum pressure differences between limits that an EoS can have when passing through a fixed point $\{\mu_0, n_0\}$. If the maximum pressure difference $\Delta p_{\rm max}$ for any EoS passing through the point is smaller than $\Delta p = p_{\rm high}-p_{\rm low}$ (fixed by low- and high-density limits), such a point would be ruled out, as no causal and stable EoS can simultaneously connect $\mu$, $n$, and $p$. Similarly, if the minimum pressure difference $\Delta p_{\rm min}$ is greater than $\Delta p$, the fixed point would be excluded.

For a fixed point $\{\mu_0, n_0\}$, the minimum area between the limits can be constructed as follows. For $\mu_\low < \mu < \mu_0$, the EoS must follow the causal line with $\cs=1$. At $\mu_0$, the EoS undergoes FOPT, where the density jumps to $n_0$. To connect $\{\mu_0, n_0\}$ to the high-density limit while minimizing the area under the curve, the EoS should follow the maximally stiff $\cs=1$ line from $\mu_0$ up to $\mu_\h$, where it exhibits another FOPT to $n_\h$. This construction, shown as a blue-dashed line in \cref{fig:mu_n}, can be expressed as:
\begin{align}
\label{eq:n1}
    n(\muB) = 
    \left\{
    \begin{array}{rl}
    n_\low \mu/\mu_\low, & \muC < \mu < \mu_0 \\
    n_0 \mu/\mu_0, & \mu_0 < \mu < \mu_\h.
    \end{array}
    \right. 
\end{align}
It is evident from the figure that any deviation from this construction would either violate causality or increase the area under the curve.

Finding where the integral of the above-described construction equals $\Delta p$ gives the following equation:
\begin{align}
\label{eq:deltapmin_0}
    \Delta p_{\rm min}(\mu_0,n_0) = \int^{\mu_0}_{\mu_\low} \frac{n_\low}{\mu_\low}\mu d\mu +  \int^{\mu_\h}_{\mu_0} \frac{n_0}{\mu_0}\mu d\mu = \Delta p
\end{align}
The solution of this equation with respect to $n_0(\mu_0)$ provides the integral constraints, shown in \cref{fig:mu_n} as the top red line. Any EoS appearing above this line would violate consistency, as $\Delta p_{\rm min} > \Delta p$. The upper integral constraints, along with the causal line, yield a maximum number density,
\begin{align}
\label{eq:nmax}
    n_{\rm max}(\muB) = 
    \left\{
    \begin{array}{ll}
     \frac{\muB^3 \nC - 
  \muC \muB(\muC \nC + 2 \Delta p)}{(\muB^2 - \muQ^2) \muC },
    &\muC \leq \muB < \muc \\
   \nQ \muB/\muQ, & \muc \leq \muB \leq \muQ,
    \end{array}
    \right. 
\end{align}
where $\muc$ is determined by the intersection of the causal line and the integral constraints,
\begin{equation}
\label{eq:muc}
\muc = \sqrt{\frac{\muC \muQ ( \muQ \nQ-  \muC\nC-2 \Delta p )}{\muC \nQ - \muQ \nC} }.
\end{equation}

Analogously, the construction that maximizes the area under the curve passing through $\{\mu_0, n_0\}$ is shown as the green dashed line in \cref{fig:mu_n}, expressed as:
\begin{align}
\label{eq:n2}
    n(\muB) = 
    \left\{
    \begin{array}{rl}
    n_0 \muB/\mu_0, & \muC < \muB < \mu_0 \\
    \nQ \muB/\muQ, & \mu_0 < \muB < \muQ.
    \end{array}
    \right.
\end{align}

The solution of $\Delta p_{\rm max} = \Delta p$, where $\Delta p_{\rm max}$ is given by
\begin{align}
\label{eq:deltapmax_0}
    \Delta p_{\rm max}(\mu_0,n_0) = \int^{\mu_0}_{\mu_\low} \frac{n_0}{\mu_0}\mu d\mu +  \int^{\mu_\h}_{\mu_0} \frac{n_\h}{\mu_\h}\mu d\mu = \Delta p,
\end{align}
provides the lower integral constraints, shown as the bottom red line in \cref{fig:mu_n}. Any EoS appearing below this line would violate consistency, as $\Delta p_{\rm max} < \Delta p$. The minimal density is given by:
\begin{align}
\label{eq:nmin}
    n_{\rm min}(\muB) = 
    \left\{
    \begin{array}{ll}
    \nC \muB/\muC, & \muC \leq \muB \leq \muc \\
   \frac{\muB^3 \nQ - 
 \muB \muQ (\muQ \nQ - 2 \Delta p)}{(\muB^2 - \muC^2) \muQ }, & \muc < \muB \leq \muQ,
    \end{array}
    \right.
\end{align}
where $\muc$ is given by \cref{eq:muc}.

The integral constraints can be intuitively understood as follows: If an EoS passes through the lower right corner (excluded by the integral constraints), it does not reach the correct area under the curve, regardless of its behavior at higher densities, as it is too small. Similarly, if an EoS passes through the upper left corner, it necessarily overshoots the correct area under the curve $\Delta p$ when it reaches $\mu_\h$. 

The global constraints on the $\mu-n$ plane are defined by $n_{\rm min}(\muB)$ and $n_{\rm max}(\muB)$. Note that these lines do not represent valid EoSs by themselves; rather, they are constructed from an infinite number of the most extreme EoSs. Since $\Delta p_{\rm min/max}$ are absolute bounds, the constraints do not place any limitation on the behavior of interpolation functions. There is no assumption on the number and strength of the FOPT. The only requirement is that the EoS must be stable, consistent, and causal. 

\subsection{Constraints in the $\mu-n-p$ space}
\label{subsec:munp}

To map the derived constraints onto the $\e$–$p$ plane, it is first necessary to extend them in the three-dimensional $\mu$–$n$–$p$ space. This can be achieved as follows. For every fixed allowed $\{\mu_0, n_0\}$, it is possible to determine the minimal and maximal pressure $p_{\rm min/max} (\mu_0, n_0)$ that a valid EoS can have at this point. Note that it is different from the construction of $\Delta p_{\rm min/max}$ described in the previous subsection, as it does not include consistency. For instance, consider the construction of $\Delta p_{\rm max}$ in \cref{fig:mu_n}. Although it is evident that $\{\mu_0, n_0\}$ is an allowed point, the green dashed line crosses the upper integral constraints, indicating a violation of consistency.

To find the minimal pressure for a given point $\{\mu_0, n_0\}$, consider two regions of the $\mu-n$ plane: $[\mu_\low, \muc]$ and $[\muc, \mu_\h]$. For the first region, it is clear from the figure that the minimal pressure corresponds to the maximally stiff causal $\cs=1$ line with a FOPT at $\mu_0$, where $n$ jumps from $n_\low \mu_0 / \mu_\low$ to $n_0$. However, for $\mu_0 > \muc$, this would violate the lower integral constraint.

For a fixed $\Delta p$, the EoS that maximizes the area between $\mu_0$ and $\mu_\h$ is the same EoS that minimizes area between $\mu_\low$ and $\mu_0$. The EoS that maximizes the area above $\mu_0>\muc$, as seen from \cref{fig:mu_n}, has a FOPT at $\mu_0$ and follows the maximally stiff EoS $\cs=1$ up to $\mu_\h$. Such an EoS, by construction, can only take one form between $\mu_\low$ and $\mu_0$ to satisfy the lower integral constraints, as it is the EoS that renders the lower integral constraints when $\Delta p_{max}(\mu_0,n_0)=\Delta p$. It starts with a FOPT at $\mu_\low$ and follows the causal line $n(\mu) = n_{\rm min}(\mu_0) \mu / \mu_0$ up to $\mu_0$.  At the point of intersection between the causal line and the lower integral constraints $n_{\rm min}(\mu_0)$, the EoS exhibits a FOPT, passing through $n_0$ and switching to the EoS that maximizes the area above $\mu_0$, as explained earlier.

The lower bound on the pressure for both cases can be written as follows: 
\begin{align}
\label{eq:pmin1}
    p_{\rm min}(\mu_0,n_0) &= p_\low + \int^{\mu_0}_{\mu_\low} n_{\rm min}(\mu_0) \frac{\mu}{\mu_0} d\mu \nonumber \\
    &= \pC + \frac{\mu_0^2-\muC^2}{2 \mu_0}n_{\rm min}(\mu_0) 
\end{align}

To find the maximal pressure for a given point $\{\mu_0, n_0\}$, the $\mu-n$ plane needs to be divided into two regions by the EoS with a constant $\cs=1$ between $\mu_\low$ and $\mu_\h$. This particular EoS connects the lowest point of the upper integral constraint with the highest point of the lower integral constraint, and can be expressed as:

\begin{equation}
\label{eq:nc}
n_c(\muB) = n_{\rm max}(\muC) \muB / \muC = n_{\rm min}(\muQ) \muB / \muQ.
\end{equation}

For number densities below this line $n < n_c(\mu)$, the maximal pressure is trivially obtained by a FOPT to the maximally stiff line $n(\mu) = n_0 \mu / \mu_0$, which leads to the following bound on the pressure:

\begin{align}
p_{\rm max}(\mu_0, n_0) 
&= p_\low + \frac{\mu_0^2 - \mu_\low^2}{2 \mu_0} n_0, \quad n < n_c(\muB).
\label{eq:pmax1}
\end{align}

However, for $n > n_c(\mu)$, this construction would lead to a FOPT at $\mu_\low$, intersecting the upper integral constraint. The same trick used for $p_{\rm min}$ can be applied here. The EoS that minimizes the area between $\mu_0$ and $\mu_\h$ provides an EoS that maximizes the area in the region of interest, namely $[\mu_\low, \mu_0]$. For any allowed point with $n > n_c(\mu)$, it is evident that the minimal area between $\mu_0$ and $\mu_\h$ can be achieved by the EoS following $\cs=1$, starting from $\{\mu_0, n_0\}$: $n(\mu) = n_0 \mu / \mu_0$. Thus, the maximal pressure for a given point can be found as the difference between $\Delta p$ and the minimal area between $\mu_0$ and $\mu_\h$.
\begin{align}
    p_{\rm max}(\mu_0,n_0)
    &= \Delta p - p_\low - \int^{\mu_\h}_{\mu_0} n_0 \frac{\mu}{\mu_0} d\mu \nonumber \\
    &= \pQ - \frac{\muQ^2-\mu_0^2}{2 \mu_0}n_0,\quad n > n_c(\muB).
    \label{eq:pmax2}
\end{align}

\begin{figure}[ht!]
    \centering
\includegraphics[width=0.9\textwidth]{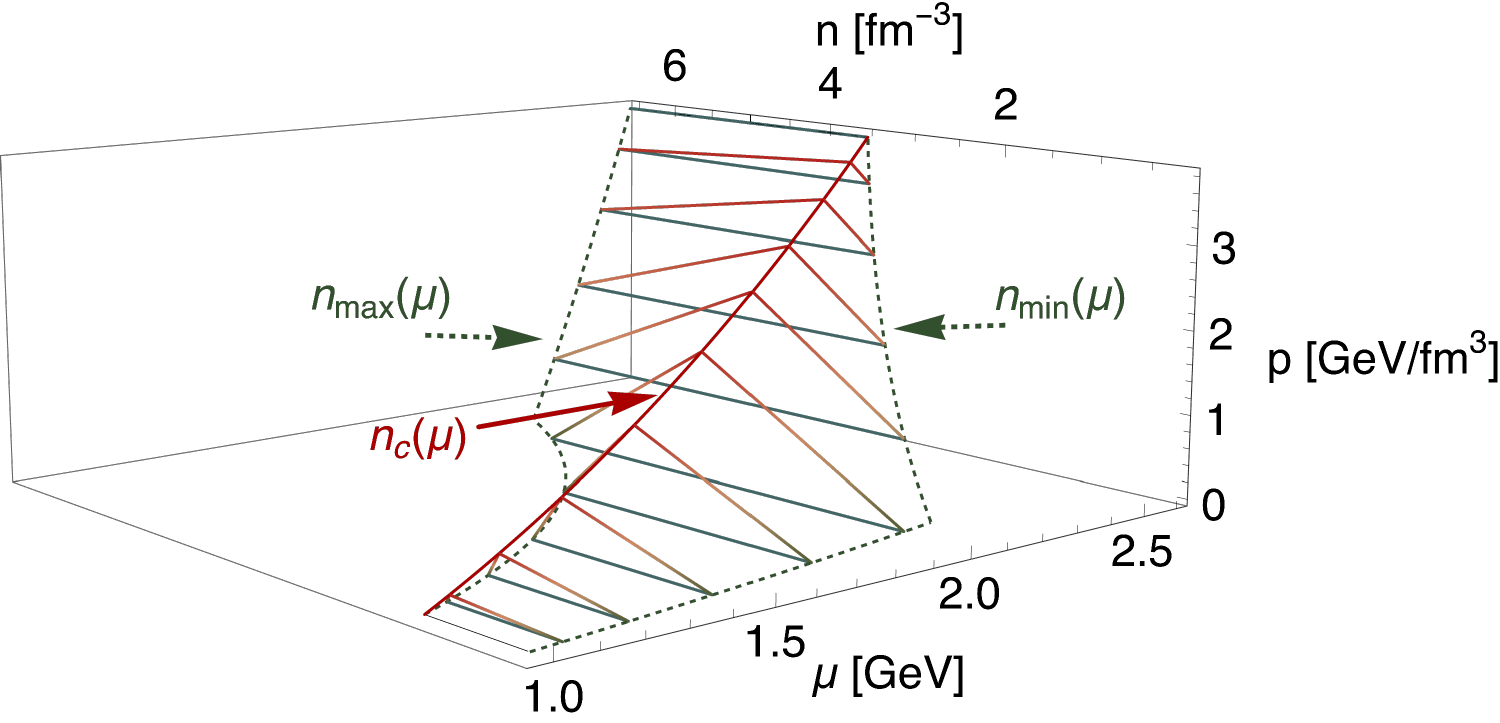}
    \caption{A three-dimensional representation of pQCD constraints in the $\mu$–$n$–$p$ space. \Cref{fig:mu_n} provides a top-view perspective of this structure. Each triangle, with its apex on the $n_c(\mu)$ line, defined in \cref{eq:nc}, represents a slice of the $p$–$n$ constraints for a fixed $\mu$.}
    \label{fig:3D}
\end{figure}

Obtaining $p_{\rm min/max}(\mu,n)$ allows to demonstrate the constraints in the $\mu-n-p$ phase space, as illustrated in \cref{fig:3D}. The $\mu-n$ plane can be recognized from the top view of the plot. Note how, for a slice of fixed $\mu$, the upper bound on the pressure changes behavior when crossing the $n_c$ line.

\subsection{Mapping constraints to $\e-p$ plane}
\label{subsec:map_to_ep}
To utilize the constraints for the hydrodynamic description of neutron-star matter, it is useful to map these constraints from the $\mu-n$ plane to the $\e-p$ plane. This mapping is not straightforward, as there is no one-to-one correspondence between points on the $\mu-n$ plane and points on the $\e-p$ plane. However, for a fixed enthalpy $h = \e + p = \mu n$, there is a correspondence between the diagonal lines $p(\e) = -\e + h$ on the $\e-p$ plane and the hyperbolas $n(\mu) = h / \mu$ on the $\mu-n$ plane. Therefore, determining the minimal and maximal pressure and the corresponding $\mu$ and $n$ along an isenthalpic line, as allowed by the constraints, provides the bounds on the $\e-p$ plane. The corresponding energy density $\e$ can be calculated using \cref{eq:Euler}.

The minimal pressure for a fixed point $\{\mu_0, n_0\}$ is given by \cref{eq:pmin1}. Note that $p_{\rm min}$ does not depend on $n_0$. This is because the EoS can always exhibit an FOPT jumping from $n_{\rm min}(\mu_0)$ to $n_0$. The minimal pressure is a monotonically increasing function of $\mu_0$, as seen from \cref{eq:pmin} or \cref{fig:3D}. Therefore, the smallest pressure is always obtained at the smallest value of $\mu$, which, in the case of an isenthalpic line, corresponds to the intersection of the hyperbola $n(\mu) = h / \mu$ with the upper constraint $n_{\rm max}(\mu)$. For a fixed $h=\e+p$ the minimal pressure corresponds to the maximal energy density. Thus, the lower bound on the $\e-p$ plane is given by $\{ \e_{\rm max}(\mu) , p_{\rm min}(\muB, n_{\rm max}(\muB)) \}$, where
 \begin{align}
 \label{eq:eqmax}
     &\e_{\rm max}(\muB) 
     =  -p_{\rm min}(\muB, n_{\rm max}(\muB)) + \muB n_{\rm max}(\muB) \\
     & =  
     \left\{
     \begin{array}{cc}
     \frac{\left(\muB^2+\muQ^2\right) \left(\muB^2 \nC+\muC(2 \pC-\muC\nC)\right)-4 \muB^2 \muC\pQ}{2 \muC(\muB-\muQ) (\muB+\muQ)}, & \muB < \muc \\
      \frac{1}{2} ((\muB^2 \nQ)/\muQ + \muQ \nQ - 2 \pQ),  & \muB > \muc.
     \end{array}
    \right. \nonumber
 \end{align}
 
The maximal pressure for a fixed point is given by \cref{eq:pmax1} and \cref{eq:pmax2}. In this case $p_{\rm max}(\mu,n)$ depends on both arguments. Substituting $n=h/\mu$ in both equations it can be shown that $p_{\rm max}$ is monotonically increasing function of $\mu$ along the isenthalpic lines. Thus, the largest pressure is obtained by the largest allowed value of $\mu$, which is given by the intersection of the hyperbola $n=h/\mu$ with $n_{\rm min}(\mu)$. Similarly, the upper bound on $\e-p$ plane is the line $\{\e_{\rm min}(\muB), p_{\rm max}(\muB, n_{\rm min}(\mu)) \}$, where
 \begin{align}
 \label{eq:eqmin}
    & \e_{\rm min}(\muB) 
      =  -p_{\rm max}(\muB, n_{\rm min}(\muB)) + \muB n_{\rm min}(\muB) \\
     & = 
     \left\{
     \begin{array}{cc}
    \frac{1}{2} ((\muB^2 \nC)/\muC + \muC \nC - 2 \pC)
     & \muB < \muc \\
     \frac{ 
   \frac{\muB^4}{\muC^2} \frac{\nQ}{\muQ} + 
 (\frac{\muB}{\muC})^2 (\muC^2 \frac{\nQ}{\muQ} - \muQ \nQ + 4 \pC - 2 \pQ) + 2 \pQ - \muQ \nQ 
    }{2 ((\frac{\muB}{\muC})^2 - 1)},  & \muB > \muc.
      \end{array}
    \right. \nonumber
 \end{align}

The constraints on the $\e-p$ plane are shown in \cref{fig:e_p_koku_global} as a green envelope constructed from dashed and solid lines. Similar to the $\mu-n$ plane, some lines arise from causality constraints, while a significant portion of the allowed area is cut by integral constraints. Both \cref{fig:mu_n} and \cref{fig:e_p_koku_global} display the projection of the three-dimensional constraints onto a two-dimensional plane. Therefore, it is possible to see stricter bounds on the $\e-p$ plane when considering a fixed density $n$.

\begin{figure}[ht!]
    \centering
\includegraphics[width=0.9\textwidth]{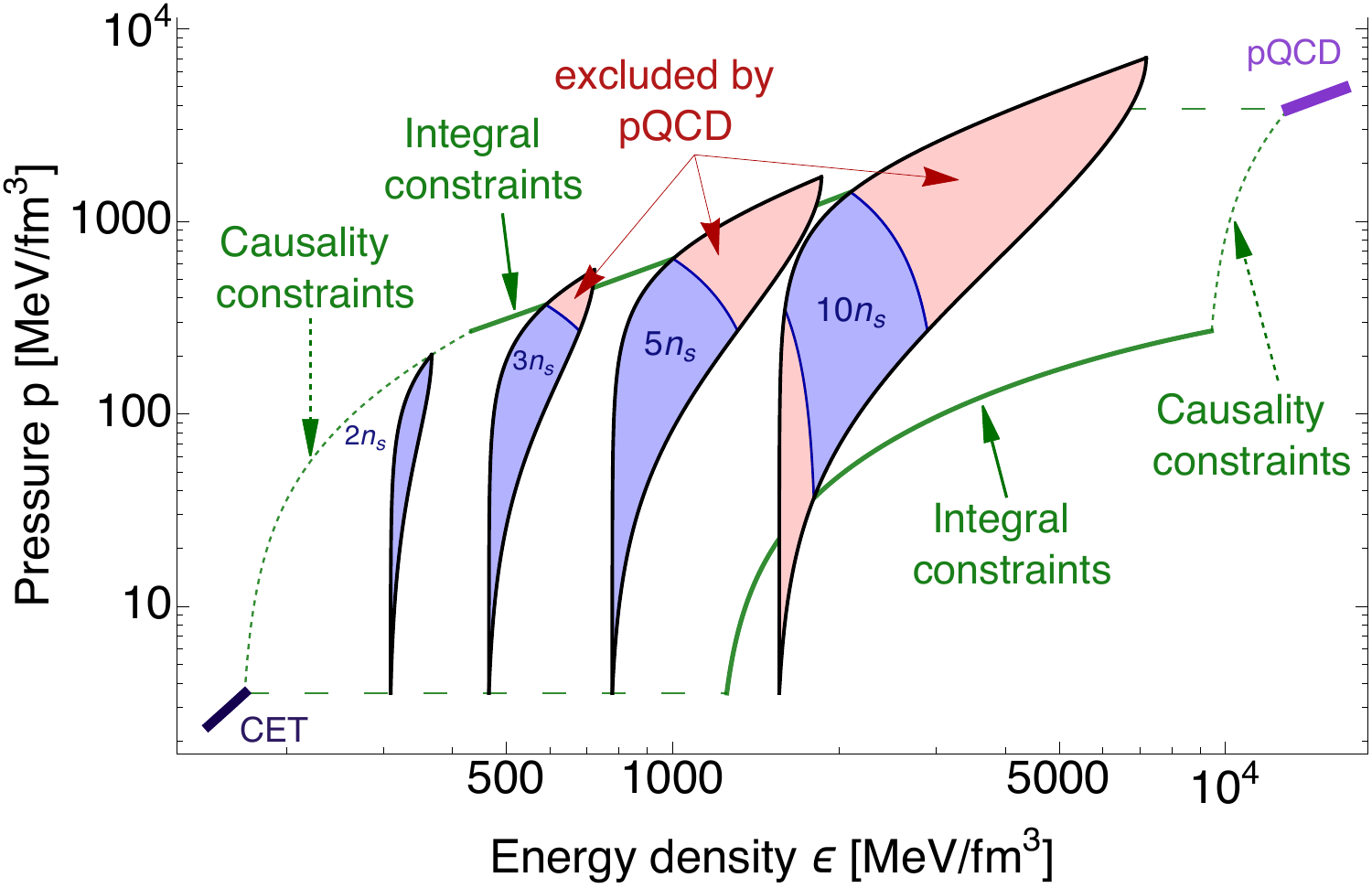}
    \caption{The pQCD constraints mapped onto the energy density–pressure plane. The green envelope corresponds to the causality and integral constraints from \cref{fig:mu_n}. The black-outlined shapes represent the allowed $p$–$\varepsilon$ region when extrapolating a causal EoS from the low-density limit up to fixed densities $n = 2, 3, 5,$ and $10\ns$. The blue regions correspond to the allowed areas where, in addition to stability and causality, consistency with the high-density limit is imposed. The red regions are explicitly excluded by the pQCD limit.}
    \label{fig:e_p_koku_global}
\end{figure}

The black-outlined shapes correspond to the allowed areas for fixed $n=2,3,5$ and $10\ns$ without the high-density limit, arising from the low-density input due to stability and causality. The blue regions represent the allowed areas if the high-density limit is additionally imposed. Thus, the red region, which is the difference between the black shape and the blue area, is explicitly excluded by the high-density limit. Strikingly, already at $5\ns$ high-density input exclude 75\% of otherwise allowed area when plotted in linear scale. For $3\ns$ and $10\ns$, this number is 32\% and 93.5\%, respectively.

\subsection{Simple check of consistency between two limits}
\label{subsec:simpler_check}

In this subsection, a simple binary check is presented to determine whether two limits can be connected by a stable, causal, and consistent EoS. This check is equivalent to verifying whether the envelope of allowed values derived from global constraints, e.g., shown as green lines in \cref{fig:e_p_koku_global}, is non-empty. As evident from the figure, cEFT and pQCD limits can indeed be connected by a valid EoS. However, the derivation is completely general and can be applied to any low- and high-density limits. This turned out to be particularly useful tool to check consistency of modeled EoS with high-density limit. 

If an EoS is modeled up to a certain termination density, $\nterm$, the endpoint of the EoS can then be treated as a new low-density limit. In this case, there is no need to check the EoS against the global constraints; instead, it is possible to verify if the endpoint of the modeled EoS can be connected to pQCD. 

Assume an EoS is known up to a certain termination density $\{\mu_{\rm term}, \nterm,$ $p_{\rm term}\}$, which is now treated as a new low-density limit. To satisfy the integral constraints, the necessary condition is $\Delta p_{\rm min} < \Delta p < \Delta p_{\rm max}$, where $\Delta p = p_\h - p_{\rm term}$. The new bounds $\Delta p_{\rm min/max}$ can be derived from \cref{eq:deltapmin_0} and \cref{eq:deltapmax_0} with the following simplification: $\{\mu_0, n_0\}$ can be substituted with either $\{\mu_\low, n_\low\}$ or $\{\mu_\h, n_\h\}$, which gives the same result by construction. This substitution is valid because the goal is to determine the absolute bounds on the pressure difference between two limits, rather than assuming the EoS passes through a specific fixed point. As a result, one of the integrals always cancels out, leaving the following expressions:

\begin{align}
    \Delta p_{\rm min}(\mu_{\rm term},\nterm) & = \frac{1}{2}
    \left( \frac{\mu_\h^2}{\mu_{\rm term}} - \mu_{\rm term} \right) \nterm,\label{eq:pmin}\\
     \Delta p_{\rm max}(\mu_{\rm term},\nterm) & = \frac{1}{2}
    \left( \mu_\h  - \frac{\mu^2_{\rm term}}{\mu_\h} \right) n_\h.\label{eq:pmax}
\end{align}

which is simply the area under the causal line $\cs=1$, starting from the low-density limit for $\Delta p_{\rm min}$ and the high-density limit for $\Delta p_{\rm max}$. 

Therefore, an arbitrary EoS can be connected to the high-density limit if it satisfies the condition:
\begin{align}
\label{eq:qcd_check}
    \Delta p_{\rm min} < \Delta p < \Delta p_{\rm max}, 
\end{align}

where $\Delta p_{\rm min/max}$ is defined by \cref{eq:pmin,eq:pmax}, and $\Delta p = p_\h - p_{\rm term}$. If the EoS is causal before reaching $\nterm$, then this requirement is fully equivalent to the EoS being within the global constraints shown in \cref{fig:e_p_koku_global}.

\begin{figure}[ht!]
    \centering
\includegraphics[width=0.75\textwidth]{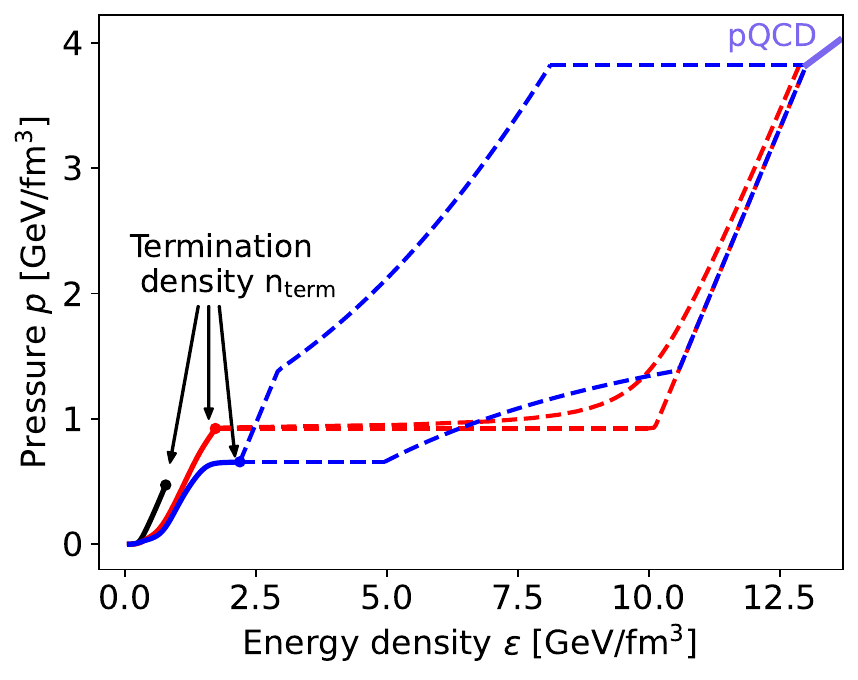}
    \caption{Three representative EoSs modeled up to a termination density, $\nterm$. The black EoS is excluded by pQCD constraints, while the red and blue EoSs are allowed. The dashed line represents the region an EoS must pass through to connect to pQCD while maintaining stability and causality (see main text). The restrictive area outlined by the red dashed line indicates that the red EoS is only marginally allowed.}
    \label{fig:ep_koku_example}
\end{figure}

As an example, in \cref{fig:ep_koku_example}, three different EoS models are represented by the black, red, and blue solid lines. Each is modeled up to a termination density $\nterm$ and checked against the integral constraints using \cref{eq:qcd_check}. The black EoS fails to meet the requirement, indicating that no causal and stable interpolation exists between its termination density and the pQCD limit. In contrast, the red and blue EoS models satisfy \cref{eq:qcd_check}, suggesting the existence of at least one valid EoS between $\nterm$ and $n_\h$.

Note that \cref{eq:pmin,eq:pmax,eq:qcd_check} does not provide the envelope of all possible EoSs between $\nterm$ and $n_{\rm high}$, depicted by the blue and red dashed lines in \cref{fig:ep_koku_example}, but rather provides a binary output indicating whether a stable, causal, and consistent interpolation exists. To determine this envelope, \cref{eq:pmin1,eq:pmax1,eq:pmax2,eq:eqmax,eq:eqmin} must be used, with the low-density limit substituted by the termination point. 

\begin{figure}[ht!]
    \centering
\includegraphics[width=0.85\textwidth]{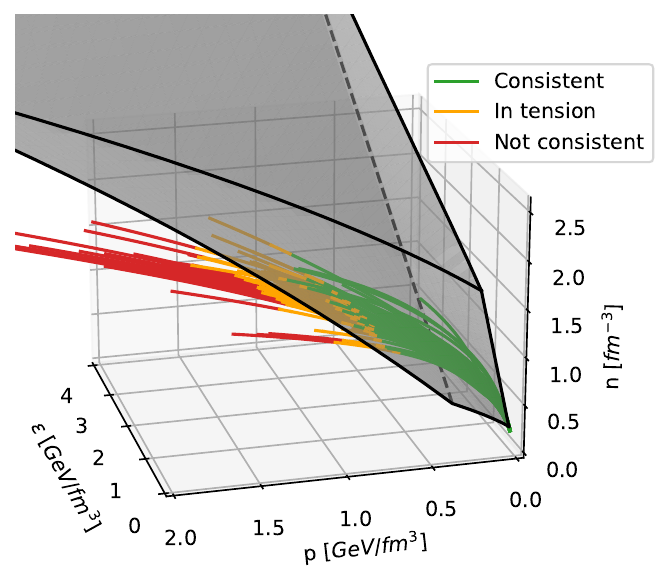}
    \caption{A three-dimensional representation of pQCD constraints in the $\e$–$p$–$n$ space. The allowed region is highlighted in gray. The colored lines represent publicly available EoSs from the CompOSE database \cite{Typel:2013rza}. The categories “consistent,” “in tension,” and “not consistent” correspond to the EoS remains consistent with pQCD input for all values of the renormalization scale parameter $X$, for some values, or for none within the range $[1/2,2]$, respectively.}
    \label{fig:3D_database}
\end{figure}

On a more practical note, it is possible to check a large number of publicly available EoSs at zero temperature in $\beta$-equilibrium from the CompOSE database \cite{Typel:2013rza} against the new pQCD constraints. The results are presented in \cref{fig:3D_database}. Notably, most of the available EoSs are inconsistent with the high-density input within the provided range. The categories “consistent,” “in tension,” and “not consistent” refer to variations in the renormalization scale parameter $X$, as discussed in more detail in \cref{sec:uncertainty}.

\mybox{Summary of \cref{sec:analytic}}{blue!20}{white!10}{
\begin{itemize}
    \item The requirement for a stable, consistent, and causal EoS between known low- and high-density limits of EoS imposes global constraints on thermodynamic quantities.
    \vspace{0.1cm}
    \item By using cEFT and pQCD as the low- and high-density limits, it is shown how these results can be used to propagate constraints from pQCD calculations from around $40\ns$ down to lower densities found in neutron stars.
    \vspace{0.1cm}
    \item Perturbative QCD constraints exclude 32\%, 75\% and 93.5\% of otherwise allowed area for fixed density of $3\ns$, $5\ns$ and $10\ns$, respectively.
    \vspace{0.1cm}
    \item Most of the publicly available EoS models are inconsistent with pQCD constraints within the provided range.
\end{itemize}
}
\clearpage

\section{Bayesian inference}
\label{sec:bayesian}

While the previous section stated that pQCD constraints affect neutron star densities, it is not clear that they offer any new information for EoS inference beyond the current astrophysical data. In this section, I demonstrate that the novel QCD input indeed provides nontrivial information to EoS inference, which is completely independent of astrophysical observations and their systematic uncertainties.

To demonstrate the effect of the QCD input, the most natural approach is to generate a large ensemble of model-agnostic EoSs. Each EoS is then conditioned on current astrophysical observations and theoretical inputs. The novelty of the approach introduced in this section is in utilizing pQCD constraints at NS densities instead of interpolating between the cEFT and pQCD limits. An EoS can be extrapolated from the low-density limit and directly conditioned with the QCD input at lower densities. This method removes prior dependence between the endpoint of the EoS and the pQCD limit. It allows for a direct assessment of the impact of the QCD input by comparing the results of inference with and without the conditioning.

The holy grail of the inference is Bayes’ theorem, which in this case can be expressed as follows: 
\begin{equation}
P({\rm EoS }\,|\, {\rm data} ) = \frac{ P({\rm EoS}) \, P(  {\rm data} \,|\, {\rm EoS} \,)}{P({\rm data})},
\label{eq:bayes}
\end{equation}
where $P({\rm EoS})$ is the prior. The particular choice of prior used throughout this thesis is based on the Gaussian processes (GP) framework, as discussed in \cref{subsec:gp}. $P({\rm data} \,|\,  \rm EoS)$ is the product of uncorrelated likelihoods of the data given an EoS, which in our context can be expressed as
\begin{align}
P({\rm data} & \,|\, {\rm EoS} ) =  P({\rm QCD} \,|\, {\rm EoS}) P({\rm Astro} \,|\, {\rm EoS}).
 \label{eq:likelihoods}
\end{align}
The astrophysical likelihoods, $P({\rm Astro} \,|\, {\rm EoS})$, are explored in \cref{subsec:astro_likelihoods}, and the QCD input with the novel QCD likelihood function $P({\rm QCD} \,|\, {\rm EoS})$ is discussed in \cref{subsec:QCD_likelihood}. Finally, the results of the inference and the effect of the QCD input are covered in \cref{subsec:inference}.

\subsection{Gaussian process prior}
\label{subsec:gp}

GP regression can be viewed as a highly flexible, nonparametric interpolation method, where the values of the regression function, $\{\phi(n_i)\}$,  are assumed to be drawn from a multivariate Gaussian distribution \cite{ebden2015gaussianprocessesquickintroduction,10.7551/mitpress/3206.001.0001}: 
\begin{equation}
\label{eq:GP_normal}
    \phi \sim \mathcal{N}\left({\bar{\phi}}, {K}(n_i,n_j) \right),
\end{equation}
where $\phi=\{\phi(n_i)\}$ represents the vector of function values, and $K(n_i,n_j)$ is a covariance matrix, the elements of which are determined by the kernel. One of the option is the standard choice of the squared-exponential kernel:
\begin{align}
\label{eq:kernel}
    k(n,n') = \sigma e^{-(n-n')^2/2 l^2} + \sigma_n \delta(n,n').
\end{align}
The kernel has two hyperparameters: the variance $\sigma$, which controls the overall magnitude of the correlations between points $n$ and $n'$, and $l$, the length scale over which this correlation occurs. Together with the mean $\bar{\phi}$, these hyperparameters shape the EoS generated by the GP. The parameter $\sigma_n$ is defined based on the uncertainties in the given data $\phi$.

The main assumption of a GP is that the function values $\phi$ and the unknown value $\phi_* = \phi(n_*)$, which is to be estimated, are sampled from a multivariate Gaussian distribution. Therefore, the expression the expression provided in \cref{eq:GP_normal} can be extended as follows:
\begin{align}
    \begin{bmatrix}
        \phi\\ \phi_*
    \end{bmatrix}
    \sim \mathcal{N}\left(\bar{\phi}, \begin{bmatrix}
{K} & {K}_*^\mathsf{T} \\
{K}_* & {K}_{**}
\end{bmatrix}\right),
\end{align}
where $K=K(n_i,n_j)$ is covariance matrix, $K_*$ is given by the vector $K(n_*,n_i)$, and $K_{**}=k(n_*,n_*)$.

The goal of GP regression is to estimate the conditional probability $p(\phi_*|\phi)$, which, given the data, predicts the unknown value and quantifies the uncertainty in the estimation. This probability follows a Gaussian distribution:
\begin{align}
\label{eq:phi_star_distribution}
   \phi_* | \phi & \sim \mathcal{N}(\bar{\phi} + K_* K^{-1} \phi,\  K_{**} - K_* K^{-1} K_*^{\mathsf{T}}),
\end{align}
where the optimal prediction and its variance are given by:
\begin{align}
    \bar{\phi}_* &= \bar{\phi} + K_* K^{-1} \phi\\
    \mathrm{var}(\phi_*) & = \  K_{**} - K_* K^{-1} K_*^{\mathsf{T}}.
\end{align}

As discussed in \cref{sec:analytic}, accessing the full thermodynamic potential is necessary to utilize the novel QCD constraints. One approach is to start with the sound speed as a function of number density and then reconstruct $\mu(n), \e(n),$ and $p(n)$. GP naturally spans the region $[-\infty,\infty]$, which can be mapped to $\cs \in [0, 1]$ using an auxiliary variable:
\begin{equation}
\label{eq:phi_cs2}
    \phi(n) = -\log\bigl( 1/c_s^{2}(n) - 1\bigr)
\end{equation}

Note that this GP regression differs from that in \cite{Landry:2018prl,Landry:2020vaw}, where GP was originally used for EoS generation.

The choice of hyperparameters determines the behavior of the EoS in regions where no data is available. To allow for a wide range of possible behaviors, the hyperparameters for each EoS generated by the GP are randomly drawn from the following distributions:
\begin{align}
\label{eq:hyper}
l \sim \mathcal{N}\bigl( 1.0 n_s, (0.25 n_s)^2 \bigr),\  
\sigma \sim \mathcal{N}(1.25, 0.2^2),\  
\bar c_s^2 \sim \mathcal{N}(0.5, 0.25^2).
\end{align}

GP for the variable $\phi(n)$ is conditioned on cEFT EoS up to 1.1$\ns$. To estimate the uncertainties in cEFT calculations, the mean was taken as the average of the “soft” and “stiff” results from \cite{Hebeler:2013nza}, with the 90\%-credible interval representing the difference between these results. The credible interval is related to $\sigma_n$ from \cref{eq:kernel}. Below $n=0.57\ns$ each EoS is merged with BPS crust EoS \cite{1971ApJ...170..299B}. GP is then used to extrapolate the EoS from the low-density limit up to $\nterm=10\ns$, a density at which it is safe to assume that all EoSs are above the TOV density \cite{Kurkela:2014vha,Hebeler:2013nza}. While this was the conventional choice at the time, this has since changed, partly due to the analysis presented in \cref{sec:termination}.

\begin{figure}[ht!]
    \centering
\includegraphics[width=1\textwidth]{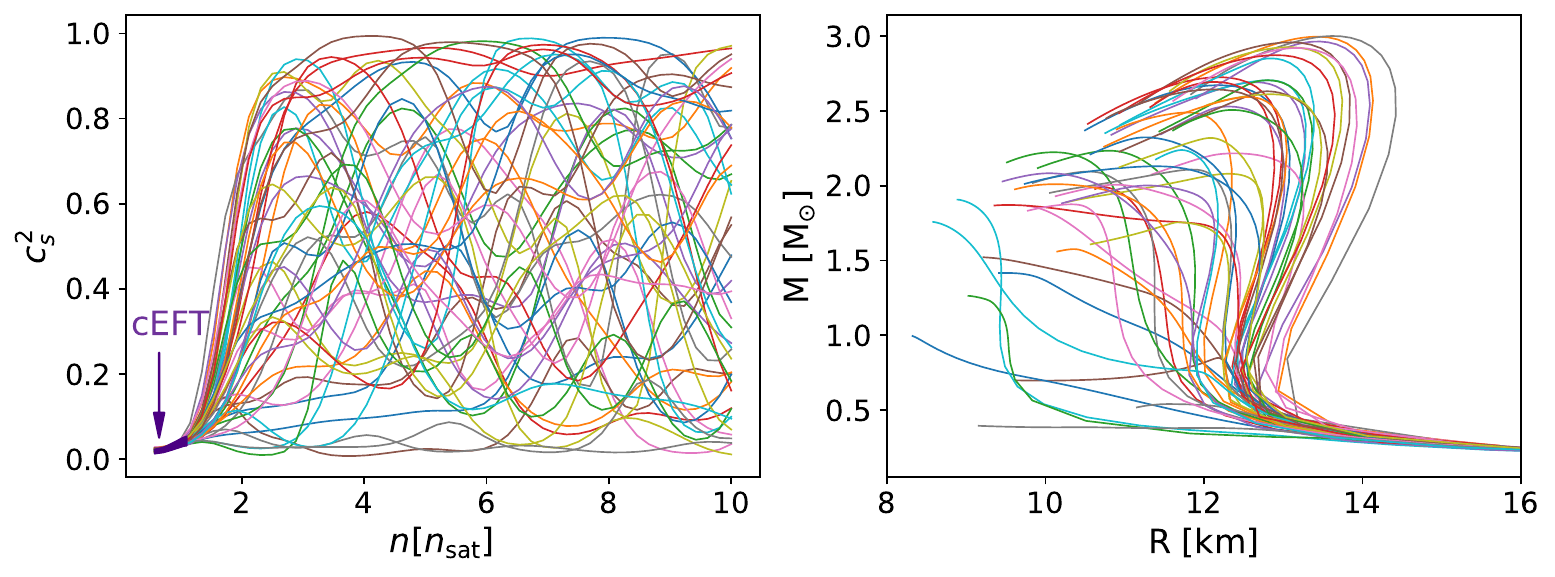}
    \caption{An illustrative sample of EoSs generated using GP and the corresponding mass-radius curves obtained by solving the TOV equation.}
    \label{fig:gp_prior}
\end{figure}

Sampling the distribution in \cref{eq:phi_star_distribution} provides EoSs that are causal (as ensured by \cref{eq:phi_cs2}) and consistent with the low-density limit used to train the GP. \Cref{fig:gp_prior} (left) shows an example of 50 different EoSs sampled from the prior. The reconstruction of thermodynamic quantities can be done as follows:
\begin{align}
    c_s^2(n)  = \frac{1}{e^{-\phi(n)}+ 1}, \quad
    \mu(n)  = \mu_0 \exp\left(\int_{n_0}^{n} dn' c_s^2(n')/n' \right), \\ \nonumber
    \epsilon( n)  = \epsilon_0 + \int_{n_0}^n d n' \mu(n'), \quad
    p(n) = - \epsilon(n) + \mu(n) n.
\end{align}

The final step in constructing the prior is to solve the TOV equation \cref{eq:TOV} to predict the mass $M$, radius $R$, and $\Lambda$ as functions of central density for each EoS as well as maximal mass, $M_{\rm TOV}$, and stable branches as explained in \cref{sec:ns}. The resulting mass-radius relationship is shown in \cref{fig:gp_prior} (right) for the same EoSs depicted in the left plot.

\subsection{Astrophysical likelihoods}
\label{subsec:astro_likelihoods}

Detailed discussions of the current astrophysical observations are provided in \cref{sec:ns}. In this subsection, I present how to implement these measurements to Bayesian framework. To study the effect of the QCD input on the EoS, the following astrophysical observations are considered: radio measurements of PSR J0348+0432 ($M=2.01 \pm 0.04 M_\odot$) \cite{Antoniadis:2013pzd} and PSR J1614-2230 ($M = 1.928 \pm 0.017M_\odot$) \cite{Demorest:2010bx}; X-ray measurements of the mass and radius of PSR J0740+6620 \cite{Fonseca:2021wxt,Miller:2021qha,Riley:2021pdl}; and multimessenger data from GW170817, including TD measurements \cite{PhysRevLett.119.161101, LIGOScientific:2018cki,LIGOScientific:2018hze} and the BH hypothesis \cite{Margalit:2017dij,Rezzolla:2017aly,Ruiz:2017due,Shibata:2017xdx,Shibata:2019ctb}. The astrophysical likelihood can be factorized as follows:
\begin{align}
P({\rm Astro} \,|\, {\rm EoS }) = P( {\rm Radio} \,|\, {\rm EoS}) P({\rm X\text{-}ray} \,|\, {\rm EoS}) P({\rm \tilde \Lambda}, {\rm BH}  \,|\, \rm EoS),
 \label{eq:astro_likelihoods}
\end{align}
where each factor is explained separately in this section.

\textbf{Radio measurements.} The most precise NS mass measurements are obtained via radio observations, which can be approximated by a normal distribution: $\mathcal{N}(M_i,\sigma_i)$, where the index $i$ refers to a specific measurement. For instance, in the case of PSR J0348+0432, $M_{i}=2.01$ and $\sigma_{i}=0.04$. The maximal mass, $M_{\rm TOV}$, can exceed the measured value; however, the likelihood function should return zero if the EoS cannot support such a mass. The likelihood function is integrated over a flat prior distribution for the masses:
\begin{equation} 
P_0(M \,|\, {\rm EoS} ) = \frac{\mathbf{1}_{[M_{\rm min}, M_{\rm TOV}]}(M)}{M_{\rm TOV} - M_{\rm min} },
\label{eq:razor}
\end{equation}
where $M_{\rm min}=0.5M_\odot$. Note that a prior choice of a flat distribution for the mass within the range $[0.5 M_\odot, M_{\rm TOV}]$ is incorporated into the likelihood function. The lower limit is chosen to cover the entire mass range relevant to the measurements. The factor in the denominator $(M_{\rm TOV} - M_{\rm min})$ is not strictly necessary, but it introduces dependence on the stellar population and selection effects (for a detailed discussion, see \cite{Miller:2019nzo,Landry:2020vaw}). Since it produces a similar effect to that of the QCD input, it has been included in the study to maintain a conservative approach. However, its effect is almost negligible and will not be considered in the following \cref{sec:termination} and \cref{chpt:cores}. 

The likelihood function for an individual measurement can be expressed as follows:
\begin{align}
\label{eq:radio_likelihood}
P({\rm Radio_i}  \,|\, {\rm EoS} )  &\propto 
 \frac{1}{\sqrt{2\pi \sigma_i}}\int^{M_{\rm TOV}}_{M_{\rm min}} dM e^{ - \frac{(M- M_i)}{2 \sigma_i }} P_0(M \,|\, {\rm EoS} )  \nonumber \\ & \approx \frac{1 }{2(M_{\rm TOV} - M_{\rm min})} \left(1 + {\rm Erf} (\frac{M_{\rm TOV} - M_i}{\sqrt{2\sigma_i } })\right).
\end{align}

The final likelihood of the radio measurements is then obtained by taking the product of the individual likelihoods:
\begin{align}
    P({\rm Radio}  \,|\, {\rm EoS} ) =  P({\rm J0348}  \,|\, {\rm EoS} )  P({\rm J1614}  \,|\, {\rm EoS} ).
\end{align}

\textbf{X-ray observations.} Mass-radius measurements provide two-dimensional posterior probability densities $P({\rm X\text{-}ray}  \,|\, M,R)$, such as the combined measurement of PSR J0740+6620 from NICER and XMM-Newton data (see fig. 1 of \cited{Miller:2021qha}). To construct likelihood function, the posterior distribution is integrated over the mass as follows:
\begin{align}
\label{eq:Xray_likelihood}
P({\rm X\text{-}ray}\,|\, {\rm EoS} ) \propto \int^{M_{\rm TOV}} dM P( {\rm X\text{-}ray} \,|\, M,R(M) ) P_0(M \,|\, {\rm EoS}),
\end{align}
where the mass-radius curve $R(M)$ is obtained for each EoS by solving the TOV equation. The factor $P_0(M \,|\, {\rm EoS})$ is the same mass prior as given in \cref{eq:razor}.

\textbf{Binary TD.} The LIGO/Virgo collaboration provide two-dimensional posterior probability densities for the event GW170817: $P({\rm GW}  \,|\, \tilde \Lambda, q)$ (see fig. 12 of \cite{LIGOScientific:2018hze}) , where $\tilde \Lambda$ is binary tidal deformability and $q$ is a mass ratio of the merged stars. The binary tidal deformability can be expressed from the tidal deformability of two stars $\Lambda_i$ and their masses $M_i$ as follows \cite{Han:2018mtj}:
\begin{align}
   \tilde{\Lambda} = \frac{16}{13} \frac{(M_1 + 12M_2)M_1^4 \Lambda_1 + (M_2 + 12M_1)M_2^4 \Lambda_2}{(M_1 + M_2)^5}. 
\end{align}

The dimensionless tidal deformability (TD), $\Lambda$, defined in \cref{sec:intro_tov}, is obtained by solving the Love number equation \cref{eq:Love} alongside the TOV equation for each EoS. The chirp mass is accurately measured in the event:  
\begin{equation}
\label{eq:chirp}
    \mathcal{M}_{\text{chirp}} \equiv \frac{(M_1 M_2)^{3/5}}{(M_1 + M_2)^{1/5}} = 1.186(1) M_\odot.
\end{equation}
Given this precise measurement, the second mass can be approximated as a function of the first mass and the chirp mass, $M_2=M_2(M_1,\mathcal{M}_{\rm chirp})$. Thus, to construct the likelihood function, the integration can be performed over the mass of the first binary component, effectively replacing the integration over $M_2$ with a delta function. Assuming $M_1>M_2$ the likelihood can be expressed as:
\begin{align}
P(\widetilde \Lambda \,|\, {\rm EoS} ) =  \int^{M_{\rm TOV}}_{1.3621M_\odot} dM_1 
P_0(M_1, M_2 \,|\, {\rm EoS}) P( {\rm GW} \,|\, \tilde \Lambda , q),
\label{eq:TD} 
\end{align}
where the lower limit of integration, $1.3621M_\odot$, corresponds to the mass ratio $q=1$ for the given chirp mass \cref{eq:chirp}. Flat prior distribution for the masses in the two-dimensional case is given by (similarly to \cref{eq:razor}):
\begin{equation}
\label{eq:TD_likelihood}
    P_0(M_1, M_2 \,|\,\rm EoS) = \frac{\mathbf{1}_{[M_2, M_\mathrm{TOV}]}(M_1) \mathbf{1}_{[M_\mathrm{min}, M_\mathrm{TOV}]}(M_2)}{1/2(M_\mathrm{TOV} - M_\mathrm{min})^2}.
\end{equation}

\textbf{BH hypothesis.} As explained in \cref{sec:ns}, current numerical simulations of NS mergers suggest that the remnant in GW170817 likely collapsed into BH. This implies that the remnant’s total baryon number, $N_\mathrm{remnant}$, exceeds the maximum baryon number, $N_{\rm TOV}$, which is computed under the assumption of a stable, non-rotating neutron star. Baryon number conservation in the merger gives the relation: 
\begin{align}
    N_1 + N_2 = N_\mathrm{remnant}(q) + N_\mathrm{ejecta}(q) = N(q),
\end{align}
where $N_1$ and $N_2$ are computed by solving the TOV equation using \cref{eq:total_baryonic}. The total baryon number of each component $N_i$ can be expressed as a function of $M_i$, which, for a fixed $\mathcal{M}_{\rm chirp}$, leads to the total baryon number of the merger being a function of the mass ratio $q$.

To obtain a conservative bound, it has been suggested to ignore small ejecta \cite{PhysRevX.12.011058}, resulting in the criterion $N(q)>N_{\rm TOV}$.  To construct the Bayesian likelihood, this criterion should be integrated over all possible mass ratios in GW170817. However, the distribution for the mass ratio in GW170817 is already incorporated in the TD likelihood, see \cref{eq:TD}. Consequently, the two likelihood functions are not independent and should be treated simultaneously. The combined likelihood can be expressed as:
\begin{align}
P&(\widetilde \Lambda, {\rm BH} \,|\, {\rm EoS} ) = \int_{M_1 > M_2} dM_1 P_0( M_1, M_2 \,|\, {\rm EoS}) \nonumber \\
&\times P( {\rm GW} \,|\, \widetilde \Lambda , q(M_1, M_2) )\mathbf{1}_{[N_\mathrm{TOV}, \infty]}\bigl( N(q) \bigr),
\label{eq:BH_with_TD}
\end{align}
where the indicator function $\mathbf{1}_{[N_\mathrm{TOV}, \infty]}$ is equivalent to the criterion $N(q) > N_{\rm TOV}$. To construct the likelihood function for the BH hypothesis alone, when TD is not used in the inference, the 2D posterior density $P( {\rm GW} \,|\, \widetilde \Lambda , q(M_1, M_2))$ should be replaced by the marginalized posterior $P( {\rm GW} \,|\, q(M_1, M_2) )$.

\subsection{QCD likelihood function}
\label{subsec:QCD_likelihood}

The conclusion of the previous section stated that global constraints arise from imposing thermodynamic relations and low- and high-density limits. An EoS modeled up to a certain density (in this case, a GP-generated EoS up to $\nterm = 10\ns$) can be checked against these constraints using a simple consistency check, as provided in \cref{eq:qcd_check}.

However, the pQCD constraints are derived under the assumption of known values for $n_\h=n_{\rm QCD}$ and $p_\h=p_{\rm QCD}$, which, in the case of perturbative calculations, depend on the renormalization scale parameter $X$ and $\mu_\h = \mu_{\rm QCD}$. Constructing the QCD likelihood function requires a Bayesian estimation of uncertainties. While the subsequent \cref{sec:uncertainty} is dedicated to exploring different methods for handling perturbative uncertainties within a Bayesian framework, it will become evident that variations in the uncertainty estimation methods do not significantly affect the inference conclusions. To first assess the impact of the QCD input, I present here a simple construction of the QCD likelihood function.

As discussed in \cref{sec:intro_eos}, the conventional approach to estimating perturbative uncertainties is to vary the renormalization scale by a factor of 2. The parameter $X$, related to the renormalization scale through \cref{eq:X}. The pQCD band shown in \cref{fig:theory_inputs} corresponds to the range $X \in [0.5, 2]$, with $X = 1$ being the central value.

One method for constructing a Bayesian likelihood based on this uncertainty estimation is the scale-averaging interpretation proposed by \cite{Duhr:2021mfd}, combined with the log-uniform distribution of the parameter $X$ suggested in \cite{Cacciari:2011ze}:
\begin{align}
\label{eq:log_x}
    w(\ln X) =  {\mathbf 1}_{[ \ln (1/2),\,\ln (2)]}(\ln X). 
\end{align}
In this case, the integration over $X$ distributes equal weight between the intervals $[1/2, 1]$ and $[1, 2]$. The QCD likelihood function can be expressed as follows 
\begin{align}
\label{eq:qcd_likelihood}
    P(\text{QCD} \mid \text{EoS})& = \int^2_{1/2} d(\ln X) w(\ln X) \mathbf{1}_{[\Delta p_{\min}, \Delta p_{\max}]}(\Delta p), 
\end{align}
where the arguments of $\Delta p_{min/max}$ are omitted. While the complete form is provided in \cref{eq:pmin,eq:pmax}, it is useful to explicitly specify the arguments of the function to highlight the dependence on $X$, over which the expression is integrated:
\begin{align}
    &\Delta p_{\rm min}  = \Delta p_{\rm min}(\mu_{\rm term}, \nterm, \mu_\h)\nonumber\\
    &\Delta p_{\rm max}  = \Delta p_{\rm max}(\mu_{\rm term}, n_\h(X), \mu_\h)\nonumber \\
    &\Delta p  = p_\h(X)-p_{\rm term}. \nonumber
\end{align}

The scale-averaging prescription is illustrated in \cref{fig:QCD_likelihood}. The left plot depicts the allowed region at $n=10\ns$ for different values of $X$, where the blue region ($X=1$) corresponds to the blue region ($10\ns$) in \cref{fig:e_p_koku_global}. The resulting QCD likelihood function, as expressed in \cref{eq:qcd_likelihood}, is shown in the right plot. 

\begin{figure}[ht!]
    \centering
\includegraphics[width=0.95\textwidth]{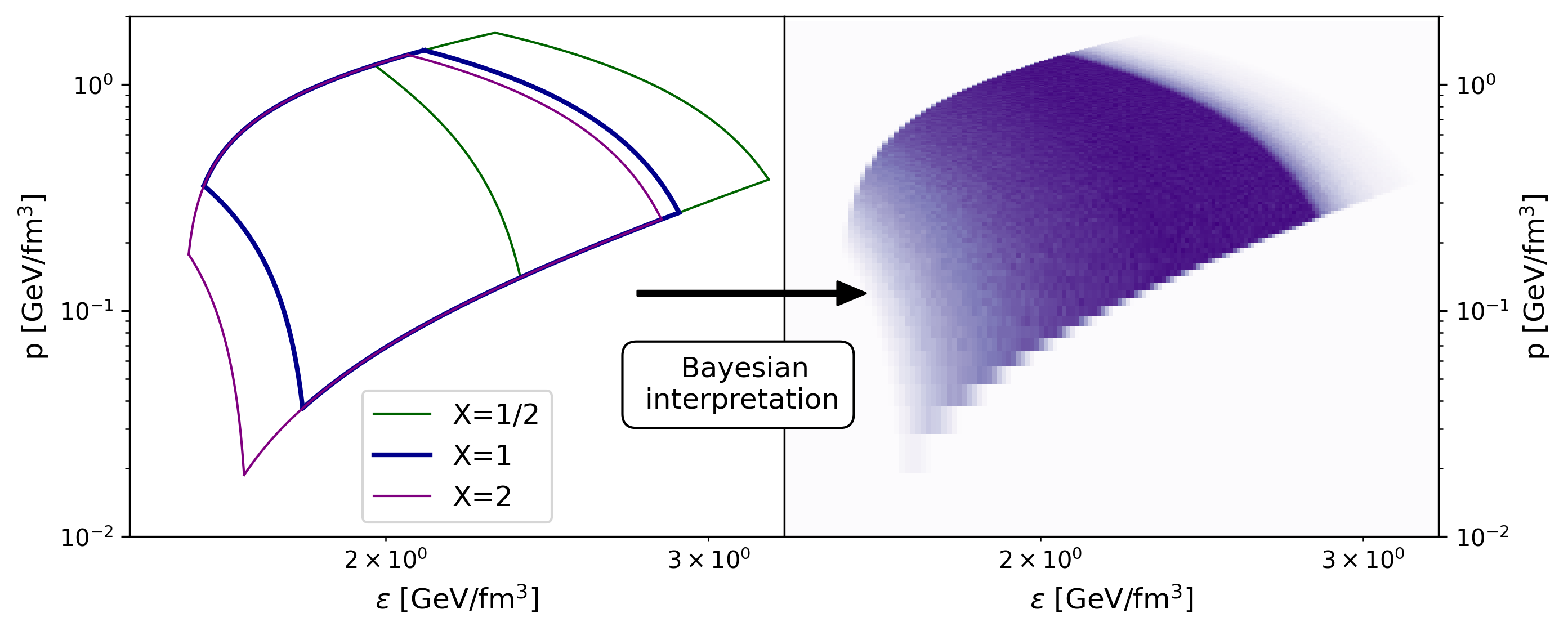}
    \caption{(Left) The allowed region in the $p-\e$ plane for three values of the parameter X =1/2, 1, and 2 at a fixed $n = 10\ns$. (Right) The resulting QCD likelihood function, obtained by scale averaging over $X$ in the range [1/2,2] according to \cref{eq:qcd_likelihood}.}
    \label{fig:QCD_likelihood}
\end{figure}

\subsection{Results of the inference}
\label{subsec:inference}

With the prior established and all likelihood functions defined, the impact of each input on the EoS inference can be examined. The upper plot in \cref{fig:individul_posteriors} shows the effect of each individual input on the $\e-p$ plane, where “Pulsar” represents the combined Radio and X-ray likelihoods. The color-coding of individual EoSs represents their likelihood, with darker shades of red indicating higher likelihood values. The likelihoods are normalized to the maximum likelihood in the ensemble. The figure clearly demonstrates that different inputs are complementary, constraining the different regions of the $\e-p$ plane. Pulsar data mainly affects the stiffness of the EoS at intermediate densities, pushing $p(\e)$ to higher values. TD measurement excludes EoSs that are either too stiff or too soft at the same densities. The BH hypothesis has a similar effect to the QCD input, softening the EoS at higher densities. However, the QCD input has a stronger impact, particularly at the highest densities reached in NSs. 

\begin{figure}[h!]
    \centering
    \includegraphics[width=1\textwidth]{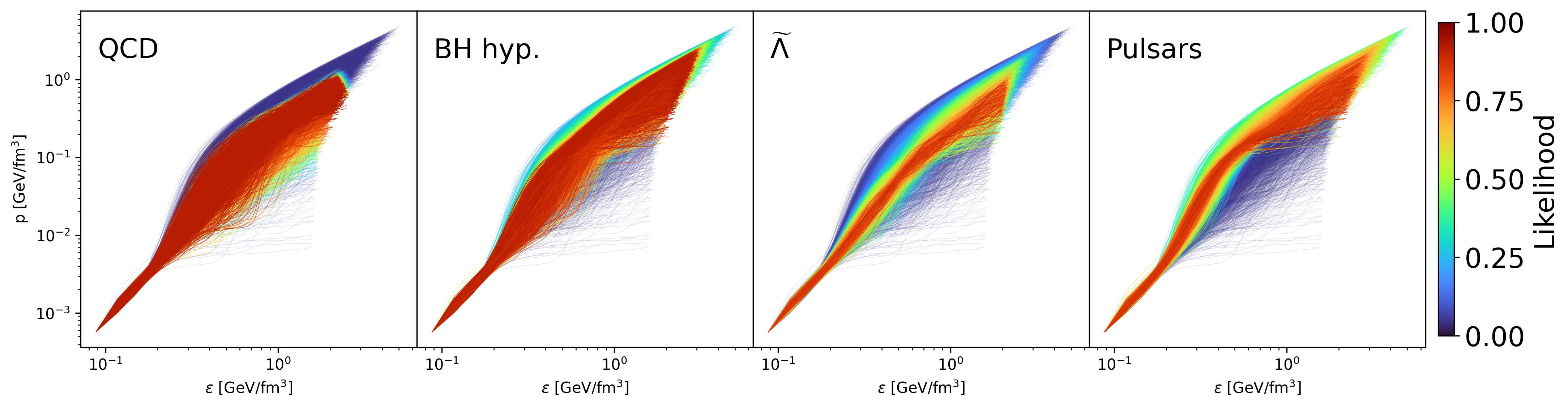} 
    \caption{The representative sample of 5k EoSs from the ensemble, conditioned on different inputs. The coloring of individual EoSs corresponds to their likelihood, with higher likelihood values indicated by darker shades of red.}
    \label{fig:individul_posteriors}
\end{figure}

\begin{figure}[h!]
    \centering
    \includegraphics[width=1\textwidth,trim={3cm 9cm 2cm 10cm},clip]{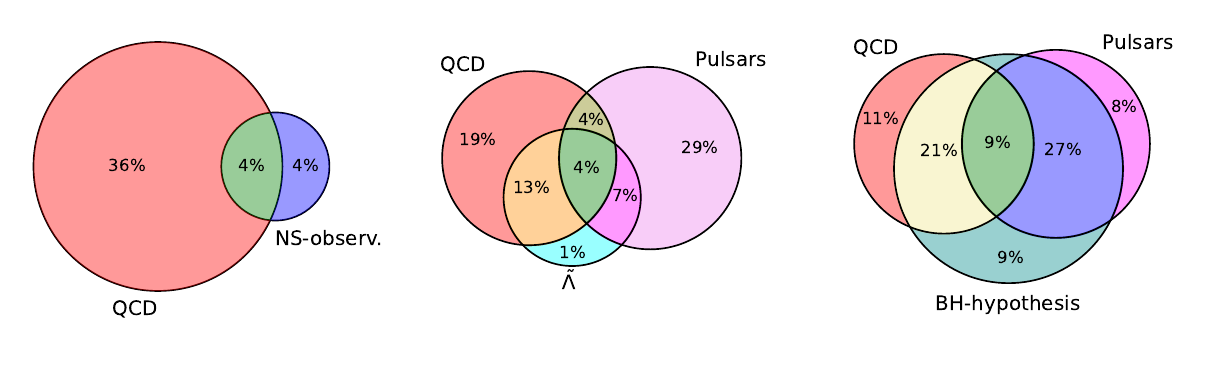} 
    \caption{Venn diagram illustrating the overlap between different inputs. The percentage indicates the fraction of EoSs accepted by the corresponding input in the resampled posterior (see main text). The astrophysical posterior, labeled “NS-observ.,” includes all the discussed observational measurements.}
    \label{fig:venn}
\end{figure}

The overlap between inputs can be quantified more explicitly. To do this, the ensemble is resampled, with each EoS assigned a binary accept/reject weight for each input, where the probability of being accepted is proportional to the normalized likelihood of that input. This approach allows to count the number of EoSs that are mutually rejected or accepted by two different inputs, thus quantifying how corroborative or complementary they are. The result is shown as a Venn diagram in \cref{fig:venn}. Only half of the resampled astrophysical posterior (labeled “NS-observ.” in the plot) is consistent with the resampled QCD input, specifically 4\% out of the 8\% accepted by the resampled astrophysical input. The most significant overlap in the resampled posterior occurs between the QCD input and the BH hypothesis, suggesting that BH formation in GW170817 is a prediction or postdiction of the pQCD results. Notably, after imposing the QCD input, the BH hypothesis does not contribute any additional information.

The results of the inference using all inputs are shown in \cref{fig:posterior_hair} for various physical observables. The coloring of the EoSs corresponds to the likelihoods according to \cref{eq:likelihoods}. The dark blue EoSs represent a versatile prior, covering a broad range of possible behaviors, as determined by the distributions of hyperparameters in \cref{eq:hyper}. The reddish EoSs indicate the regions with the highest likelihoods, representing the most probable behavior based on the current data. A prominent feature of the inference is the peak in the speed of sound, followed by the softening of the EoS. This is clearly visible in the $\e-p$ and $\mu-n$ plane as well, as a change in the slope of the function $p(\e)$. The physical interpretation of this behavior is discussed in \cref{chpt:cores} (spoiler alert: it can be interpreted as a crossover to quark matter cores).

\begin{figure}[h!]
    \centering
    \includegraphics[width=1\textwidth]{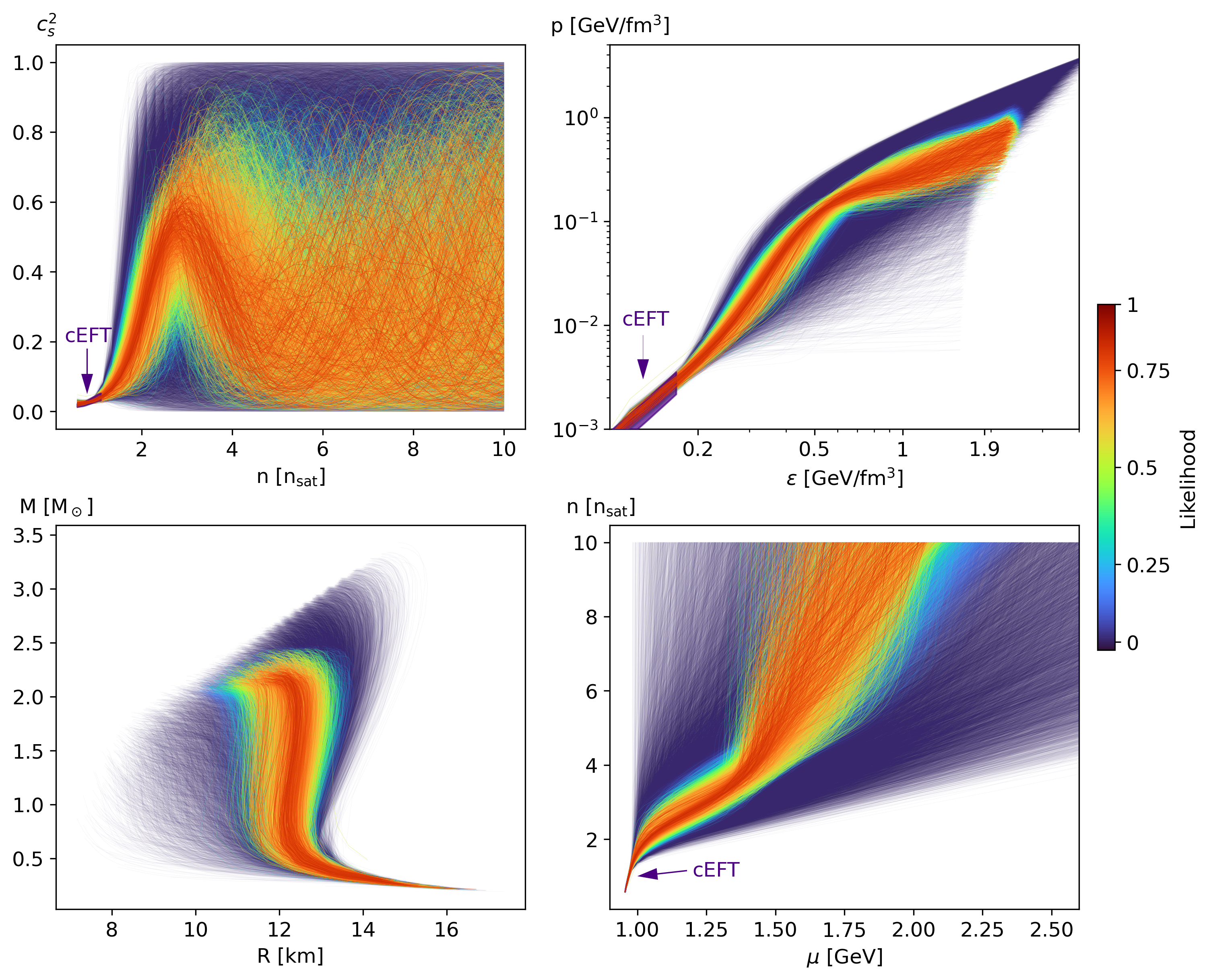} 
    \caption{The representative sample of 10k EoSs from the ensemble for different quantities. The coloring of individual EoSs represents the likelihood, obtained by incorporating all the discussed inputs, as determined by \cref{eq:likelihoods,eq:astro_likelihoods}, with higher likelihood values represented by darker shades of red.}
    \label{fig:posterior_hair}
\end{figure}

To clearly illustrate the effect of the QCD input on the inference, the 68\% confidence intervals (CIs) are shown in \cref{fig:posterior_CIs}. The gray dotted lines represent the prior distribution, while the blue dashed-dotted and purple regions correspond to the astrophysical input with and without the BH hypothesis, respectively. The green CI is obtained by imposing the QCD input on top of the astrophysical data. From the left plot, it is evident that the previously observed softening in studies incorporating pQCD results is indeed a robust prediction of QCD, rather than a limitation of extrapolating between two orders of magnitude in energy density. The gray vertical band represents the 68\% CI for the maximum central densities. Notably, the softening occurs in the stable branch of NSs, starting around 750 MeV/fm$^3$.

\begin{figure}[h!]
    \centering
    \includegraphics[width=0.49\textwidth]{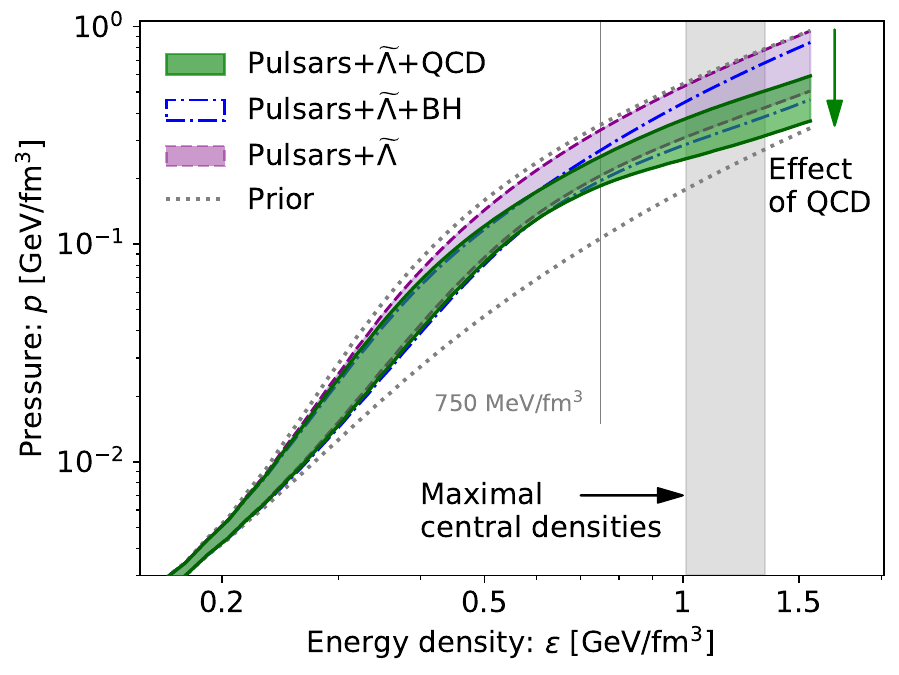} 
    \includegraphics[width=0.49\textwidth]{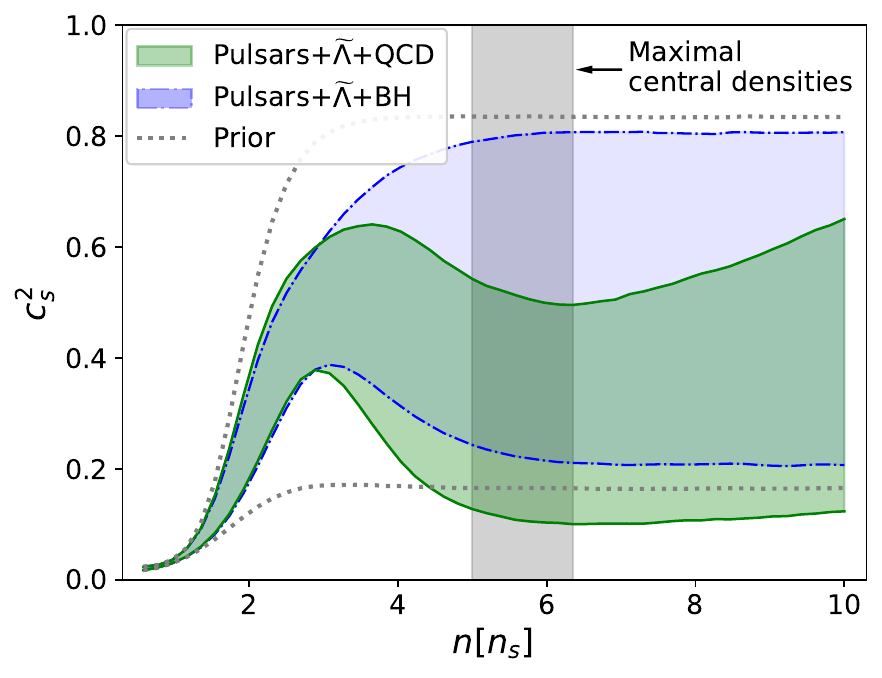} 
    \caption{The impact of the QCD input on the EoS is shown for the $p-\e$ and $c_s^2-n$ planes. The bands represent 68\%-credible intervals conditioned on the different inputs. The label “pulsars” refers to the combined radio and X-ray posterior. The gray band indicates the 68\%-credible interval for the maximum densities reached in stable, non-rotating NSs.}
    \label{fig:posterior_CIs}
\end{figure}

The right plot of \cref{fig:posterior_CIs} emphasizes a crucial feature: the nature of peak in the speed of sound. The plot shows that astrophysical data leads to a rapid stiffening of the EoS, primarily due to mass constraints, as shown in \cref{fig:individul_posteriors}. However, after reaching approximately 2$M_\odot$, these astrophysical constraints are relaxed. Subsequently, driven solely by the QCD input, the EoS softens at the maximal central densities. Thus, the peak structure of $\cs$ in the EoS is a consequence of the interplay between mass constraints and the QCD input.

The distribution of the maximal mass of NS is depicted in \cref{fig:posteriors_KDEs} (left). Astrophysical data suggests that $M_{\rm TOV}$ lies approximately in the range [2 ,2.5]$M_\odot$. The lower limit of the maximal mass, resulting from imposing radio observation constraints, requires that $M_{\rm TOV}\gtrsim2M_\odot$. The sharp cutoff in the upper limit of $M_{\rm TOV}$ with the BH hypothesis, compared to the “Pulsar + $\tilde{\Lambda}$”, arises from the fact that the remnant in GW170817 would not collapse into a BH if the maximal mass of a stable NS exceeded approximately $2.5M_\odot$. Incorporating the QCD inputs further reduces the maximal mass, as a softer EoS leads to slower mass gain with increasing density. Except for the reduction in $M_{\rm TOV}$, the QCD input has only a minor impact on the $M-R$ plane.

The right plot in \cref{fig:posteriors_KDEs} shows the distribution of the sound speed at the maximal central density. The prior and posterior distributions, when considering only astrophysical inputs, are nearly flat between 0 and 1. However, when the QCD input is included, the distribution is no longer prior-driven. The QCD input favors smaller values of sound speed in the cores of NS, leading to the softening of the EoS toward the conformal value of the sound speed, $\cs=1/3$.

\begin{figure}[h!]
    \includegraphics[width=0.44\textwidth]{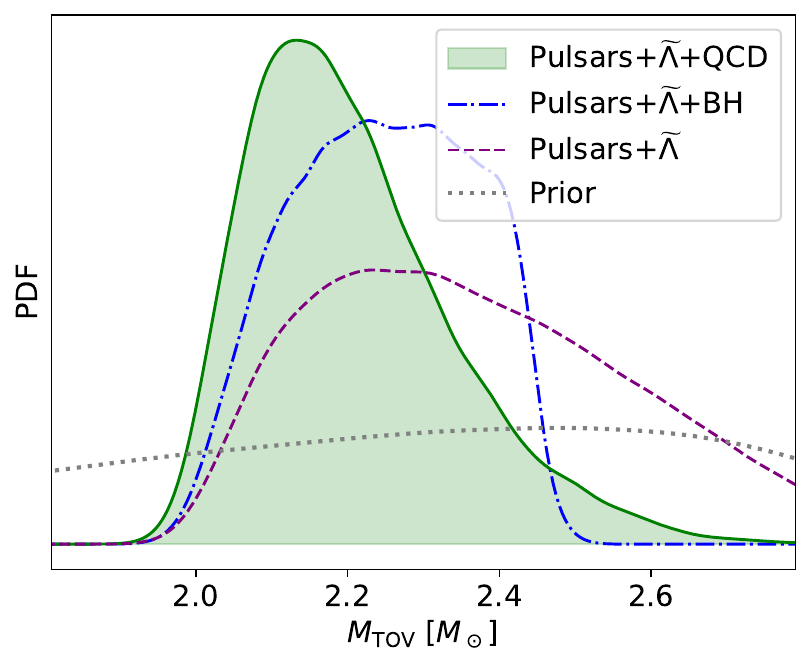}
    \includegraphics[width=0.44\textwidth]{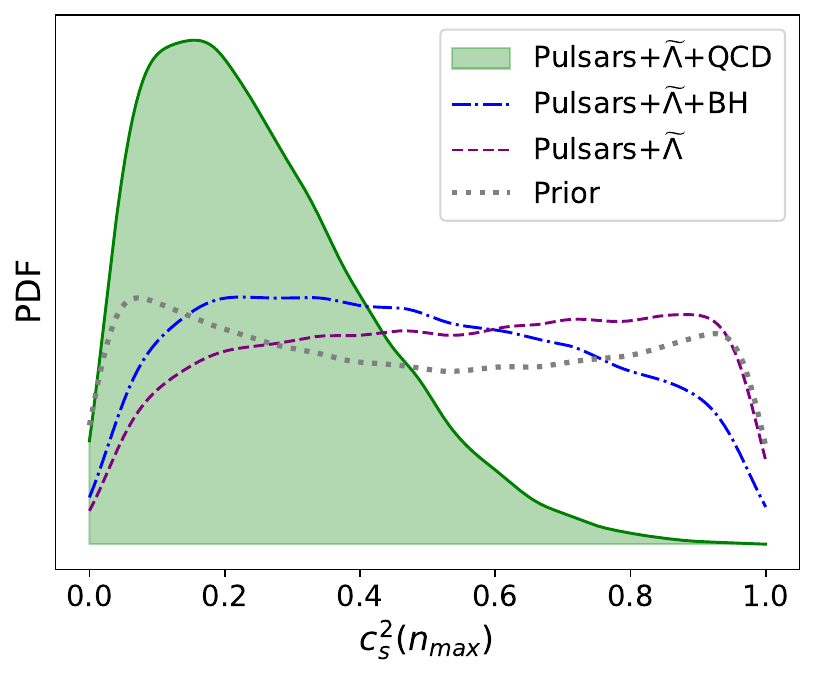}
    \caption{Kernel density estimate of the distributions of the maximal mass and the sound speed at the maximal central density. The shift from the blue dash-dotted line (representing the astrophysical posterior) to the filled green area illustrates the impact of the QCD input.}
    \label{fig:posteriors_KDEs}
\end{figure}

An interesting consequence of the QCD input is that it can be treated as a prediction for the BH formation in GW170817. This argument can be generalized to different chirp masses for any future binary NSs mergers. The probability of BH formation for different chirp masses is shown in \cref{fig:BH}. The prior probability for the BH formation in GW170817 is around 50\%. Astrophysical inputs increase the posterior probability, as the factor $(M_{\rm TOV} - M_{\rm min})^{-1}$ in \cref{eq:radio_likelihood,eq:Xray_likelihood,eq:TD_likelihood} disfavors higher TOV masses. The QCD input further increases the probability, resulting in the prediction of the BH formation in the majority of realistic BNS mergers. For equal mass binary components, $q=1$, the probability of collapsing into a BH exceeds 95\% when the chirp mass is greater than 1.2$M_\odot$, corresponding to to the mass of the binary component greater than 1.38$M_\odot$.

\begin{figure}[h!]
\centering
    \includegraphics[width=0.75\textwidth]{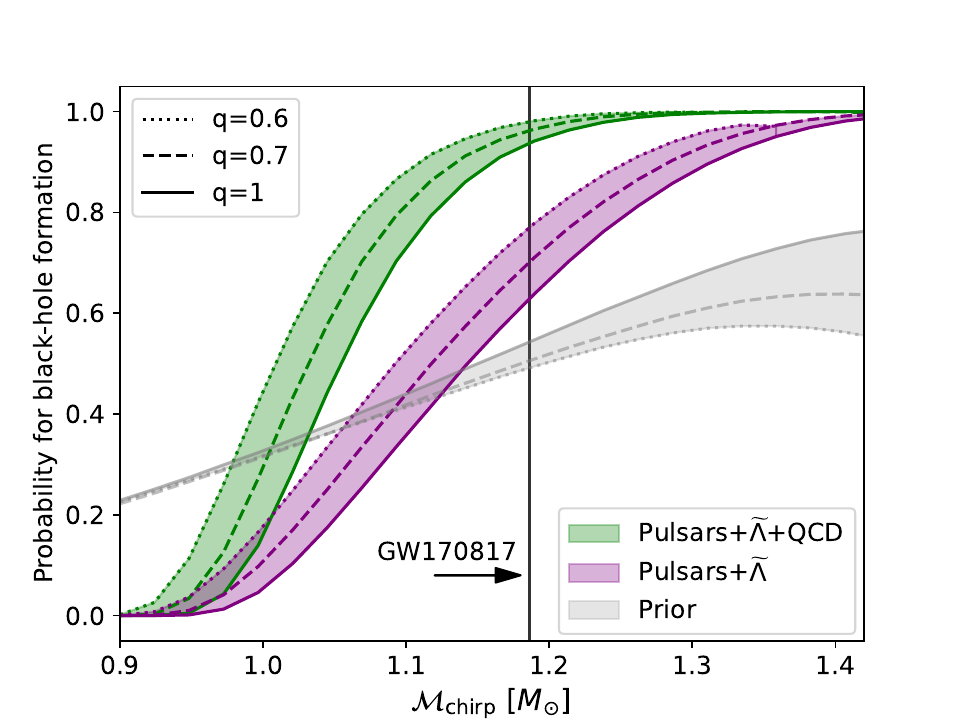}
    \caption{The posterior probability of black hole formation in a BNS merger as a function of chirp mass. Three different mass ratios are considered. The inclusion of the QCD input significantly increases the likelihood that the remnant in GW170817 collapsed into a black hole.}
    \label{fig:BH}
\end{figure}

\mybox{Summary of \cref{sec:bayesian}}{blue!20}{white!10}{
\begin{itemize}
    \item The novel QCD input provides significant constraints on the EoS of neutron stars, going beyond current astrophysical observations.
    \vspace{0.1cm}
    \item Perturbative QCD constraints soften EoS, starting at the energy density of around 750 MeV/fm$^{-3}$.
    \vspace{0.1cm}
    \item The peak structure in the sound speed arises due to the interplay between mass constraints and the QCD input. 
    \vspace{0.1cm}
    \item The QCD input is the only constraint affecting the region [2$M_\odot, M_{\rm TOV}$] apart from the BH hypothesis, with QCD giving a substantially stronger constraint.
    \vspace{0.1cm}
    \item The QCD input increases the posterior probability of BH formation for the majority of realistic BNS mergers. 
\end{itemize}
}
\clearpage

\section{Bayesian interpretation of perturbative uncertainty}
\label{sec:uncertainty}

State-of-the-art pQCD calculations at $T=0$ in $\beta$-equilibrium are discussed in \cref{sec:intro_eos}. The uncertainties in these perturbative calculations come from the truncation of the series at a finite order, which excludes missing higher-order (MHO) terms. This truncation introduces a dependence on the unphysical renormalization scale $\bar{\Lambda}$. In the limit where the series is fully resumed, the dependence on $\bar{\Lambda}$ vanishes. The standard approach to estimating the impact of MHO terms is to vary the scale by a factor of two around a central value. This approach was the basis for constructing a simple QCD likelihood function in \cref{sec:bayesian}. However, this method provides only limited statistically interpretable error estimates for the pQCD results, and the choice of the reference chemical potential $\mu_\h$ at which pQCD is imposed is, in principle, arbitrary.

In this section, I employ machine-learning techniques to provide a Bayesian interpretation of the uncertainties, a method previously applied in LHC physics predictions. The goal of this section is to determine QCD likelihood function, analogous to previous section, with various and more sophisticated uncertainties estimation. This approach allows for a more rigorous study of the impact of these uncertainties on the EoS inference.

The QCD likelihood function is proportional to the posterior distribution:
\begin{equation}
 P(\e_{\rm term}, p_{\rm term} \,|\, \nterm, \bm{p}^{(k)},\bm{n}^{(k)} ),
 \end{equation}  
where the endpoint of the modeled EoS is given by the triplet $\{\e_{\rm term}$, $ p_{\rm term}$, $ \nterm\}$, and $\textbf{p}^{(k)}=\{p^{0},...,p^{k}\}$, $\textbf{n}^{(k)}=\{n^{0},...,n^{k}\}$ are the vectors of the first $k+1$ terms being summed in the perturbative series for the pressure and the number density. 

The evaluation of the posterior distribution is conducted within the framework of the \texttt{MiHO} code \cite{Duhr:2021mfd, mihogit}, with its main concept working as follows. First, for fixed values of $X$ and $\mu_{\h}$, the perturbative coefficients of the asymptotic series are assumed to be independent draws from a statistical model of a convergent series. This assumption allows for a Bayesian analysis of the model parameters. By determining the distribution of these parameters with the given $k+1$ terms of the series, it becomes possible to estimate the next $k+2$-th term, thereby quantifying the MHO terms in a statistically interpretable manner.

Let $p_\h$ and $n_\h$ denote the predictions for the first unknown order in the series. The probability distribution for $p_\h$ and $n_\h$, given first $k+1$ order, is denoted as:
\begin{equation}
    P_{\rm MHO}( p_\h, n_\h | \bm{p}^{(k)}(\mu_\h , X), \bm{n}^{(k)}(\mu_\h , X)).\label{eq:jointPnp}
\end{equation}
The construction of this term and the potential statistical models are discussed in \cref{subsec:fixed_scale}.

Consequently, the obtained distribution for $p_\h$ and $n_\h$ can be integrated over various ranges of the scale $X$ with an integration weight $P_{\text{sa/sm}}$: 
\begin{align}
\label{eq:sm/sa}
  P( p_\h, & n_\h | \bm{p}^{(k)}(\mu_\h), \bm{n}^{(k)}(\mu_\h)) \\ & = \int\! dX\,  P_{\rm MHO}( p_\h, n_\h | \bm{p}^{(k)}(\mu_\h , X), \bm{n}^{(k)}(\mu_\h , X)) \nonumber \\
  & \quad\quad\quad \times P_{\text{sa/sm}}( X |\bm{p}^{(k)}(\mu_\h), \bm{n}^{(k)}(\mu_\h)),\nonumber
\end{align}
One such prescription for the integration weight, called scale-averaging (SA) \cite{Bonvini:2020xeo} and denoted by $P_{\text{sa}}$, was introduced in \cref{sec:bayesian}(c). Another approach, known as scale-marginalization (SM) \cite{Duhr:2021mfd} and denoted by $P_{\text{sm}}$, is discussed in detail in \cref{subsec:scale_marg}. Similar to the marginalization over the scale $X$, the marginalization over $\mu_\h$, where the perturbative results are used, 
is introduced in \cref{subsec:mu_marg}. In this case, the posterior distribution is given by:
\begin{align}
\label{eq:sm/sa_with_marg_mu}
  P( p_\h, & n_\h | \bm{p}^{(k)}, \bm{n}^{(k)}) \\ & = \int\! dX d\mu_\h\,  P_{\rm MHO}( p_\h, n_\h | \bm{p}^{(k)}(\mu_\h , X), \bm{n}^{(k)}(\mu_\h , X)) \nonumber \\
  & \quad\quad\quad \times  P_{\rm sm}(\mu_\h, X |\bm{p}^{(k)}, \bm{n}^{(k)}).\nonumber
\end{align}

Since the goal is to assess the impact of uncertainty estimation on EoS inference, the final component to include in the integral is the QCD input, which propagates the constraints from high-density calculation to lower densities, as introduced in \cref{sec:bayesian}(d). The QCD input checks whether the  endpoint of the EoS, $\{\e_{\rm term}, p_{\rm term}, n_{\rm term}\}$, can be connected to the high-density limit, $\{\mu_\h,p_\h, n_\h\}$, by any stable, causal, and consistent (SCC) EoS. This can be expressed as
\begin{align}
P_{\rm SCC}(\e_{\rm term}&, p_{\rm term}|\nterm, \mu_\h, p_\h, n_\h)= \\ \nonumber
&\mathbf{1}_{[\Delta p_{\min}, \Delta p_{\max}]}(\Delta p) / A(n_\h, \mu_\h, \nterm),
\end{align}
where the indicator function $\mathbf{1}_{[\Delta p_{\min}, \Delta p_{\max}]}$ from \cref{eq:qcd_likelihood} is additionally divided by the so-called area weight $A(n_\h, \mu_\h, \nterm)$, which is explained and derived in detail in \cref{subsec:mu_marg}.

Therefore, the posterior distribution is given by
\begin{align}
\label{eq:master_formula_full}
P(&\e_{\rm term}, p_{\rm term} | n_{\rm term}, \bm{p}^{(k)},\bm{n}^{(k)}) = \int d\mu_\h dp_\h dn_\h dX \, \nonumber  \\
& \quad \times P_{\rm SCC}(\e_{\rm term}, p_{\rm term}|\nterm, \mu_\h, p_\h, n_\h)\, \nonumber  \\
& \quad \times P_{\rm sa/sm}(\mu_\h, X |\bm{p}^{(k)}, \bm{n}^{(k)})  \nonumber  \\
& \quad \times P_{\rm MHO}(p_\h, n_\h| \bm{p}^{(k)}(\mu_\h , X),\bm{n}^{(k)}(\mu_\h , X)).
\end{align}

In principle, the simple likelihood function in \cref{eq:qcd_likelihood} can be derived from \cref{eq:master_formula_full} up to a constant factor by replacing $P_{\rm MHO}$ with the product $\delta(p^{(k)}(\mu_\h,X) - p_\h)\times$ $\delta(n^{(k)}(\mu_\h,X) - n_\h)$, thereby constraining $p_\h$ and $n_\h$ to the values of the last known order in the series, and setting $P_{\rm sa}(X)$ to be $w(\log X)$ from \cref{eq:log_x}. The area weight can be neglected since, for fixed values of  $\mu_\h$, it remains nearly constant. 

The primary source of uncertainty arises from the pressure $\bm{p}^{(k)}$, as the number density $\bm{n}^{(k)}$ converges significantly faster (see \cref{fig:pQCD_p_n} in the appendix). This allows for a well-justified simplification of \cref{eq:master_formula_full}. Instead of using the joint probability distribution, the assumption can be made that the distributions of $p_\h$ and $n_\h$ are independent. This simplifies the process, as the model implemented in the \texttt{MiHO} code currently does not support computing the joint probability of two variables. Consequently, by assuming that the distributions of $p_\h$ and $n_\h$ are independent, and approximating the distribution of the number density with a delta function, the final result is:
\begin{align}
\label{eq:master_formula}
    P(\e_{\rm term}&, p_{\rm term} | n_{\rm term}, \bm{p}^{(k)}) = \int d\mu_\h dp_\h dn_\h dX    \\  
    & \quad  \times \mathbf{1}_{[\Delta p_{\min}, \Delta p_{\max}]}(\Delta p)  / A(n_\h,\mu_\h,\nterm) \, \nonumber  \\
& \quad \times P_{\rm sa/sm}(\mu_\h, X |\bm{p}^{(k)}) \nonumber \\
    &  \quad \times   P_{\rm MHO}(p_\h| \bm{p}^{(k)}(\mu_\h , X)) \, \delta(n^{(k)}(\mu_\h,X)-n_\h).  \nonumber
\end{align}

\subsection{Estimating missing-higher-order terms}
\label{subsec:fixed_scale}

This section focuses on determining $ P_{\rm MHO}(p_\h | \bm{p}^{(k)}(\mu_\h , X)) $. As outlined earlier, the central assumption is that each order in the perturbative series is treated as a draw from a statistical model. Two models are considered: the geometrical model and the \textit{abc} model. In the simplest case, which is the geometrical model \cite{Bonvini:2020xeo}, the perturbative coefficients normalized to the LO term,
\begin{equation}
 \delta_k(\mu,X)= \frac{p^{(k)}(\mu,X)}{p^{(0)}(\mu,X)},
\end{equation}
are assumed to be draws from flat prior distributions: 
\begin{equation}
    P_\text{geo}(\delta_k|a,c) \equiv \frac{1}{2a^kc}\theta\left(c-\frac{|\delta_k|}{a^k}\right).\label{eq:geo}
\end{equation}
The two parameters of the model, $c$ and $a$, control the width of the uniform distribution and the rate at which this width decreases with increasing order ($0<a<1$).

The statistical model can be modified to use an asymmetric prior distribution:
\begin{equation}
b-c \leq \frac{\delta_k}{a^k} \leq b+c,
\end{equation}
where $-1<a<1$, allowing $a$ to take negative values to capture both alternating and non-alternating series. This modification is referred to as the \textit{abc} model \cite{Duhr:2021mfd}:
\begin{equation}
    P_{abc}(\delta_k|a,b,c) \equiv \frac{1}{2|a|^kc}\theta\left(c-\left|\frac{\delta_k}{a^k}-b\right|\right)\label{eq:abc}.
\end{equation}

The goal is to perform Bayesian inference on the parameters of the model, given the first $k+1$ order $\bm{\delta}_k=(\delta_0,...,\delta_k)$: 
\begin{align}
\label{eq:miho_bayes}
    P(a,c| \bm{\delta}_k) = 
    \frac{P( \bm{\delta}_k | a, c) P_0(a)P_0(c)}
    {P(\bm{\delta}_k)},
\end{align}

The priors for the parameters were extensively analyzed in \cite{Duhr:2021mfd}, with the conclusion that different choices of priors have only a mild effect on the results. The judiciously chosen priors can be summarized as follows. For the \textit{geo} model:
\begin{align}
    &P^{geo}_0(a)\equiv(1+\omega)(1-a)^\omega \theta(a)\theta(1-a),\nonumber\\
    &P^{geo}_0(c)\equiv\frac{\epsilon}{c^{1+\epsilon}}\theta(c-1),\nonumber\\
    &(\epsilon, \omega)^{geo}=(0.1,1).
\end{align}
And for the \textit{abc} model, these priors are adjusted accordingly:
\begin{align}
    &P^{abc}_0(a)\equiv\frac{1}{2}(1+\omega)(1-|a|)^\omega \theta(1-|a|),\nonumber\\
    &P^{abc}_0(b,c)\equiv\frac{\epsilon \eta^\epsilon}{2\xi c^{2+\epsilon}}\theta(c - \eta)\theta(\xi c-|b|),\nonumber\\
    &(\epsilon, \omega, \xi, \eta)^{abc}=(0.1,1,2,0.1).
\end{align}

By inputting the perturbative coefficients and priors into \cref{eq:miho_bayes}, the posterior distribution for the parameters $P(a,c| \bm{\delta}_k)$ is obtained. This allows for the estimation of the posterior distribution of the next order in the series:
\begin{align}
P(\delta_{k + 1}| \bm{\delta}_k) =
\int da dc P(\delta_{k+1}| a, c ) P(a, c | \bm{\delta}_k).
\end{align}

One of the assumptions of the method is that the full sum of the statistical model for a convergent series can be approximated by the partial sum up to order $k+2$ (with $k+1$ known terms and an estimated $k+2$-th term). According to \cite{Duhr:2021mfd}, the posterior probability for the partial sum up to the $k+2$ order is given by:
\begin{align}
    \label{eq:PMHO}
    P_{\rm MHO}(p | \bm{p}^{(k)}(\mu,X)) \approx \frac{1}{p^{(0)}} P\left( p = p^{(0)}(\delta_{k+1}+\sum_{i=0}^k \delta_i)| \bm{\delta}_k\right).
\end{align}

\Cref{fig:miho_distr} shows the posterior distributions $P_{\rm MHO}(p | \bm{p}^{(k)}(\mu,X))$ for different orders $k$ at a fixed $\mu_{\h}=2.6$ GeV and a central scale $X=1$. At LO, no information regarding the convergence of the series is available, so the distribution simply follows the prior. The $abc$ model exhibits an asymmetric distribution at NLO due to a negative correction to LO, as an alternating series, with a positive correction expected next. However, with another negative correction at N$^2$LO, the distribution becomes symmetric again. 

\begin{figure}[ht!]
    \centering
\includegraphics[width=1\textwidth]{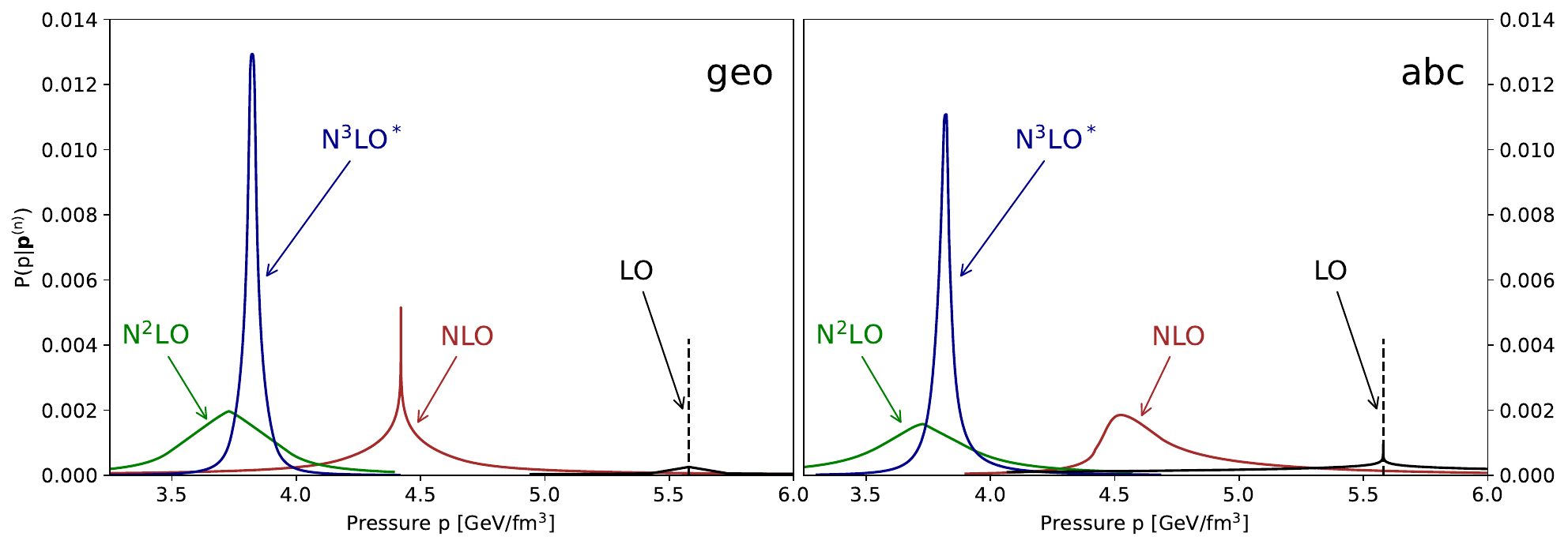}
    \caption{Estimates of the missing higher-order uncertainties for the pressure, based on \cref{eq:PMHO}, at a fixed $\mu_{\rm QCD}=2.6$GeV and central scale $X=1$, using two different statistical models (geometrical and \textit{abc}) for the prior distribution of the perturbative coefficients.}
    \label{fig:miho_distr}
\end{figure}

Throughout this section, the last fully computed order, N$^2$LO, is used for estimating MHO terms. The partially computed next order, N$^3$LO$^*$, is used only in \cref{fig:miho_distr} and provided for reference in other figures. However, it is used with scale-averaging prescription, without MHO estimation, as was done in \cref{sec:bayesian}.

\subsection{Scale dependence}
\label{subsec:scale_marg}

The denominator of \cref{eq:miho_bayes} represents the evidence, which can be obtained by marginalizing the numerator over the model parameters for a given scale $X$. This marginalized likelihood (or evidence) can be used to incorporate scale dependence in uncertainty estimation. The evidence for a given $X$ is expressed as:
\begin{align}
    P(\bm{\delta}_k(X)) \equiv \int da dc P(\bm{\delta}_k(X)| a, c) P_0(a)P_0(c).\label{eq:evidence}
\end{align}
This provides a quantitative measure of how well the model reproduces the known input data. In \cref{fig:miho_evidence}, the evidence is shown as the black dashed line. It demonstrates that the model better reproduces the results for larger values of $X$, corresponding to faster-converging series. Similarly, for smaller values of $X$, the series converges more poorly, as indicated by the lower marginalized likelihood. The green bands correspond to the 1$\sigma$ and 2$\sigma$ CI for N$^2$LO pQCD input, while the red and blue lines represent the NLO and N$^3$LO$^*$ results for reference. Notably, N$^3$LO$^*$ lies well within the 1$\sigma$ CI for N$^2$LO.

\begin{figure}[ht!]
    \centering
\includegraphics[width=1\textwidth]{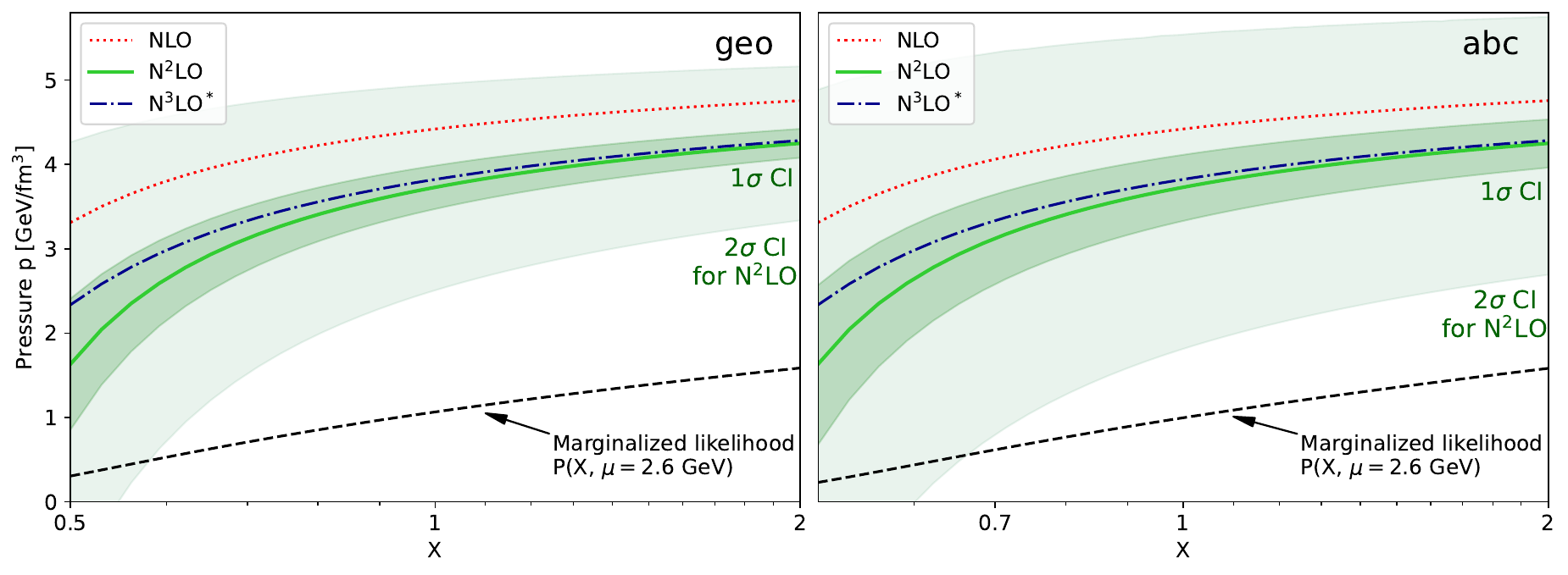}
    \caption{The $1\sigma$ and $2\sigma$-CI estimates of the MHO, as predicted by the geometrical and \textit{abc} models using N$^2$LO pQCD results for the pressure at fixed $\mu_{\rm QCD}$=2.6 GeV as a function of $X$ (\cref{eq:PMHO}). The black dashed line represents the evidence (\cref{eq:evidence}), which is used to marginalize over the renormalization scale, plotted as a function of $X$ while keeping $\mu_{\rm QCD}$ fixed.}
    \label{fig:miho_evidence}
\end{figure}

To incorporate the uncertainties arising from renormalization scale dependence and obtain the integrand used in \cref{eq:sm/sa}, $ P_{\rm sm}(X |\bm{p}^{(k)})$, the marginalized likelihood is integrated with $P_0(X)$, which defines the range of scale variation (for the standard range [1/2,2], see \cref{eq:p0_sa}):

\begin{align}
\label{eq:sm}
    P_{\rm sm}(X |\bm{\delta}_k)=\frac{ P_0(X) P(\bm{\delta}_k(X))}{\int dX P_0(X) P(\bm{\delta}_k(X))}.
\end{align}

The scale independent distribution, which incorporates both scale-marginalization and the estimate for the MHO, can be expressed as:
\begin{equation}
\label{eq:scale_ind_P_sm}
    P^{\rm sm}_{X}(\delta_{k+1}|\bm{\delta_k})=\int dX P_{\rm sm}(X|\bm{p}^{(k)}(X)) P_{\rm MHO}(p^{(k)}(X)| \bm{p^{(k)}}(X)),
\end{equation}
where the argument of $P_{\rm sm}(X|\bm{\delta_k})$ from \cref{eq:sm} can be trivially substitute by $P_{\rm sm}(X|\bm{p^{(k)}})$. 

To obtain a scale-independent distribution using the scale-averaging prescription, the integration weight for $X$, $P_{\rm sm}(X|\bm{\delta_k})$, is replaced with $P_{\rm sa}(X)$, which can be expressed according to \cref{eq:log_x}: 

\begin{align}
\label{eq:p0_sa}
    P_{\rm sa}(X)=P_0(X)= \frac{1}{2X\ln2}\theta \left(\ln 2-|\ln X|\right).
\end{align}

The quantity $P^{\rm sm/sa}{X}$ is shown in \cref{fig:miho_SM_SA} for both the SM and SA prescriptions using N$^2$LO pQCD input. Different ranges for $X$ are considered to assess sensitivity to the choice of range, with central values given by $X_{\text{central}} = \{1, 1.5, 0.75, 1.2\}$ and corresponding variation factors of $\{2, 2, 2, 4\}$. These choices result in different ranges: $X \in [1/2,2]$, $[0.75,3]$, $[0.375,1.5]$, and $[0.3,4.8]$, respectively.

For the SA case, the peak of the distribution lies approximately between $X=1$ and $X=2$, due to faster convergence of the series at higher values of $X$ (as evident from \cref{fig:miho_evidence}) and the log-uniform distribution. The peak is more pronounced for SM, as the marginalized likelihood shown in \cref{fig:miho_evidence} gives greater weight to larger values of $X$. Overall, the dependence on the choice of the renormalization scale parameter range is mild, suggesting that it has a limited impact on the results.

\begin{figure}[ht!]
    \centering
\includegraphics[width=1\textwidth]{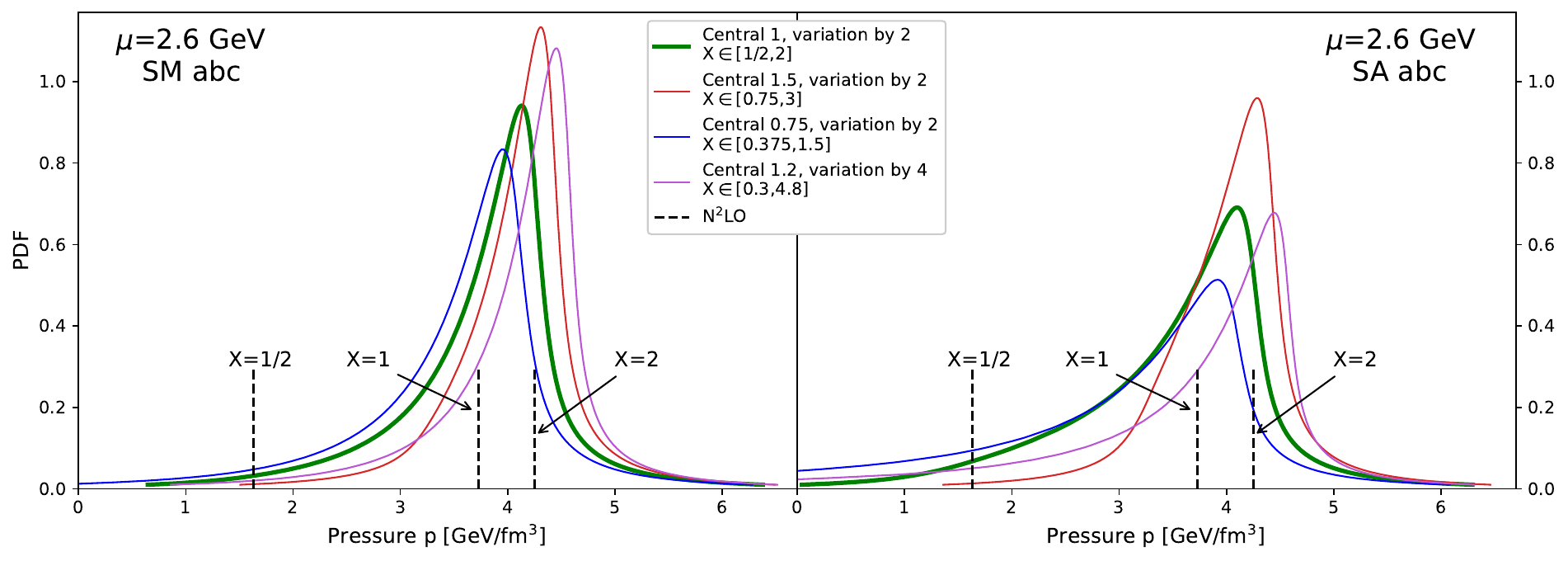}
    \caption{The scale-independent distribution, incorporating both the estimate for the MHO and either scale marginalization (left) or scale averaging (right), is constructed according to \cref{eq:scale_ind_P_sm}, with \cref{eq:p0_sa} substituted in the case of SA. The sensitivity to the choice of the parameter range for $X$ and the variation factor is shown.}
    \label{fig:miho_SM_SA}
\end{figure}

\subsection{Marginalization over $\mu_\h$}
\label{subsec:mu_marg}

The conventional choice of $\mu_{\h} = 2.6$ GeV is made to roughly match the relative uncertainties of cEFT calculations of \cite{Hebeler:2013nza} at 1.1$\ns$, as mentioned in the \cref{sec:intro_eos}. While imposing pQCD at a higher reference chemical potential reduces uncertainties in the perturbative calculations, due to faster convergence of the series, it also weakens the constraints when propagated to lower densities. Therefore, quantifying this interplay would enable a more accurate interpretation of the uncertainties.

The perturbative coefficients depend on both $\mu_\h$ and $X$. As a result, the evidence in \cref{eq:evidence} depends on $\mu_\h$, which quantifies the convergence of the series for different values of $X$ and $\mu_\h$. In a similar manner to the previous section, simultaneous marginalization over the renormalization scale and chemical potential can be introduced:
\begin{align}
\label{eq:sm_full}
    P_{\rm sm}(\mu_\h,X |\bm{\delta}_k)=\frac{ P_0(X) P_0(\mu_\h) P(\bm{\delta}_k(\mu_\h,X))}{\int dX d\mu_\h P_0(\mu_\h) P(\bm{\delta}_k(\mu_\h,X))} .
\end{align}
The marginalized likelihood over $X$ for a given $\mu_\h$ can be defined as follows:
\begin{align}
\label{eq:marg_likelihood_mu}
    P(\mu_\h)=\int dX P_0(X) P(\bm{\delta}_k(\mu_\h,X)).
\end{align}
This quantity is depicted in \cref{fig:miho_fig1} as a black dashed line, representing the relative weight for different reference chemical potentials. For smaller values of $\mu$, the perturbative uncertainties increase rapidly, reducing confidence in the results, as reflected by the marginalized likelihood. The green bands represent the 1 and 2$\sigma$ CIs of the posterior distribution for normalized pressure obtained using the $abc$ model and marginalization over scale $X$. The hatched area indicates the standard scale-variation error estimation for $X\in[1/2,2]$. The colored lines correspond to the EoSs inferred in the previous section with the most sophisticated QCD likelihood function, which is introduced later in \cref{subsec:miho_inference}.

\begin{figure}[ht!]
    \centering
\includegraphics[width=0.75\textwidth]{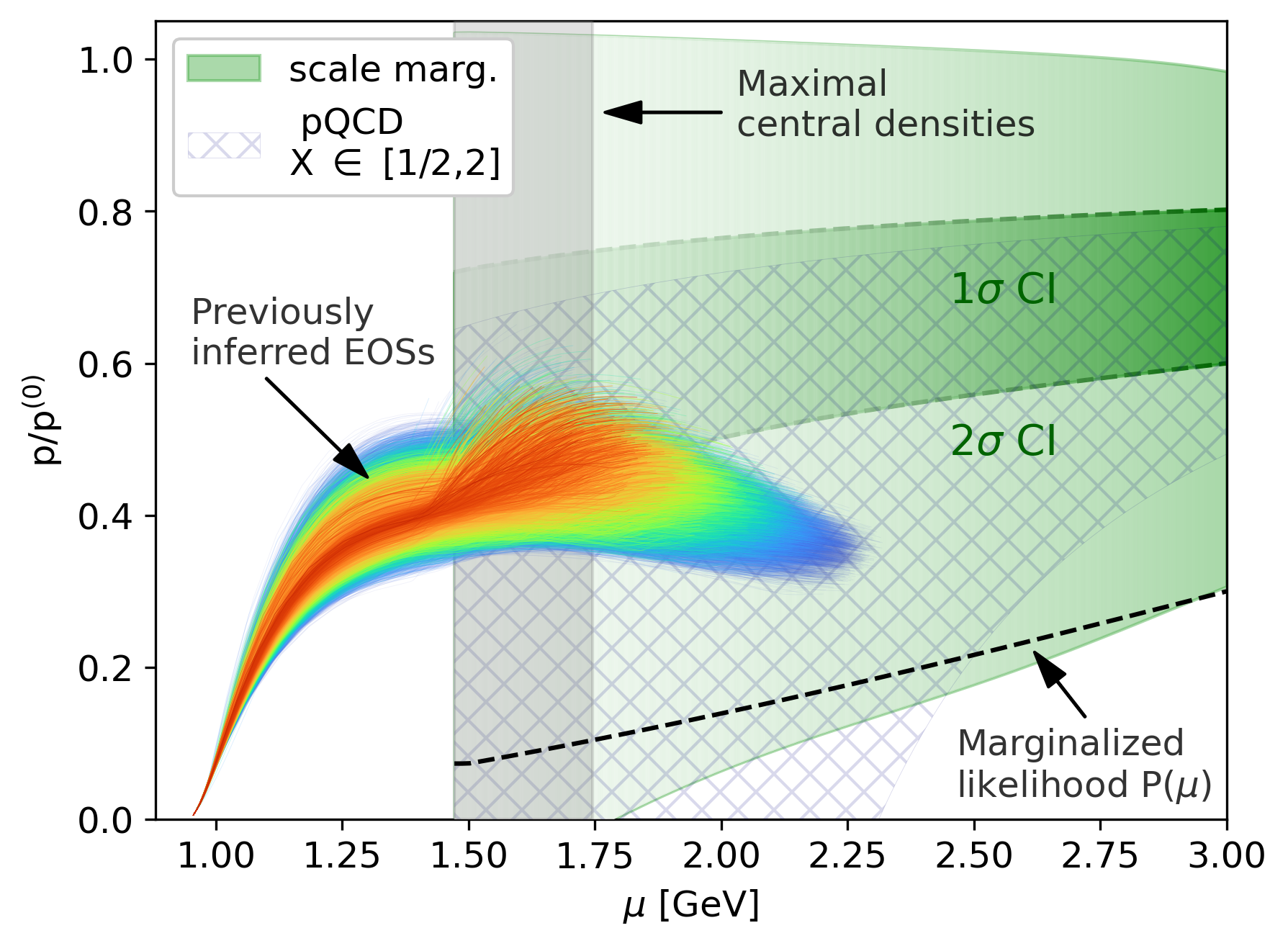}
    \caption{The green bands correspond to the $1\sigma$ and $2\sigma$ confidence intervals (CIs) for MHO uncertainty estimates using the \textit{abc} model with the scale-marginalization prescription for the pressure, normalized to that of a free Fermi gas of quarks, as a function of chemical potential. The hatched purple area represents the standard uncertainty estimate, obtained by varying the scale parameter $X$ by a factor of 2. The marginalized likelihood over $X$ as a function of $\mu$, given by \cref{eq:marg_likelihood_mu}, is shown as a black dashed line. The colored lines represent the EoSs inferred in \cref{sec:bayesian}, where the QCD input is replaced with the QCD likelihood function from the middle panel in the second row of \cref{fig:miho_panels} (see main text).}
    \label{fig:miho_fig1}
\end{figure}

Imposing pQCD at a higher value of $\mu_\h$ increases the allowed region in the $p$-$\e$ plane, thereby reducing the constraining power. The spread of this area can be quantified as a function of $\mu_\h$, $n_\h$, and $\nterm$. To determine the allowed region that can be connected to QCD through a stable, consistent, and causal EoS for a fixed $\nterm$, the $\e-p$ values are checked against the criteria for the modeled EoS derived in \cref{eq:qcd_check}. Note that this area differs from the red or blue areas in \cref{fig:e_p_koku_global}, as it depends solely on the QCD results without incorporating cEFT input. Finding solutions to $\Delta p_{\rm min/max} = p_\h - p_{\rm min/max}(\e)$ with $\mu_{\rm term} \nterm = p_{\rm min/max}(\e) + \e$ provides bounds on $p$ as a function of $\e$.
\begin{align}
\label{eq:pqcd_allowed_area}
    \Delta p_{\rm min}& = p_\h - p_{\rm min}(\e) = \frac{\nterm}{2}\left( \frac{\mu_\h^2\nterm}{p_{\rm min}(\e)+\e} - \frac{p_{\rm min}(\e)+\e}{\nterm}\right),\\
    \Delta p_{\rm max}& = p_\h - p_{\rm max}(\e) = \frac{n_\h}{2}\left( \mu_\h - \left(\frac{p_{\rm max}(\e)+\e}{\nterm}\right)^2\frac{1}{\mu_\h}\right).\nonumber
\end{align}
\begin{figure}[ht!]
    \centering
\includegraphics[width=0.7\textwidth]{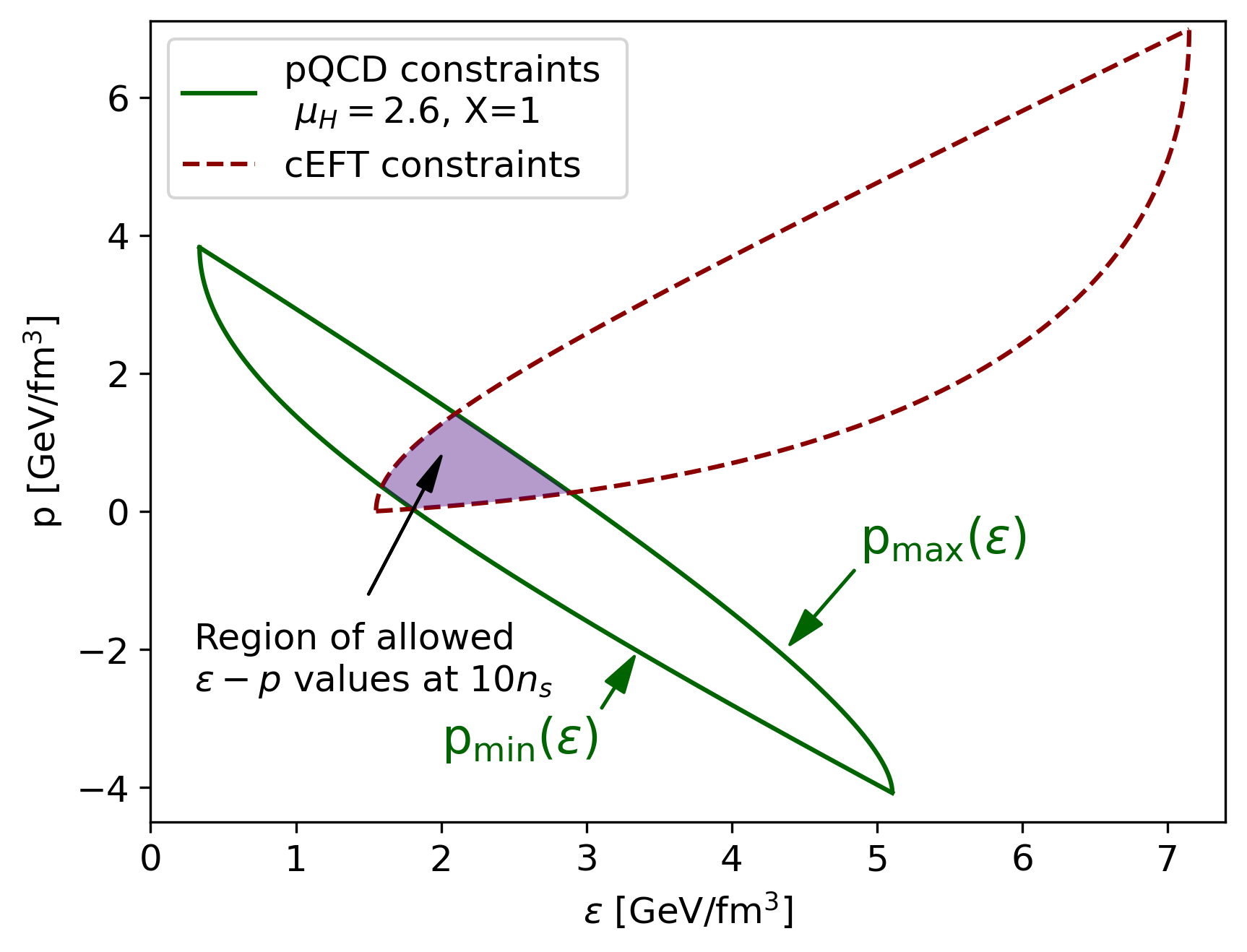}
    \caption{The allowed regions at $10\ns$ for causal and stable EoSs, extrapolated either from cEFT (red dashed line) or from pQCD (green solid line), are shown according to \cref{eq:pqcd_allowed_area}. The intersection of these regions is used to construct the likelihood function presented in \cref{fig:QCD_likelihood,fig:miho_panels}.}
    \label{fig:miho_pqcd_ceft}
\end{figure}
The analytic solutions to these equations, $p_{\rm min/max}(\e)$, are given by:
\begin{align}
\label{eq:p_bounds}
   p_{\min}(\e)& = p_\h - \sqrt{\e^2 + 2 \e p_\h - \mu_\h^2 \nterm^2 + p_\h^2} , \\
    p_{\max}(\e) &= \frac{\nterm \sqrt{\mu_\h \left(-2 \e n_\h + \mu_\h \nterm^2 + \mu_\h n_\h^2 - 2 n_\h p_\h \right)}}{n_\h} \\ &- \e + \frac{\mu_\h \nterm^2}{n_\h}. \nonumber
\end{align}
The corresponding equations for propagating cEFT constraints to higher densities are provided in Appendix B of \cite{Gorda:2023usm}.

These solutions are depicted in \cref{fig:miho_pqcd_ceft} for fixed $\nterm=10\ns$. Each region individually represents the $\e-p$ values that can be connected to the corresponding high- or low-density limit. For example, this includes negative pressure values for the pQCD limit, which, in principle, can be connected to the high-density limit, although such scenarios may be excluded for other reasons. The intersection of the two regions, highlighted in purple, represents the allowed region at the 10$\ns$, as also shown in \cref{fig:e_p_koku_global,fig:QCD_likelihood}.

The area bounded by the green lines  $p_{\rm min/max}(\e) $ is used to quantify how the constraining power diminishes as $\mu_\h$ increases. This area can computed as:
\begin{align}
\label{eq:area}
   & A(n_\h,\mu_\h,\nterm)  \equiv \int^{\e_{\max}}_{\e_{\min}} d\e \left( p_{\rm max}(\e)- p_{\rm min}(\e) \right)\\
    &=\frac{\mu_\h^2 \nterm}{12 n_\h^2} \left(4 n_\h^3-3 n_\h^2 \nterm-6 n_\h^2 \nterm \log \left(\frac{n_\h}{\nterm}\right)-\nterm^3\right),\nonumber
\end{align}
where $\e_{\rm min/max}$ are the points of intersection of $p_{\rm min/max}$.

For fixed $\nterm$ it can be assumed that the probability density is uniformly distributed within the allowed region on the $\e-p$ plane. Hence, the differential probability is constant:
\begin{align}
	\frac{d^2 P(\e, p | \mu_\h,n_\h, \nterm)}{d \e \, d p} = \textrm{const} = 1/A(\mu_\h,n_\h, \nterm).
\end{align}

As a result, the previously derived QCD likelihood function $\mathbf{1}_{[\Delta p_{\min}, \Delta p_{\max}]}(\Delta p)$ is additionally multiplied by  $1/A(\mu_\h,n_\h, \nterm)$ in \cref{eq:master_formula} to account for the spread of the area as $\mu_\h$ varies. This factor was neglected in \cref{sec:bayesian}, as, for fixed $\mu_\h$, the area depends only on $n_\h(X)$ and $\nterm$. With $\nterm$ fixed, the dependence on $n_\h(X)$ has a negligible effect on the final result.

As shown in \cref{fig:miho_fig1}, the marginalized likelihood $P(\mu_\h)$ increases with $\mu_\h$, in contrast to the $1/A$ factor. This interplay between the marginalization of the chemical potential and the area weight identifies an optimal range in $\mu_\h$ where the constraining power is maximized while keeping the perturbative uncertainties under control.

\subsection{Impact on the Bayesian inference}
\label{subsec:miho_inference}

With all the integrands of \cref{eq:master_formula} introduced and derived, it is possible to evaluate the posterior probability  $P(\e_{\rm term}, p_{\rm term} | n_{\rm term}, \bm{p}^{(k)})$  for a fixed $\nterm = 10 \ns$. This posterior probability is treated as the QCD likelihood function used for EoS inference. \Cref{fig:miho_panels} presents a panel of nine different likelihoods, each calculated using various prescriptions and reference chemical potential $\mu_\h$. The range for scale $X\in[1/2,2]$ is adopted for all the likelihoods. 

\begin{figure}[ht!]
    \centering
\includegraphics[width=0.95\textwidth]{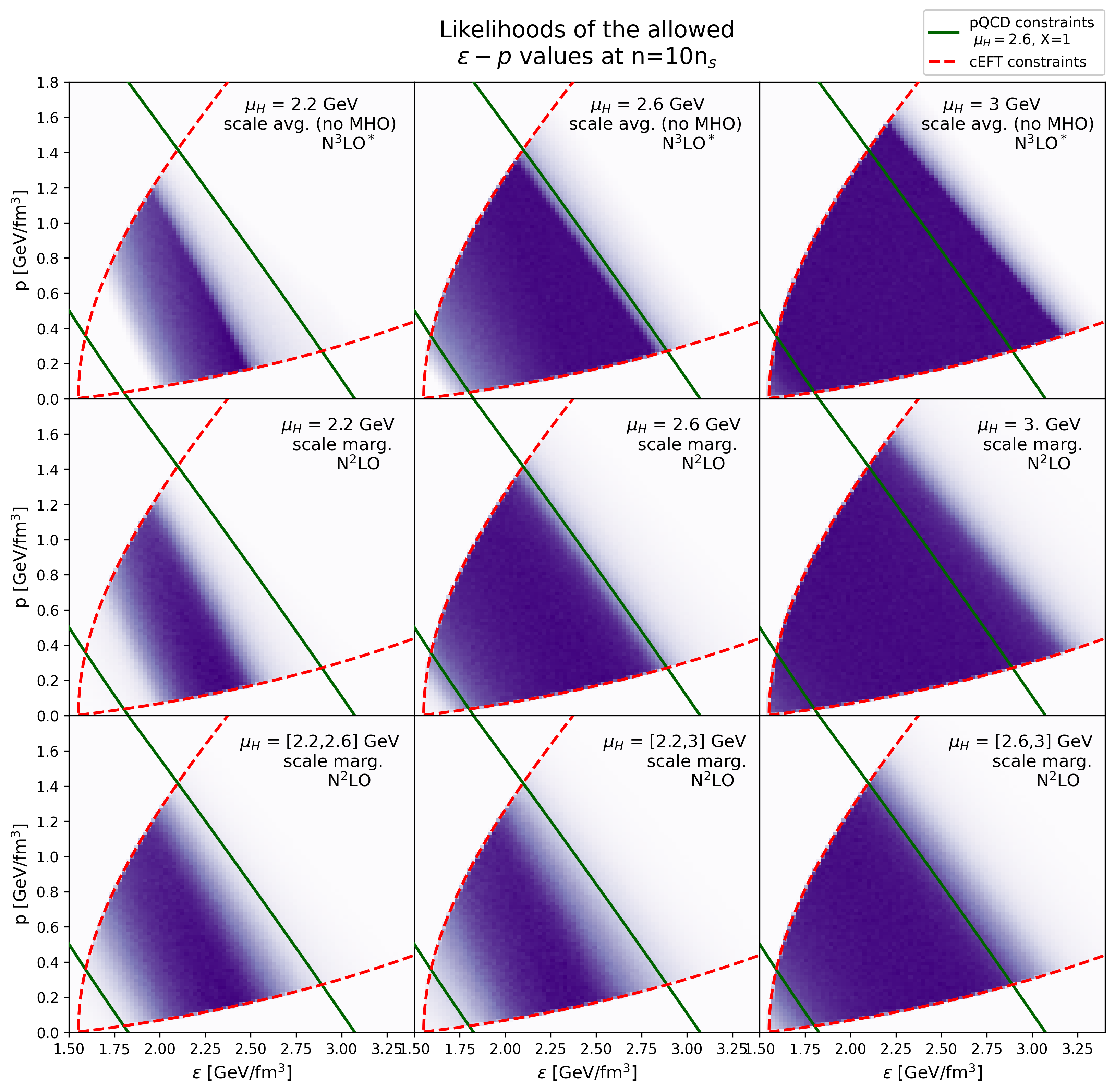}
    \caption{The panel displays different likelihood functions for the allowed $p-\e$ values at $10\ns$, using various prescriptions for estimating perturbative uncertainties. The first row corresponds to the SA prescription with a log-uniform weight for $X \in [1/2,2]$ at different reference chemical potentials, $\mu_{\rm high}$. The middle plot of the first row represents the previously used QCD likelihood function from \cref{sec:bayesian}. The second and third rows show uncertainty estimates based on the \textit{abc} model with scale marginalization over $X$, either for a fixed $\mu_{\rm high}$ or with marginalization over $\mu_{\rm high}$ within the specified range.}
    \label{fig:miho_panels}
\end{figure}

The first row represents the scale-averaging prescription using the N$^3$LO$^*$ input without MHO uncertainty estimation for fixed $\mu_\h = 2.2$, 2.6, or 3 GeV, which corresponds to $n_\h \approx$ 23, 40, and 64$\ns$. The middle subplot reproduces the likelihood plot from \cref{fig:QCD_likelihood} (displayed in linear scale). It is evident that the allowed region shrinks as the QCD input is imposed at lower chemical potentials, bringing it closer to the reference density, $10\ns$.

The MHO estimate is introduced in the second row, using the $abc$ model and SM prescription for the renormalization scale parameter $X$, with fixed $\mu_\h$ and N$^2$LO pQCD input, corresponding to the last fully computed order. For the middle subplot, the pressure distribution at $\mu_{\rm high} = 2.6$ GeV, which is used to propagate constraints to $10\ns$, is shown as a solid green line in the left plot of \cref{fig:miho_SM_SA}.

The third row demonstrates the further addition of simultaneous marginalization over both $X$ and $\mu_\h$, based on the same input as in the second row, but for three distinct ranges. The main effect of this marginalization is an additional blurring of the boundaries defined by $p_{\rm min/max}$, reducing the sensitivity to the specific choice of $\mu_\h$. Importantly, the plot shows that the previously used likelihood at $\mu_\h = 2.6$ GeV approximately reproduces a conservative choice, similar to marginalizing over the range [2.6, 3] GeV in chemical potential.

Using the likelihoods presented in \cref{fig:miho_panels}, the corresponding effects on the EoS inference are shown in \cref{fig:miho_final}. The previously inferred CI, labeled ‘Astro’, corresponds to the ‘Pulsars+$\tilde\Lambda$’ from \cref{fig:posterior_CIs}. For the QCD input, two choices of SA are considered: a fixed $\mu_\h = 2.6$ GeV (green hatched area) and $\mu_\h = 3$ GeV (red dash-dotted line), compared to SM over $\mu_\h \in [2.2, 3]$ GeV, represented by the blue dashed line. In all cases, even with the most conservative choice, it is evident that the inclusion of the QCD input leads to a softening of the EoS. The previously used likelihood SA for a fixed $\mu_\h = 2.6$ GeV is indistinguishable from the most sophisticated and agnostic likelihood computed in this section, namely SM with $\mu_\h \in [2.2, 3]$ and $X \in [1/2, 2]$, obtained using the $abc$ model. This likelihood function is also used to constrain the ensemble shown in \cref{fig:miho_fig1}, where most of the EoSs with high posterior weight (indicated by a more reddish color) are nearly aligned with the 1$\sigma$ band of the posterior distribution for the pressure.

\begin{figure}[ht!]
    \centering
\includegraphics[width=0.69\textwidth]{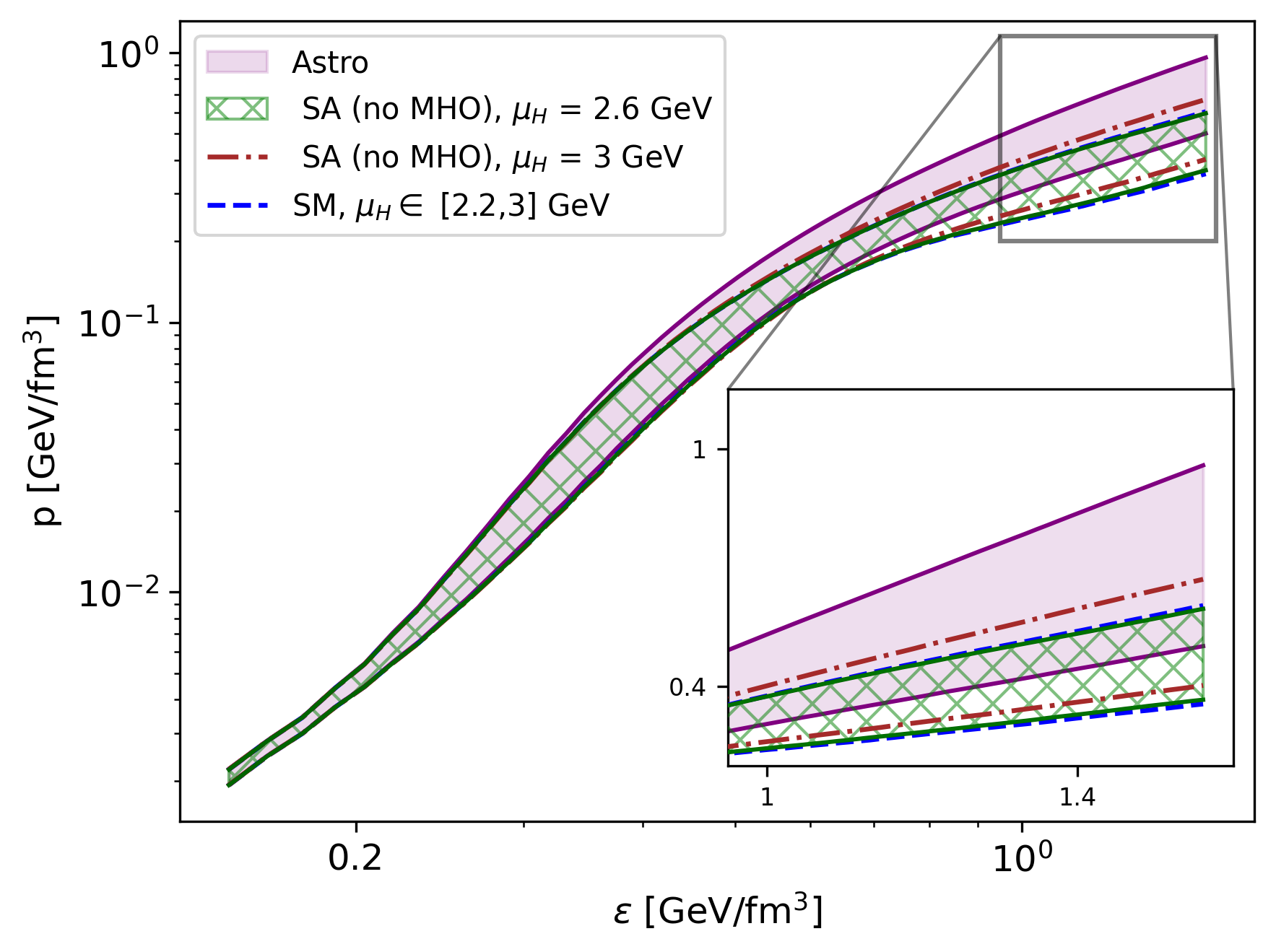}
    \caption{The impact of the QCD input with different prescriptions for uncertainty estimation on the EoS inference. The bands represent the 68\%-CI, all conditioned on astrophysical inputs and different QCD likelihood functions. The pink and green hatched bands, identical to those in \cref{fig:posterior_CIs}, correspond to “Pulsar+$\tilde{\Lambda}$” and “Pulsar+$\tilde{\Lambda}$+QCD”, respectively. Different prescriptions correspond to the likelihoods from \cref{fig:miho_panels}, with the green hatched and red dashed-dotted bands referring to the middle and last plots of the first row, while the blue dashed line corresponds to the middle plot of the last row.}
    \label{fig:miho_final}
\end{figure}

\mybox{Summary of \cref{sec:uncertainty}}{blue!20}{white!10}{
\begin{itemize}
    \item The QCD input is only mildly sensitive to the choices made in estimating perturbative uncertainties, such as the prescription used to estimate MHO, the range of the renormalization scale, and the reference density.
    \vspace{0.1cm}
    \item The softening of the EoS is a robust prediction of the QCD input against perturbative uncertainties.
\end{itemize}
}
\clearpage

\section{EoS termination density}
\label{sec:termination}

Another somewhat arbitrary aspect of NS modeling is the choice of the termination density, $\nterm$, of the EoS — that is, up to what density the EoS is extrapolated or modeled. For neutron star phenomenology, a conservative choice for the termination density is the maximum central density of a non-rotating, stable star, the TOV density. Modeling the unstable branch of neutron stars introduces additional prior dependence. However, in certain phenomenological applications, such as binary neutron star mergers or differentially rotating neutron stars, the maximum central density can exceed the TOV density \cite{Ujevic:2023vmo,10.21468/SciPostPhys.13.5.109,Cassing:2024dxp}. Furthermore, as will become clear in this section, the further pQCD constraints are propagated from $\mu_\h$ (i.e., the smaller the termination density), the less significant the impact becomes. This occurs because prior assumptions about the EoS extend only up to $\nterm$, with an abrupt change of prior just above the termination density allowing for a wide range of extreme behaviors. Thus, it is unclear what the optimal choice for the termination density is. This issue, along with its impact on EoS inference, is thoroughly examined and addressed in this section.

The conflicting conclusions regarding the constraining power of the QCD input on EoS inference were originally presented in \cite{Gorda:2022jvk} and \cite{Somasundaram:2022ztm}. In our study, with a termination density of $10\ns$, the effect of QCD, as illustrated in \cref{fig:posterior_CIs}, is significant. In contrast, in \cite{Somasundaram:2022ztm}, where the termination density was set to the TOV density, the impact appeared marginal. This discrepancy motivated us to combine efforts and thoroughly investigate the constraining power of the QCD input as a function of termination density in \cite{Komoltsev:2023zor}, which forms the basis of this section.

First, in \cref{subsec:power}, the results of EoS inference for different termination densities are presented for various observables, highlighting the strong dependence of QCD constraining power on the choice of termination density. In the next \cref{subsec:IQCD}, I address the question of which EoSs are accepted by QCD constraints at TOV densities but become incompatible at higher densities. This analysis clarifies the discrepancies observed in the results for different termination densities, revealing that the sensitivity in constraining power comes from EoSs with a unique behavior. I explicitly demonstrate the possible extensions of these unique EoSs beyond the TOV density required to be in agreement with the QCD input.

Finally, in \cref{subsec:marg_qcd}, I construct a QCD likelihood function that addresses the issue of an abrupt change in prior assumptions at $\nterm$, by penalizing extreme behavior beyond the termination density. This is achieved by marginalizing over a broad range of prior models for possible extensions between $\nterm$ and the pQCD limit. Moreover, this approach enables the incorporation of additional information from the well-converged pQCD sound speed at high densities into the QCD likelihood function.

Throughout this section, the ensemble obtained in \cref{sec:bayesian} is used, which includes the GP prior, various astrophysical inputs, and a scale-averaged prescription without MHO estimation for the QCD likelihood function. The latter is chosen for simplicity, given the previous section’s conclusion that the QCD input is only mildly sensitive to choices made in estimating perturbative uncertainties.

\subsection{Constraining power of the QCD input}
\label{subsec:power}

It is evident from \cref{fig:e_p_koku_global} that the area explicitly excluded by the QCD input on the $\e-p$ plane diminishes with decreasing number density. A similar dependence is observed in the effect of the QCD input on the posterior densities obtained by incorporating astrophysical constraints for different termination densities. \Cref{fig:koku_post} shows the posterior probability for fixed termination densities of $5\ns$ and $10\ns$. The purple-outlined region represents the posterior based solely on astrophysical inputs, while the green hatched area illustrates the additional constraints imposed by the QCD input. The blue lines outline the allowed regions (as in \cref{fig:e_p_koku_global}) resulting from simultaneous constraints from cEFT and pQCD ($X$=1, $\mu_\h = 2.6$ GeV), which arise from thermodynamic stability, consistency, and causality at fixed number densities.

\begin{figure}[ht!]
    \centering
\includegraphics[width=0.8\textwidth]{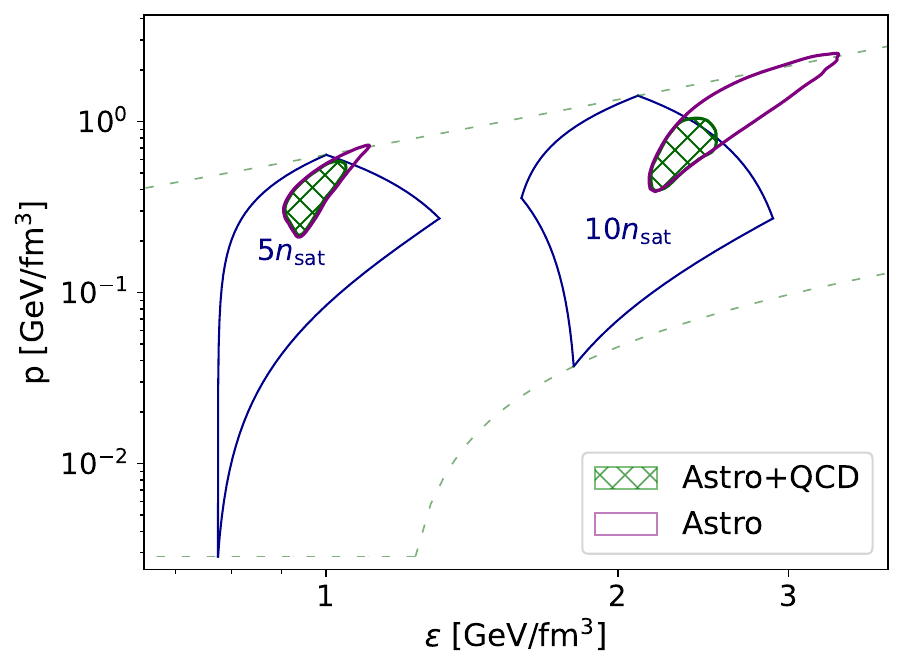}
    \caption{The propagated pQCD constraints at a fixed number density. The purple and hatched green areas represent the 68\% credible regions of the posterior density, conditioned on astrophysical and the QCD inputs ($\mu_{\rm high} = 2.6$ GeV with the SA prescription), imposed at fixed termination densities of $\nterm = 5\ns$ and $10\ns$.}
    \label{fig:koku_post}
\end{figure} 

Although the reduction in constraining power is evident when comparing the extent to which the QCD input excludes otherwise allowed area from the astrophysical posterior at $5\ns$ and $10\ns$, this effect can be quantified. The left subplot of \cref{fig:fraction_cut_QCD} shows the fraction of evidence (marginalized likelihood) that the QCD input removes from the evidence based solely on astrophysical inputs. This fraction is defined as follows:
\begin{align}
\label{eq:qcd_fraction}
    1 - \frac{\sum_i w_i^{\text{astro}} \cdot w_i^{\text{QCD}}}{\sum_i w_i^{\text{astro}}},
\end{align}
where $w_i^{\text{astro}}$ represents the likelihood assigned to an EoS with index $i$, based on the astrophysical inputs, and $w_i^{\text{QCD}}$ is the likelihood obtained from the QCD likelihood function (normalized by construction). The summation over $i$ is performed over all EoSs in the ensemble. For the resampled posterior (as introduced in \cref{sec:bayesian,fig:venn}), this quantity simply corresponds to the number of EoSs disallowed by the QCD input but allowed by the astrophysical inputs.

\begin{figure}[ht!]
    \centering
\includegraphics[width=0.95\textwidth]{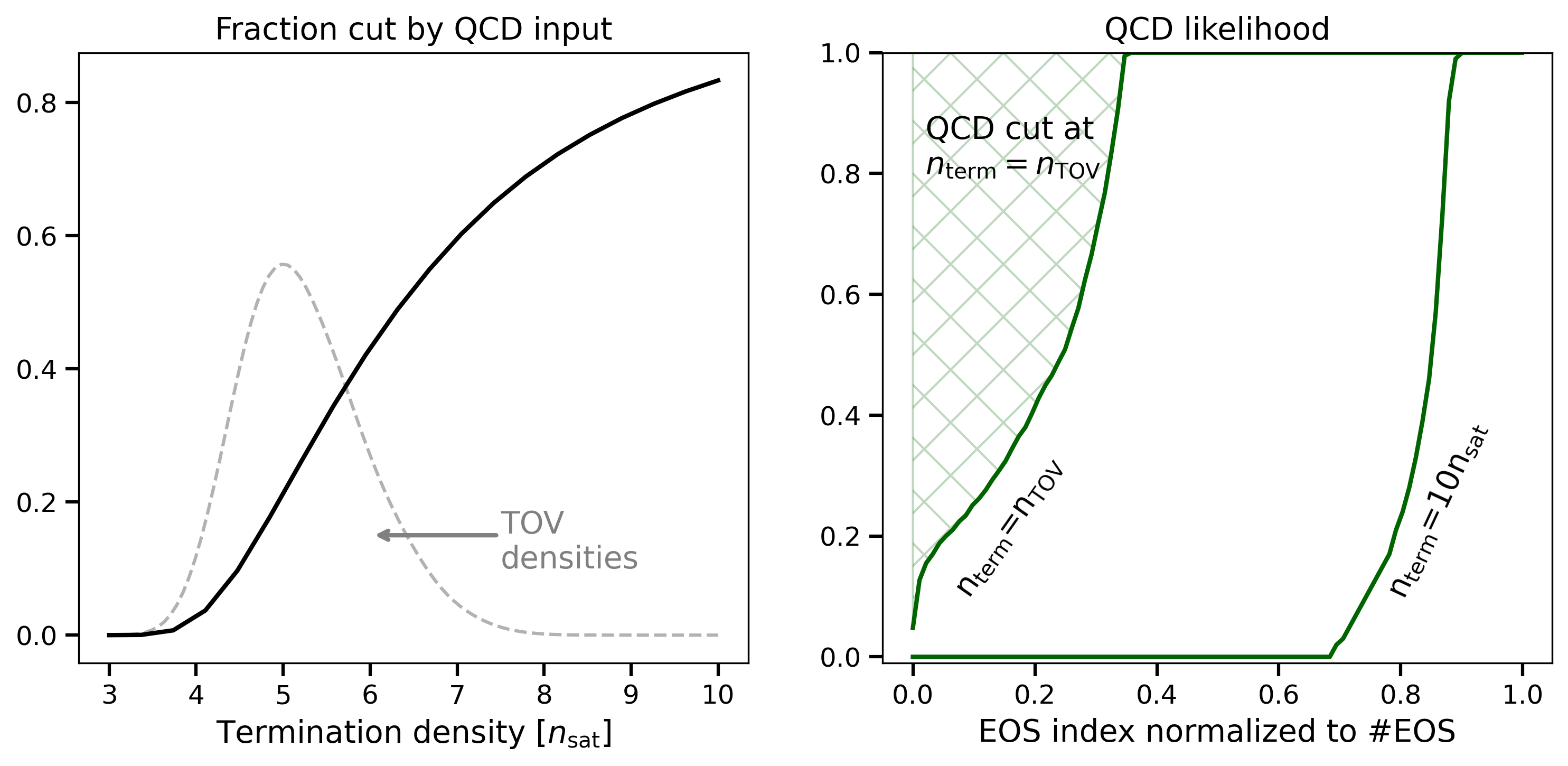}
    \caption{(Left) The fraction of the evidence removed by the QCD input, as determined by \cref{eq:qcd_fraction}. In the resampled posterior, this corresponds to the fraction of EoSs that are inconsistent with the QCD input. (Right) The sorted QCD likelihood function imposed at $\nterm=n_{\rm TOV}$, arranged according to the QCD likelihood of each EoS, as a function of the index representing individual EoSs from the resampled ensemble conditioned on astrophysical data.}
    \label{fig:fraction_cut_QCD}
\end{figure} 

The gray dashed line represents the distribution of the maximal central density of neutron stars. For EoSs with smaller TOV densities in the range of 3–4$\ns$, the impact of the QCD input is marginal, but it increases rapidly with higher termination densities. At 5$\ns$, the QCD input already excludes approximately 20\% of the marginalized likelihood, as also illustrated in \cref{fig:koku_post} (note the logarithmic scale). By 7$\ns$, which roughly corresponds to the maximal TOV density, the effect reaches around 60\%. The termination density used in \cref{sec:bayesian} is $10\ns$, where the QCD cut is approximately 80\%.

While the fraction cut defined in \cref{eq:qcd_fraction} is useful for quantifying the effect of the QCD input, it is possible, in principle, that the QCD likelihood function uniformly reduces the Bayesian weight of each EoS, leading to an overall reduction in the marginalized likelihood. In such a scenario, the information provided by the QCD input would be trivial, offering no novel constraints on the inference of the EoS. To counter this argument, the QCD likelihood function is plotted in \cref{fig:fraction_cut_QCD} (right subplot) as a function of the EoS index, ordered by increasing QCD likelihood values. The EoSs used in the analysis are taken from the resampled ensemble conditioned on astrophysical data. In other words, an EoS index with its corresponding QCD likelihood function appears in the figure only if the EoS is accepted by astrophysical constraints. The probability of an EoS being accepted is proportional to the normalized likelihood of the combined astrophysical inputs.

This plot demonstrates that the QCD input does not uniformly decrease the likelihood across all EoSs, but rather excludes some EoSs while allowing others. The EoSs with a QCD likelihood of 1 correspond to those that satisfy the criteria in \cref{eq:qcd_check} for any value of $X \in [1/2, 2]$, whereas a likelihood of 0 indicates that there is no value of $X$ for which the EoS can be connected to the pQCD limit with a stable and causal EoS. The two green lines in the right subplot of \cref{fig:fraction_cut_QCD} represent the QCD likelihood imposed at the termination densities of $n_{\rm TOV}$ (additionally highlighted by the green hatched area) and $10\ns$. The fraction cut by the QCD input at $n_{\rm TOV}$ is approximately 20\%, which can be calculated as the area of the green hatched region (this fraction increases to approximately 40\% at just 1.2$n_{\rm TOV}$).

\begin{figure}[ht!]
    \centering
\includegraphics[width=0.8\textwidth]{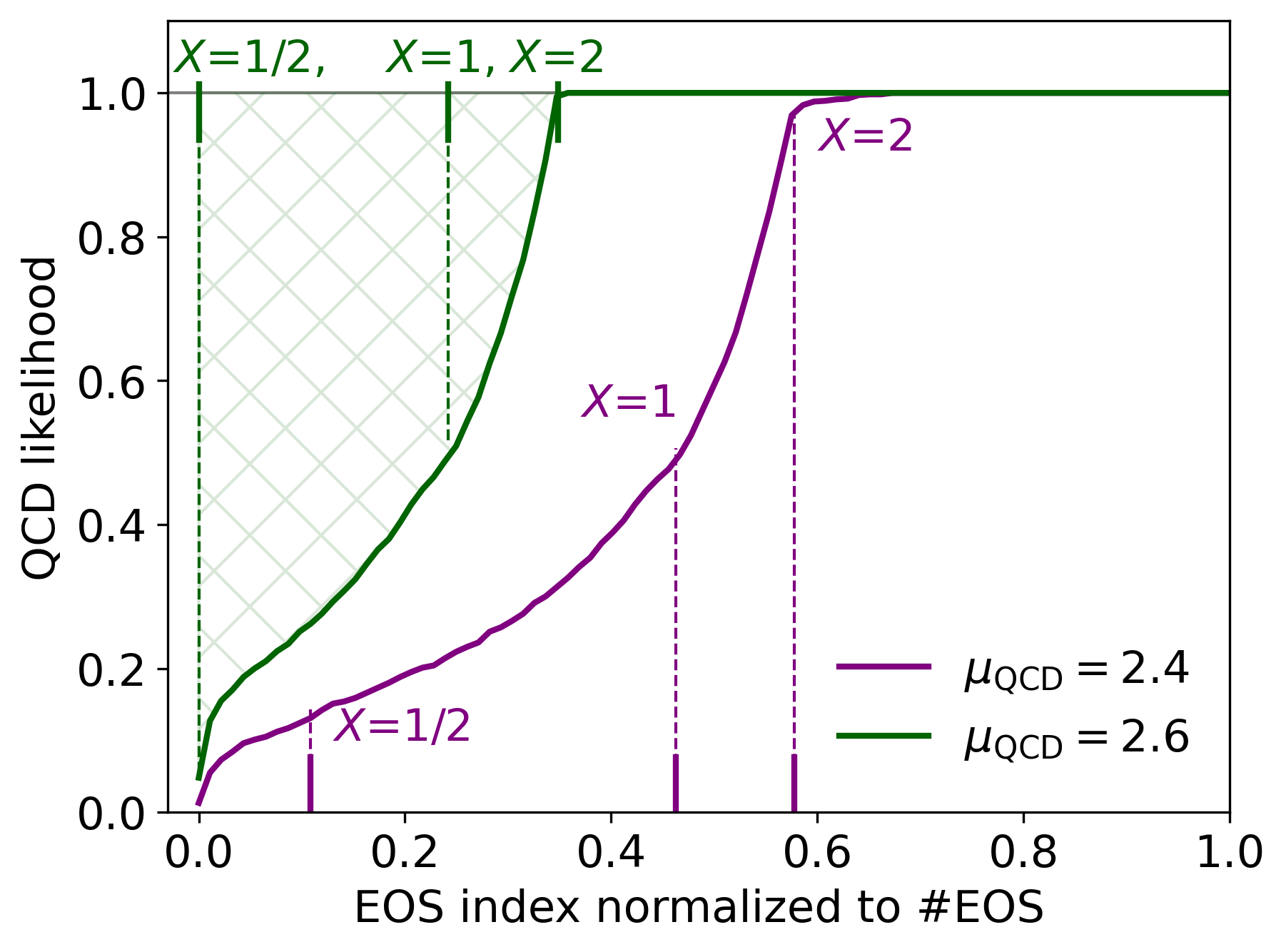}
    \caption{The sorted QCD likelihood function imposed at $\nterm = n_{\rm TOV}$ for two values of $\mu_{\rm high}$, 2.4 and 2.6 GeV, using the scale-averaging prescription. For a fixed value of $X$, the QCD likelihood function takes the form of a step function, as indicated by the dashed lines.}
    \label{fig:sorted_X_likelihood}
\end{figure}

The dependence of the sorted QCD likelihood function on the renormalization parameter $X$ is illustrated in \cref{fig:sorted_X_likelihood}. For a fixed value of $X$, the sorted QCD likelihood function behaves as a step function, assigning a likelihood of 1 if the EoS can be connected to the pQCD limit for that specific $X$, and 0 otherwise. The dashed lines represent these step functions for $X = 1/2$, 1, and 2, corresponding to two different values of $\mu_\h$, namely 2.4 GeV (shown in purple) and 2.6 GeV (shown in green). Notably, for $X = 1/2$ and $\mu_\h$, the QCD likelihood does not exclude a significant number of EoSs from the resampled ensemble. As discussed in detail in \cref{sec:uncertainty}, $X = 1/2$ obtains lower weight in the Bayesian quantification of perturbative uncertainties using the scale-marginalization prescription due to the slow convergence of the series, indicating low confidence in the pQCD calculations for small values of $X$.

Now turning to the effect of termination density on the allowed $\e-p$ values at TOV density, as shown in \cref{fig:e_p_kde_panel}, where the 68\%-credible regions are displayed for different values of $\nterm$. The purple line represents the posterior if the EoSs are conditioned only on astrophysical inputs. The green hatched, red dashed-dotted, and blue regions illustrate the posterior distributions when the EoSs are additionally conditioned with the QCD input, for $\nterm = n_{\rm TOV},\ 1.2n_{\rm TOV}$, and $10\ns$, respectively. Note that for $1.2n_{\rm TOV}$ and $10\ns$, the EoS is used beyond the density shown, with the QCD input imposed at the termination density. As discussed above and additionally shown in \cref{fig:koku_post}, the main effect of the QCD input is a softening of the EoS, disfavoring the stiffest EoSs that populate the upper-left corner of the posterior distribution of $\e_{\rm TOV} - p_{\rm TOV}$. The effect is consistent with \cref{fig:e_p_kde_panel} and \cref{fig:fraction_cut_QCD}, namely, that the constraining power of the QCD input increases significantly with higher termination density.

\begin{figure}[ht!]
    \centering
\includegraphics[width=0.8\textwidth]{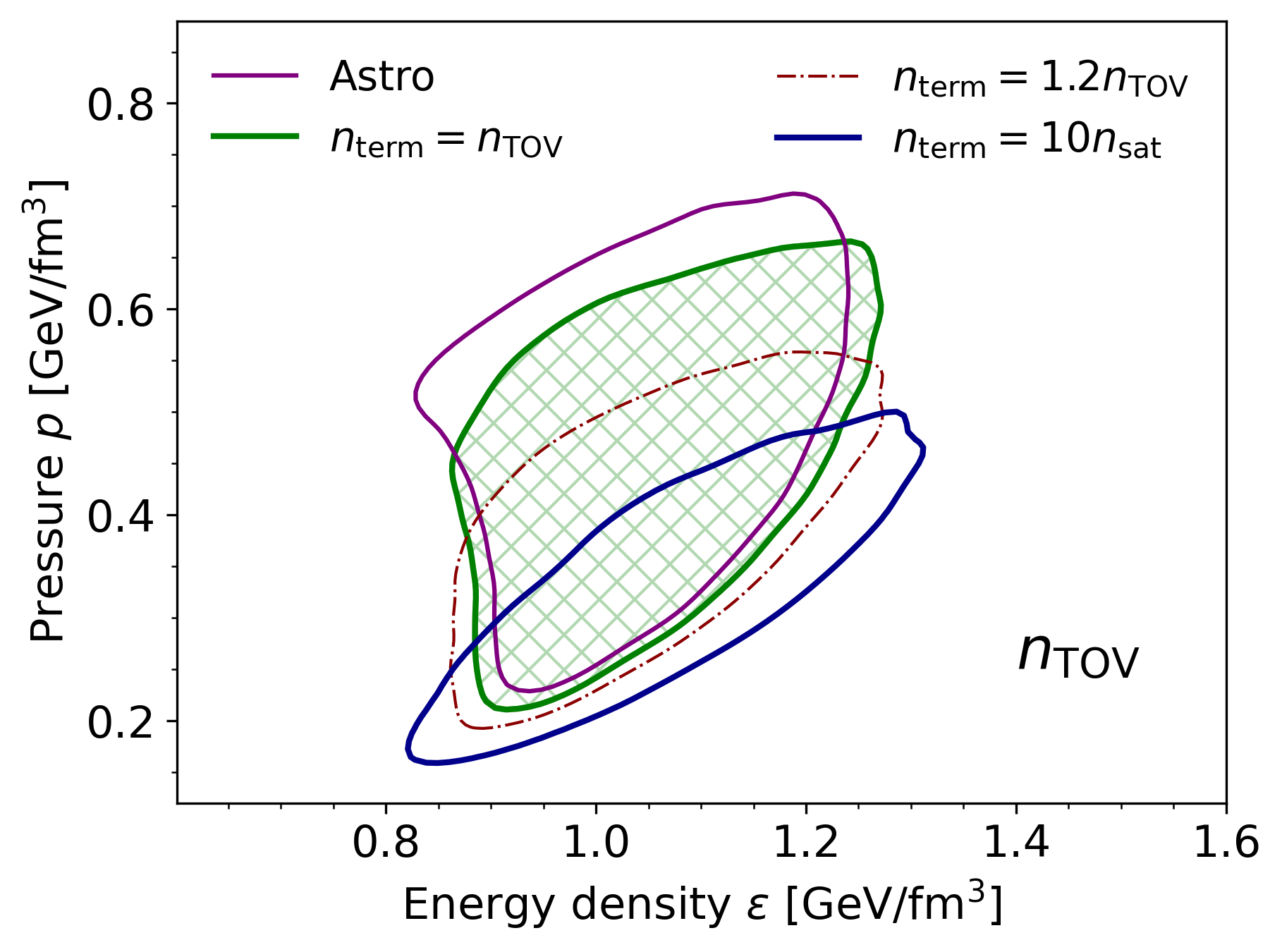}
    \caption{The 68\% credible regions of the posterior probability density for $p_{\rm TOV}-\e_{\rm TOV}$, conditioned on all astrophysical data and the QCD input, imposed at different termination densities.}
    \label{fig:e_p_kde_panel}
\end{figure}

\subsection{Which EoSs are allowed at $\nterm=n_{\rm TOV}$ but excluded at higher densities?}\label{subsec:IQCD}

To explain the effect of the termination density on the QCD input, it is important to explore how EoSs that are allowed at smaller $\nterm$ become excluded at higher densities. This analysis refers back to the criteria for the modeled EoSs in \cref{eq:qcd_check}, which can be rewritten as

\begin{align}
\label{eq:Ipqcd}
0 \leq \iqcd \equiv \frac{\Delta p - \Delta p_{\rm min}}{\Delta p_{\rm max} - \Delta p_{\rm min}} \leq 1,
\end{align}

where the new quantity $\iqcd$ is introduced, the pQCD tension index. It represents how close the EoS is to the exclusion bound. If $\iqcd \in [0,1]$, the EoS is accepted, and the allowed region through which the EoS can be extended beyond the termination density is defined by \cref{eq:eqmax,eq:eqmin}. However, if $\iqcd$ is outside the range [0,1], the EoS is excluded by the QCD input.

From \cref{fig:ep_koku_example}, it is clear that the allowed region on the $\e$-$p$ plane above the termination density varies significantly for different EoSs. To study the EoSs allowed at the TOV but excluded shortly afterward, it is important to understand how the gap in the allowed $\e-p$ values, arising from the QCD input, gradually closes. 

Once astrophysical data is imposed, the relevant bound is typically the upper bound of $\iqcd$, close to one. This occurs because the EoS needs to be stiff to satisfy astrophysical constraints, pushing it toward the lower integral constraints shown in \cref{fig:mu_n} (corresponding to the upper bound of $\iqcd$). 

\begin{figure}[ht!]
    \centering
\includegraphics[width=0.98\textwidth]{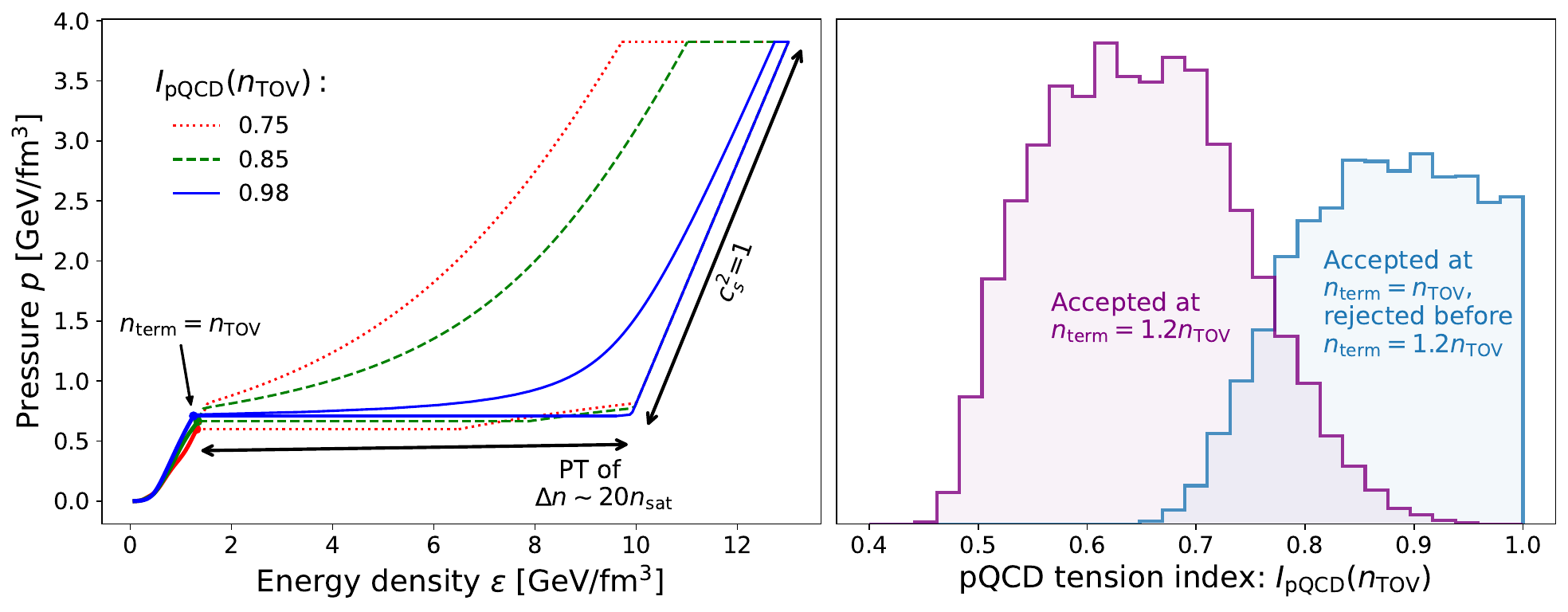}
    \caption{(Left) The allowed region an EoS must pass through to connect to pQCD ($X=1$, $\mu_{\rm QCD} = 2.6$ GeV) while maintaining stability and causality.The regions are shown for three different EoSs, terminated at $\nterm=n_{\rm TOV}$, with representative pQCD tension indices. A higher tension index results in a more restrictive allowed area, converging toward a specific EoS shape with a large first-order phase transition of $\Delta n \sim 20\ns$ and a segment with $c_s^2 = 1$. (Right) The distribution of the pQCD tension index for two sets: EoSs accepted by the QCD input at $1.2 n_{\rm TOV}$, and EoSs accepted at $n_{\rm TOV}$ but rejected before reaching $1.2 n_{\rm TOV}$. The distributions are normalized to the total number of EoSs in each set.}
    \label{fig:iqcd}
\end{figure} 

The allowed $\e-p$ values are shown in \cref{fig:iqcd} (left) for three EoSs with different pQCD tension indices at $\nterm=n_{\rm TOV}$: $\iqcd=0.75$, 0.85, and 0.98. As evident from the figure, the allowed area quickly degenerates into a very specific shape of the EoS as $\iqcd$ increases. For $\iqcd = 1$, the EoS must exhibit distinct behavior—a strong FOPT at the termination density, followed by another phase transition, with the sound speed jumping to the speed of light. This construction is derived and detailed in \cref{sec:analytic} (including the construction for $\iqcd=0$). In \cref{fig:cs2_cartoon}, the same construction for $\iqcd=1$ is shown in terms of $\cs$ and $n$, emphasizing the extreme FOPT with $\Delta n \sim 20\ns$ and the segment of $\cs=1$ approaching the pQCD limit at $\mu_{\rm QCD}=2.6$ GeV. The latter appears inconsistent with the well-converged series for the sound speed at lower chemical potentials. This fact will be utilized later in \cref{subsec:marg_qcd} to construct a new QCD likelihood function, independent of the termination density.

Note that a strong FOPT in the case of $\iqcd(n_{\rm TOV}) = 1$ does not destabilize the stars and, consequently, does not determine the location of the TOV density. This behavior requires fine-tune models, as the FOPT happens to occur just above the TOV density (within the unstable branch of NS), which is chosen as the termination density and serves as a reference point for imposing the QCD input.

\begin{figure}[ht!]
    \centering
\includegraphics[width=0.75\textwidth]{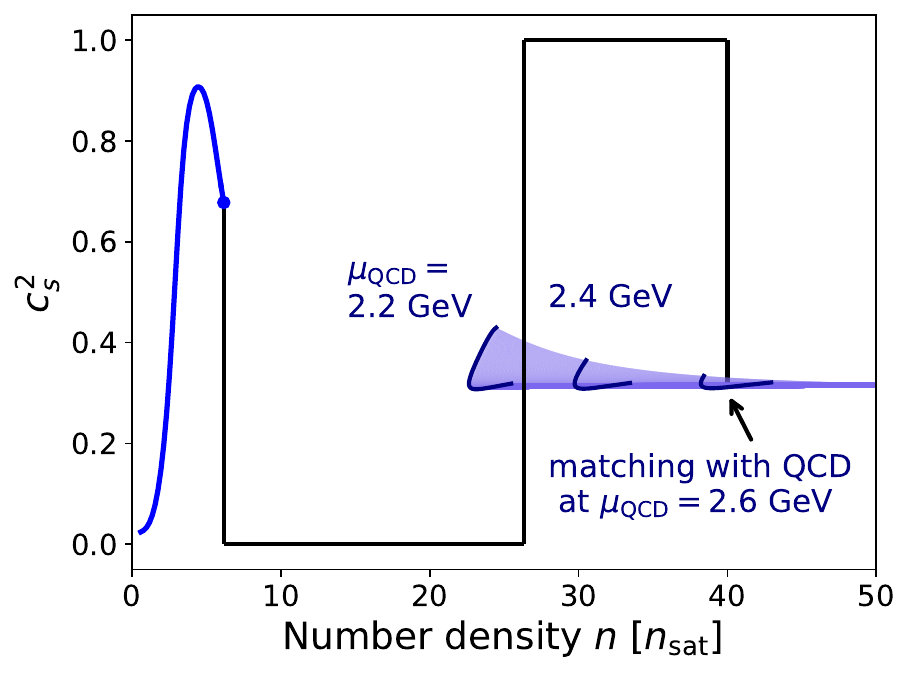}
    \caption{The EoS with $I_{\rm pQCD}=1$ at $n_{\rm TOV}$, shown in blue, must follow a specific shape above TOV density (black line) to connect to the pQCD limit at $\mu_{\rm QCD}=2.6$ GeV. This constraint forces the EoS to exhibit a large FOPT, followed by a subsequent $c^2_s=1$ segment that is inconsistent with the well-convergent N$^3$LO$^*$ pQCD calculation of the speed of sound. The latter is represented by the purple band with X variation within the range $[1/2,2]$.}
    \label{fig:cs2_cartoon}
\end{figure} 

The distributions of the pQCD tension index are shown in \cref{fig:iqcd} (right) for two mutually exclusive sets of EoS: EoS accepted at $n_{\rm TOV}$ and rejected at $1.2n_{\rm TOV}$, and EoS accepted at $1.2n_{\rm TOV}$. As mentioned earlier, the fraction cut by the QCD input is approximately 20\% at $n_{\rm TOV}$ and 40\% at 1.2$n_{\rm TOV}$. The figure indicates that EoSs contributing to this 20\% difference—those accepted at TOV but subsequently rejected between $n_{\rm TOV}$ and 1.2$n_{\rm TOV}$—tend to have a large $\iqcd$ between 0.7 and 1.

The closer the EoSs are to exclusion (i.e., the higher the pQCD tension index), the more constrained their shape becomes, approaching the construction shown in \cref{fig:cs2_cartoon}. The possible extensions of the EoS beyond TOV density for three different tension index are depicted in \cref{fig:extension}. For each EoS with $\iqcd = 0.5$, 0.75, and 0.85 at TOV density, 5000 extensions are generated between $n_{\rm TOV}$ and $15\ns$ using GP regression. These extensions are displayed in the figure only if the pQCD tension index at $15\ns$ remains below one (indicating acceptance by the QCD input), ensuring the EoS stays within the allowed envelope shown in \cref{fig:iqcd}. The color of each extension represents the $\iqcd (15\ns)$. 

\begin{figure}[ht!]
    \centering
\includegraphics[width=0.9\textwidth]{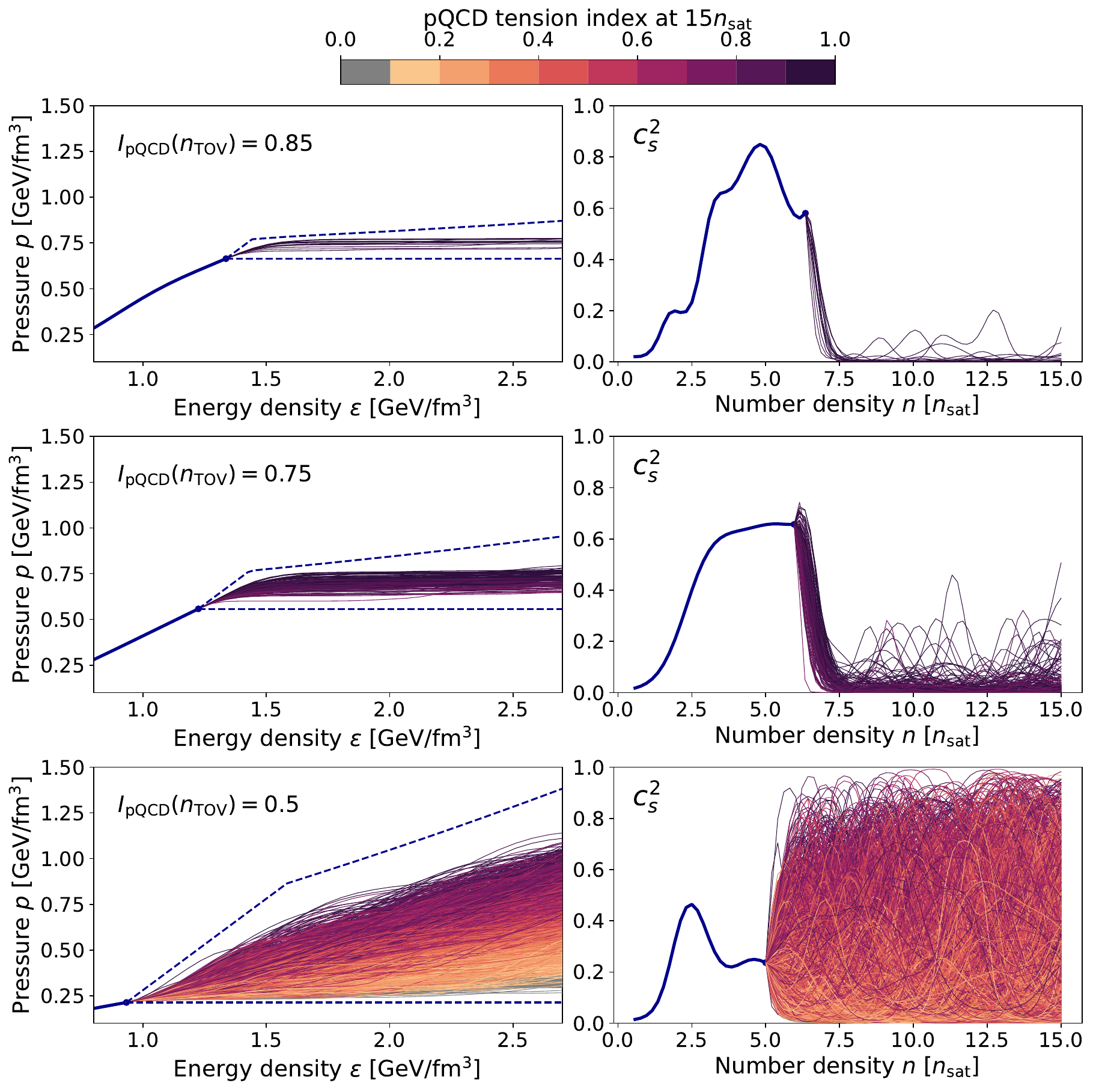}
    \caption{Possible extensions of three different EoSs with representative values of $I_{\rm pQCD}(n_{\rm TOV}) = 0.85, 0.75$, and $0.5$ beyond the TOV density. All EoSs in these extensions satisfy the pQCD constraint at $15\ns$, with color-coding indicating the value of the pQCD tension index at the last point, $15\ns$.}
    \label{fig:extension}
\end{figure} 

For $\iqcd(n_{\rm TOV}) = 0.98$, no valid EoS extensions are found in the GP prior. For $\iqcd(n_{\rm TOV}) = 0.85$, the extensions exhibit drastic softening with FOPT-like behavior up to $15\ns$. All valid samples have $\iqcd$ values near unity at $15\ns$, indicating that the EoS should resemble the extension shown in \cref{fig:cs2_cartoon} beyond $15\ns$. For $\iqcd(n_{\rm TOV}) = 0.75$, a similar trend is observed, with a large pQCD tension index at $15\ns$ but slightly higher allowed sound speed values. Finally, for $\iqcd(n_{\rm TOV}) = 0.5$, the extensions are not significantly constrained beyond $n_{\rm TOV}$, resulting in a wide range of tension indices at $15\ns$.

To quantify and compare the degree of softening beyond TOV density, the average $\cs$ of the extensions can be used. \Cref{fig:average_cs2} shows the distributions of the average sound speed for possible extensions of 100 EoSs drawn from the posterior, each with a fixed value of $\iqcd(n_{\rm TOV})$. For each EoS, the sound speed is averaged over 1000 possible extensions within the density interval $[8,15]\ns$. The distributions for $\iqcd = 0.85$ and 0.75 are heavily shifted toward lower values, with $\cs < 0.03$ and $\cs < 0.11$ at 95\% credibility, respectively. This suggests a drastic softening beyond the TOV density. Combined with a large tension index at 15$\ns$, it highlights the extremity of such EoSs. In contrast, for $\iqcd = 0.5$, the distribution resembles the prior, indicating that no significant constraints are being imposed on the EoS in the range $n\in[8,15]\ns$.

\begin{figure}[ht!]
    \centering
\includegraphics[width=0.75\textwidth]{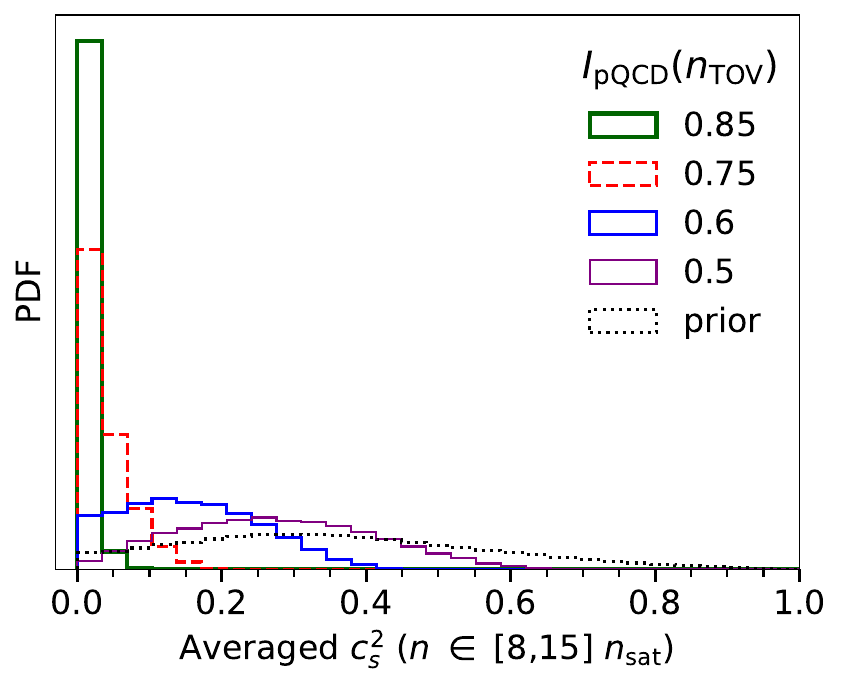}
    \caption{The distribution of the averaged speed of sound for 1000 possible extensions of 100 EoSs, drawn from an ensemble with a fixed pQCD tension index. Each EoS is used up to $\nterm = n_{\rm TOV}$, with extensions reaching up to $15n_s$, while the sound speed is averaged over the range $n \in [8,15]n_s$.}
    \label{fig:average_cs2}
\end{figure} 

The analysis in this section clarifies the sensitivity of the QCD input to the termination density. Note that the QCD input must be imposed at the highest modeled density, i.e., the termination density. Using an EoS beyond this density is incorrect, as it fails to ensure consistency with the pQCD limit. The number of EoSs excluded based on the QCD input grows rapidly with increasing $\nterm$. The difference in constraining power arise from EoSs with particular shapes. Most EoSs that are accepted by QCD constraints at the TOV density but are rejected shortly afterward tend to have a high pQCD tension index. These EoSs exhibit extreme behavior, characterized by drastic softening followed by a segment with high sound speed $\cs$, as shown in \cref{fig:cs2_cartoon} and \cref{fig:extension}.

While such behavior cannot be ruled out by the QCD input at TOV density, it represents a substantial change from prior behavior below TOV density. These EoSs are not penalized by the QCD input at TOV density and cannot be excluded solely based on thermodynamic consistency, causality, and stability. However, they may be inconsistent with the pQCD sound speed at higher densities. As of now, no microphysical model suggests this kind of behavior.

\subsection{Marginalization over EoS extensions}
\label{subsec:marg_qcd}

The final section of this chapter addresses the asymmetry of the prior below and above $\nterm = n_{\rm TOV}$. While extreme behavior of the EoSs below the TOV density is penalized by the QCD input, behavior above this density can potentially exhibit an unlimited number of FOPTs and allow the sound speed to approach the speed of light nears the pQCD limit. By addressing this asymmetry, it becomes possible to construct a QCD likelihood function that is less dependent on the termination density.

This can be achieved by introducing marginalization over a set of possible EoS extensions beyond the TOV density. Using GP regression, these extensions are constructed starting from the high-density limit and extrapolated down to lower densities. This approach explicitly models the EoS between $\nterm$ and the pQCD limit. While it introduces some model dependence to the results, it also enables the incorporation of additional information from the pQCD limit, particularly the well-convergent series of the sound speed.

Two options are considered: the first option, referred to as “prior”, involves conditioning the GP ensemble with the pQCD limit at 40$\ns$; the second, called “conditioned”, involves conditioning over a larger density range [25, 40]$\ns$ with the pQCD sound speed. The hyperparameters for the new GP ensembles anchored to the pQCD limit are chosen as follows:
\begin{equation}
\ell \sim \mathcal{U}(1\ns, 20\ns )\, , \, 
\eta \sim \mathcal{N}(1.25, 0.25^2)\, , \, 
\bar{c}_s^2 \sim \mathcal{N}(0.3, 0.3^2)\, . \, 
\end{equation}
This differs from the hyperparameters used to generate the NS EoS \cref{eq:hyper}, as it allows for a larger correlation length $\ell$ due to the broader extrapolation interval. Additionally, the mean sound speed is set to the conformal value of 1/3, but with a higher standard deviation, allowing a broad range of different EoSs. For the “conditioned” model, the standard deviation of the training data, $\sigma_n$, is set to twice the scale-averaged uncertainty of the pQCD calculation.

A sample of EoSs for the “prior” model (brown dashed lines) and the “conditioned” model (magenta solid lines) is shown in \cref{fig:extensions_hierarchical} (left). The primary difference between these two sets lies in the range [25,40]$\ns$, where the “conditioned” EoSs additionally incorporate information from the pQCD sound speed. 

\begin{figure}
\centering
  \includegraphics[height=0.36\textwidth]{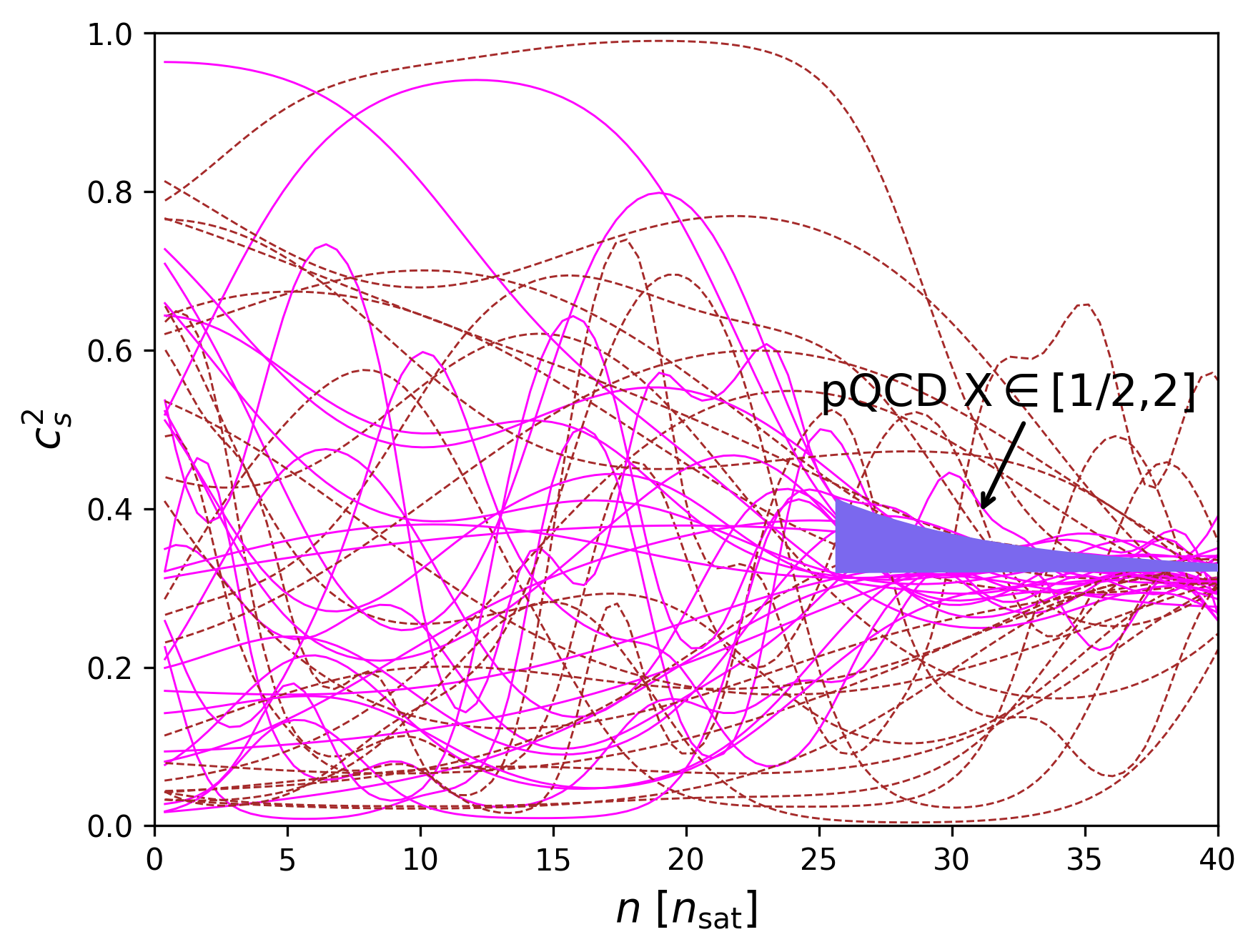}
  \includegraphics[height=0.36 \textwidth]{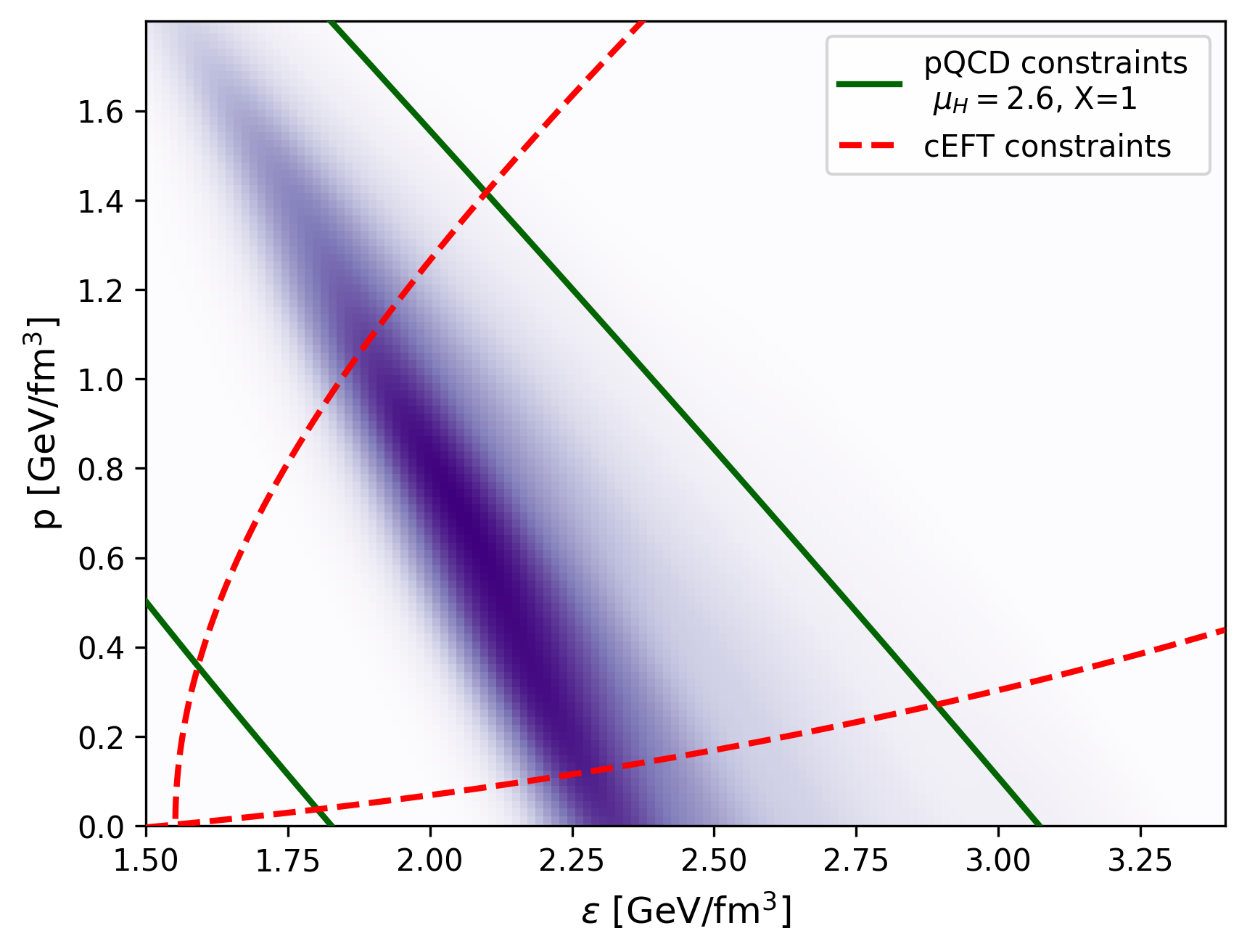} 
\caption{(Left) A sample of EoSs extrapolated from the pQCD limit using GP. The dashed brown EoSs are conditioned on pQCD results above $40\ns$, while the pink EoSs are additionally conditioned on pQCD $c^2_s$ in the range $[25,40]\ns$. (Right) The QCD likelihood function, obtained by marginalizing over the EoSs generated by the conditioned GP (shown in pink in the left plot) at $10\ns$.}
\label{fig:extensions_hierarchical}
\end{figure}

For each fixed slice of number density, EoSs extrapolated from the high-density limit form a prior distribution of $\e-p$ values. This distribution provides an estimate of how easily different endpoints of low-density NS EoSs can be connected to the pQCD limit. By applying kernel-density estimation to this distribution, it can be interpreted as a QCD likelihood function. This construction marginalizes over all possible extensions of EoSs generated using the hierarchical model between $\nterm$ and pQCD limit. 

As an example, the QCD likelihood function is shown in \cref{fig:extensions_hierarchical} (right) for a fixed $\nterm = 10\ns$ (cf. \cref{fig:miho_panels}). The resulting likelihood function can be applied to the low-density NS EoS generated in \cref{sec:bayesian} at $\nterm = n_{\rm TOV}$ and corresponding  $\e_{\rm term}-p_{\rm term}$. The marginalized QCD likelihood reflects the number of EoSs extrapolated from the pQCD limit that pass near $\e_{\rm term}-p_{\rm term}$ for a fixed $\nterm$, thereby contributing to the kernel density estimation. If the low-density EoS can only connect to the pQCD limit through a limited number of extreme EoSs (e.g., those with a high tension index), it is penalized, as these extreme EoSs are not represented in the GP prior. The marginalized QCD likelihood function is publicly available with an easy-to-use Python implementation \cite{komoltsev_2025_15407795}.

Note that this approach differs from the procedure used previously in \cref{sec:analytic} and \cref{sec:uncertainty}, where weights were assigned based on scale-averaging (or marginalization) over the renormalization scale parameter $X$. However, if an EoS does not have a stable, causal, and consistent extension between the termination density and the pQCD limit, it would obtain zero weight from any QCD likelihood function used.

The posterior distributions of the $\e-p$ regions at $n_{\rm TOV}$ with the marginalized QCD input is shown in \cref{fig:extensions_hierarchical2}. The overall effect of marginalization is similar to imposing the standard QCD input at a higher termination density, resulting in a softening of the EoS. \change{Incorporating additional information from the pQCD sound speed into the conditioned GP results in a slightly stronger softening of the EoS compared to the prior GP, though the two remain very similar. The marginalized QCD likelihood function is less sensitive to the termination density than the standard QCD input, but $\nterm = 10n_{\rm TOV}$ still provides stronger constraints.}

\change{This residual dependence on termination density when using the marginalized QCD likelihood arises from two factors. First, the prior still changes due to different hyperparameters used for the low- and high-density priors. To remain conservative, the hyperparameters used for the extensions allow for a broader variety of EoSs, including longer correlation lengths, as they span a wider density range. Second, any QCD likelihood function imposed at the TOV point introduces a discontinuity in the speed of sound\footnote{\change{In the case where $c_s^2$ drops to a lower value, it can be interpreted as a second-order phase transition.}}, where the EoS switches from the low-density prior to the higher-density extensions over which the marginalization is performed. The discontinuity allows more abrupt changes in the sound speed, which results in a stiffer EoS at TOV compared to the smooth prior (e.g., a GP extended up to $10\ns$). Note that imposing the marginalized QCD likelihood function at $10\ns$ also introduces such a discontinuity at that density. However, the EoSs at TOV densities are largely insensitive to this effect, as most EoSs soften before reaching $10,n_s$.}

\begin{figure}
\centering
\includegraphics[height=0.5 \textwidth]{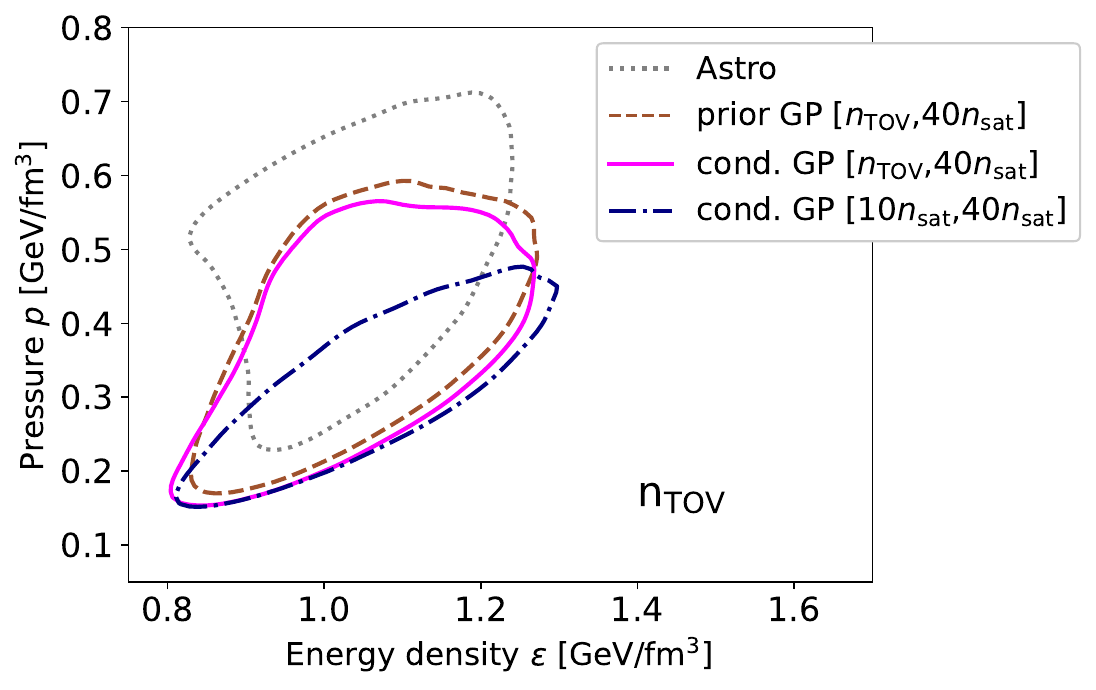} 
\caption{The 68\% credible regions of the posterior probability density for $p_{\rm TOV} - \e_{\rm TOV}$, conditioned on all astrophysical data and the marginalized QCD input. The labels ‘cond’ and ‘prior’ indicate whether the GP is additionally conditioned on pQCD $c^2_s$.}
\label{fig:extensions_hierarchical2}
\end{figure}

Lastly, the comparison of different QCD inputs is illustrated in \cref{fig:softening}. The key point is that, regardless of the chosen prescription, the overall effect remains the same—disfavoring high pressures in the most massive NSs. The sensitivity to different prescriptions arises from the stiff EoSs with high pQCD tension index. Such EoSs must exhibit a specific behavior beyond the termination density to remain consistent with the high-density limit, as discussed in \cref{subsec:IQCD}. Specifically, this behavior includes strong softening over a large density range of $\Delta n \sim 20\ns$, followed by a segment with a high speed of sound approaching the pQCD limit. Currently, no microphysical model supports such an abrupt change in prior at exactly $\nterm=n_{\rm TOV}$. 

The constraining power depends on how these EoSs are penalized. For $\nterm = n_{\rm TOV}$, such EoSs are only marginally penalized. While it is a conservative choice for modeling the NS EoS, as it requires no additional assumptions about the unstable branch, an even more conservative approach could limit the EoS inference to the heaviest observed mass, around 2.1$M_\odot$. This avoids assumptions about the density range between 2.1$M_\odot$ and the TOV mass, where no astrophysical data is available.

\begin{figure*}
\centering
\includegraphics[width = 0.42\textwidth]{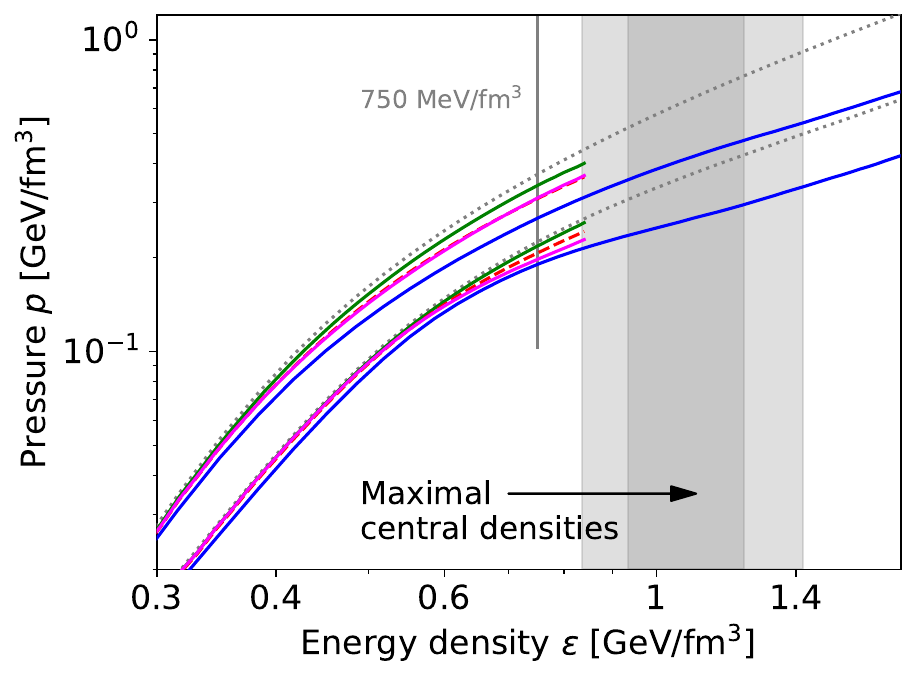}
\includegraphics[width = 0.56\textwidth]{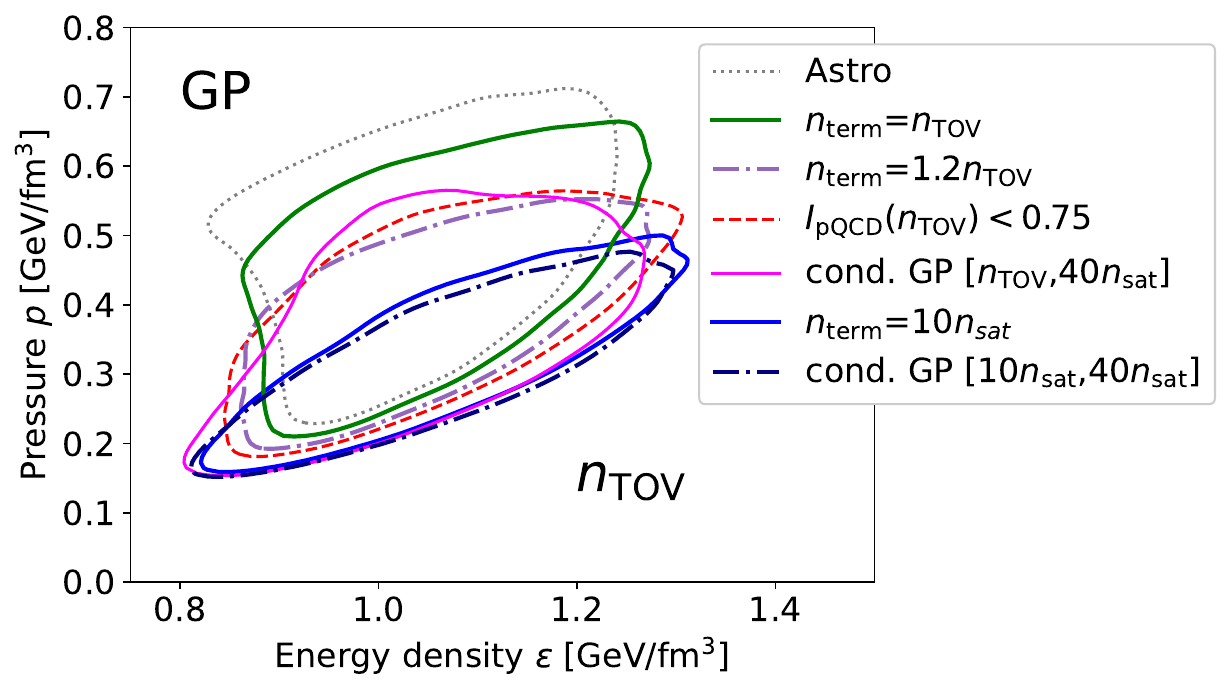}
\caption{The effect of the QCD input on EoS inference using different prescriptions for penalizing extreme behavior above TOV density. (Left) The 68\% CI for pressure as a function of energy density, extending up to the lower bound of the 95\% CI of $\e_{\rm TOV}$. For prescriptions utilizing the EoS above TOV density, the posterior probability is accessible across the entire plotted range. However, if the QCD input is imposed at $\nterm = n_{\rm TOV}$, only the conditional probability $P(p|\e, \e < \e_{\rm TOV})$ can be accessed. (Right) The 68\% credible regions of the posterior probability density for $p_{\rm TOV} - \e_{\rm TOV}$, conditioned on all astrophysical data and different QCD inputs.}
\label{fig:softening}
\end{figure*}

By introducing reasonable model dependence, such as extending the EoS modeling slightly beyond the TOV density, one can leverage the QCD input effectively. EoSs, like the one shown in \cref{fig:cs2_cartoon}, are excluded from the ensemble shortly after the TOV density and do not contribute to the posterior, thereby softening the overall results. As evident from \cref{fig:softening} (right), with $\nterm=1.2n_{\rm TOV}$, the overall impact is similar to using the \change{marginalized} QCD likelihood function. Additionally, these results effectively correspond to excluding EoSs with a high pQCD tension index, $I_{\text{pQCD}} < 0.75$, at $n_{\text{TOV}}$, as indicated by the red dashed line.

\change{Using a higher termination density, such as $10\ns$, introduces an even stronger effect for both marginalized and standard QCD inputs. As discussed in this subsection, the reason is that any QCD input imposed at $\nterm = n_{\rm TOV}$ introduces a discontinuous behavior of the sound speed, allowing stiffer EoSs with high $\iqcd$ at $n_{\rm TOV}$ — which are excluded by continuous priors.}

\mybox{Summary of \cref{sec:termination}}{blue!20}{white!10}{
\begin{itemize}
    
    \item The constraining power of the QCD input strongly depends on the termination density of EoS. Sensitivity arises from EoSs with high $\iqcd(n_{\rm TOV})$, requiring drastic softening beyond $n_{\rm TOV}$, followed by a high sound speed segment to match the high-density limit.
    \vspace{0.1cm}
    \item These EoSs are not penalized by the conservative QCD input choice at $n_{\rm TOV}$ but excluded shortly after. The choice of $\nterm=n_{\rm TOV}$ introduces \change{a discontinuity of $c^2_s$} and abrupt change in prior.
    \vspace{0.1cm}
    \item A marginalized QCD likelihood function (available in \cite{komoltsev_2025_15407795}) addresses prior asymmetry below and above $\nterm$, incorporates additional information from the pQCD sound speed, \change{but remains sensitive to a discontinuity in $\cs$ at $n_{\rm TOV}$}. 
    \vspace{0.1cm}
    \item Approaches to penalize extreme behavior above the TOV density — such as choosing a higher $\nterm$, excluding EoSs with high $\iqcd$, or employing a marginalized QCD likelihood — introduce additional but distinct model dependencies, yet ultimately result in a similar softening of the EoSs.
    \vspace{0.1cm}
    \item Any application of the EoS beyond the TOV density (e.g., in BNS merger) requires modeling an unstable branch. The QCD input must be imposed at the highest density used.
    
\end{itemize}
}

\clearpage}


%
\mainfont{\chapter{Cores of neutron stars}
\label{chpt:cores}

As mentioned in the introduction, the physics of NS cores remains largely unexplored. At such extreme densities — the highest in the universe — protons and neutrons may dissolve into their constituent particles, quarks and gluons, forming a new phase known as quark matter (QM). A phase transition from hadronic matter to QM is expected to occur at an intermediate density between the low- and high-density limits, described by cEFT and perturbative QCD, respectively. However, there is currently no theoretical framework to determine where and how this PT happens. It remains a possibility that such a PT could occur within the density range of the stable branch of neutron stars, resulting in the formation of QM cores. Alternatively, the PT could manifest as a discontinuous density jump in the case of first-order PT, leading the collapse of the neutron star into a black hole.

It was originally proposed in \cite{Annala:2019puf} that softening of the EoS might indicate a phase change to quark matter. The conclusions from previous sections suggest that this softening is a robust prediction of the novel QCD input.\footnote{In the original paper \cite{Annala:2019puf}, the authors interpolated across two orders of magnitude between cEFT and pQCD, so the softening is obtained through explicit modeling of the EoS between $\ntov$ and $n_{\rm QCD}$.} This conclusion holds when accounting for perturbative uncertainties. While the choice of termination density $\nterm = n_{\rm TOV}$ results in less pronounced softening, the approaches considered in the previous chapter to penalize extreme behavior above the TOV density result in significant softening and a change in EoS behavior that can be interpreted as a phase transition.

In this chapter, I explored in detail the physical interpretation of the softening by reproducing the results from \cite{Annala:2019puf} within a fully Bayesian framework, incorporating state-of-the-art astrophysical and theoretical inputs. Using the previously generated GP ensemble, along with piecewise-polytropic and piecewise-linear-$\cs$ EoSs, I first examine the possibility of a crossover, i.e. a smooth transition, to quark matter. The definition used to identify the crossover to QM, along with the results of the Bayesian inference, is detailed in \cref{sec:qm} and is based on \cite{Annala:2023cwx}. A limitation of this study, as well as most Bayesian inferences, is the absence of explicit first-order phase transitions in the smooth prior. Any arbitrarily rapid crossovers that mimic FOPT-like behavior are exponentially suppressed in the prior. This is addressed in \cref{sec:fopt} (based on \cite{Komoltsev:2024lcr}), where I extend the GP ensemble with explicit modeling of FOPTs and compare Bayesian factors for different scenarios of the phase transition.

\section{Crossover to quark matter cores}
\label{sec:qm}

The objective of this section is to quantify the posterior probability of a crossover — a gradual, smooth transition to QM. The process involves two steps. First, a conformal behavior of the EoS is established in the cores of the most massive NSs using the criteria introduced in \cref{subsec:conformal}. While hadronic matter breaks scale invariance due to chiral symmetry breaking, QM at high-densities is nearly conformal, with this conformality only mildly broken by the small masses of up, down, and strange quarks, as well as loop effects. The second step is to check the consistency of the conformal matter with deconfinement behavior by analyzing the active degrees of freedom (DOF). The transition, from being described by individual protons and neutrons to quarks and gluons, is accompanied by a rapid increase in active DOF, as hadronic matter inherently has fewer.

The phase change leaves a clear imprint on the thermodynamic properties of the EoS. These signatures can be studied using Bayesian inference, with the softening of the EoS serving as a notable example. In \cref{subsec:QMposterior}, the results of Bayesian inference for various quantities, along with the posterior probability of the crossover to QM based on the previously introduced criteria, are presented.

Here, I briefly outline the various astrophysical inputs used throughout this and the next section. The technical details regarding the implementation of these inputs are covered in \cref{sec:bayesian}(b). A summary of pulsar observations, including radio mass measurements and X-ray mass-radius measurements, is provided in \cref{table:astro} (the mass distributions are illustrated in \cref{fig:mass_measurments}). Details of the models used for NICER, X-ray bursts, and quiescent low-mass X-ray binaries can be found in \cite{Annala:2023cwx}. The mass priors are flat and specified in the table for each pulsar measurement. However, the factor $1/(m_{\rm TOV} - m_{\rm min})$ introduced in \cref{eq:razor} is omitted. Along with pulsar measurements, binary TD and BH hypotheses are imposed following \cref{sec:bayesian}. For the GP ensemble, the QCD likelihood function is used with the standard scale-averaging prescription and $\nterm = 10\ns$. 

\begin{table}[ht!]
\centering
\begin{NiceTabular}{lcc}[code-before = \rowcolor{blue!10}{2,4,7,15}]

\toprule
\textbf{Name} & \textbf{Mass prior [$M_\odot$]}&\textbf{Ref.}\\
\addlinespace[0.3em]
\multicolumn{3}{c}{Radio measurement} \\
\addlinespace[0.3em]
PSR J0348+0432 &$\mathcal{N}$(2.01,0.04$^2$) & \cite{Antoniadis:2013pzd}\\
\addlinespace[0.3em]
\multicolumn{3}{c}{NICER pulsars} \\
\addlinespace[0.3em]

PSR J0030+0451&$\mathcal{U}$(1.0,2.5) &\cite{Miller:2019cac,Riley:2019yda} \\
PSR J0740+6620&$\mathcal{N}$(2.08,0.07$^2$) & \cite{Fonseca:2021wxt,Miller:2021qha,Riley:2021pdl}\\
\addlinespace[0.3em]
\multicolumn{3}{c}{qLMXB systems} \\
\addlinespace[0.3em]

M13&$\mathcal{U}$(0.8,2.4)& \cite{Shaw:2018wxh}\\
M28&$\mathcal{U}$(0.5,2.8)& \cite{Steiner:2017vmg}\\
M30&$\mathcal{U}$(0.5,2.5)& \cite{Steiner:2017vmg}\\
$\omega$ Cen&$\mathcal{U}$(0.5,2.5)& \cite{Steiner:2017vmg}\\
NGC 6304&$\mathcal{U}$(0.5,2.7) & \cite{Steiner:2017vmg}\\
NGC 6397&$\mathcal{U}$(0.5,2.0)& \cite{Steiner:2017vmg}\\
47 Tuc X7 &$\mathcal{U}$(0.5,2.7)& \cite{Steiner:2017vmg}\\
\addlinespace[0.3em]
\multicolumn{2}{c}{X-ray bursters} \\
\addlinespace[0.3em]
4U 1702-429&$\mathcal{U}$(1.0,2.5) & \cite{Nattila:2017wtj}\\
4U 1724-307&$\mathcal{U}$(0.8,2.5)& \cite{Nattila:2015jra} \\
SAX J1810.8-260&$\mathcal{U}$(0.8,2.5)& \cite{Nattila:2015jra}\\
\bottomrule
\end{NiceTabular}
\caption{A summary of radio mass measurement and X-ray mass-radius measurements considered in this chapter.}
\label{table:astro}
\end{table}

This section introduces two prior ensembles: the GP ensemble generated in \cref{sec:bayesian} and a parametric interpolation approach. The latter uses piecewise-polytropic or piecewise-linear-$\cs$ EoSs with a varying  number of intermediate segments between the low- and high-density limits, denoted e.g., $c^2_{s,4}$ for four segments \cite{Annala:2019puf}. For the parametric interpolation, the parameter space is sampled using a Markov-Chain-Monte-Carlo (MCMC) method implemented with the \texttt{emcee} sampler \cite{Foreman-Mackey_2013}. As will be demonstrated later, the inference results are nearly independent of the choice of prior; therefore, this section primarily focuses on the previously used GP ensemble. Detailed information about the implementation of parametric interpolation between cEFT and pQCD limits, as well as Monte Carlo sampling, can be found in \cite{Annala:2019puf,Annala:2023cwx}.

\subsection{Conformality criteria}
\label{subsec:conformal}

The conformal symmetry leaves a distinct signature on the quantities of the EoS, such as the speed of sound $\cs$, the polytropic index $\gamma$, the normalized trace anomaly $\Delta=1/3-p/\e$ and its logarithmic derivative $\Delta’$, and the pressure normalized to the Fermi–Dirac free pressure $p/p_{\rm free}$. These quantities can be expressed as functions of pressure and energy density:
\begin{align}
    &\cs =  dp/d\e, \ \ \ \ \ \ \ \ \ \ \ \gamma = d\ln p / d\ln \e, \\
    &\Delta = 1/3-\cs/\gamma,\ \ \ \ \ \ \Delta'=d\Delta/d\ln \e = \cs(1/\gamma-1). 
\end{align}
In \cref{table:conformal}, these quantities are summarized for various density regions, including those calculated within cEFT up to around nuclear saturation density, characteristic properties of dense nuclear matter (NM) averaged over nuclear matter models in the region where most agree (up to $\sim 3\ns$), the pQCD limit at high densities above 40$\ns$, conformal field theory (CFT), and FOPT. The properties of dense NM are analyzed using hadronic models publicly available in the CompOSE database \cite{Typel:2013rza} and are presented in \cref{fig:models,fig:ppfree_models}. 

\begin{table}[t]
\centering

\begin{NiceTabular}{lccccc}[code-before = \rowcolor{blue!10}{2,4,6,8}]

\toprule

	& CEFT & Dense NM & Pert.~QM & CFTs & FOPT\\
\midrule
$\cs$	& $\ll 1$ &$[0.25,0.6]$ &$\lesssim 1/3$	&1/3  &0		\\
$\Delta$ & $\approx 1/3$	& $[0.05,0.25]$ &$[0,0.15]$	&0  & $1/3-p_\mathrm{PT}/\e$		\\
$\Delta'$ & $\approx 0$	& $[-0.4,-0.1]$ &$[-0.15,0]$	&0  & $1/3-\Delta$		\\
$d_{\mathrm{c}}$ & $\approx 1/3$	& {$[0.25,0.4]$} &$\lesssim 0.2$	&0  & $\geq 1/(3\sqrt{2})$		\\
$\gamma$ &$\approx 2.5$	& {$[1.95,3.0]$} &$[1,1.7]$	&1  &0		\\
$p/p_\text{free}$	&$\ll 1$ & {$[0.25,0.35]$} & $[0.5,1]$	& --- & $p_\mathrm{PT}/p_\text{free}$		\\
\bottomrule
\end{NiceTabular}
\caption{Characteristic values of various dimensionless quantities for strongly interacting matter across different density regions: cEFT is reliable up to nuclear saturation density, while dense nuclear models (NM) refer to densities above cEFT but below approximately $3\ns$, where most models still agree (see \cref{fig:models,fig:ppfree_models}). Pert. QM refers to pQCD calculations, which are reliable in the region $n \gtrsim 40\ns$. Additionally, characteristic properties are summarized for conformal field theories (CFTs) in 3+1 dimensions and systems exhibiting FOPTs.}
\label{table:conformal} 
\end{table}

While all of these quantities are used to draw conclusions about EoS behavior, $\Delta$ and $\Delta’$ have proven particularly useful for defining what is termed a crossover to quark matter. These two parameters can be combined into a single measure of conformality:
\begin{equation}
\label{eq:criteria}
    \dc \equiv \sqrt{\Delta^2 + (\Delta')^2} < 0.2
\end{equation}
Small values of $\Delta$ indicate that the polytropic index $\gamma$ and the speed of sound $\cs$ are close to their conformal values, 1 and 1/3, respectively. Small values of the logarithmic derivative $\Delta'$ ensure that the EoS remains conformal at higher densities, approaching the pQCD limit. An appropriate cutoff for the value of the $\dc$ parameter must be chosen to quantify the posterior of conformal matter inside NSs. The value of 0.2 can be well justified based on \cref{table:conformal}. First, for conformal field theory, the value of $\dc$ is zero. For FOPT, $\dc$ is given by the expression $\dc = \sqrt{\Delta^2 + (1/3 - \Delta)^2}$, which has a minimum at $\Delta = 1/6$, yielding a lower bound for $\dc$ of $1/(3\sqrt{2}) \approx 0.236$. From this consideration, a cutoff of 0.2 is sufficiently small to ensure that the EoS behaves similarly to the pQCD limit, with $\dc$ close to a near-conformal value, while also excluding FOPT. Furthermore, most nuclear models in the regime where they agree predict $\dc$ values within the range [0.25, 0.4].

Admittedly, the specific value of 0.2 is somewhat arbitrary, serving more as an indication that the EoS begins and remains close to the conformal behavior. However, the qualitative conclusions of this chapter are largely insensitive to variations in the cutoff for $\dc$ within a reasonable range.

While matter may be nearly conformal with $\dc < 0.2$, not all near-conformal matter is deconfined matter. Another useful quantity that can be used is the normalized pressure, $p/p_{\rm free}$, which is not fixed by conformal symmetry. This quantity is connected to the effective number of active degrees of freedom, denoted as $N_{\rm eff}$. In both weakly and strongly interacting CFTs, the pressure scales as $p \propto N_{\rm eff} p_{\rm free}$. For weakly coupled systems, Dalton’s law states that the total pressure is the sum of partial pressures, leading to the relationship $N_{\rm eff} = N_f N_c \, p / p_{\rm free}$, where $N_f$ is the number of flavors and $N_c$ is the number of colors. Regardless of the interaction strength, the normalized pressure $p / p_{\rm free}$ remains approximately constant and is sensitive to the number of active degrees of freedom.

Turning to specific values of active DOF, in finite-temperature QCD, where it is used to characterize the quark-gluon plasma (QGP) phase, the normalized pressure, $p/p_{\rm free}$, takes a constant value of approximately 0.8, as determined by nonperturbative lattice field theory calculations. At high densities, $N_{\rm eff}$ can be computed within the perturbative QCD framework and is approximately 0.6. This reduction in $N_{\rm eff}$ is due to perturbative corrections. For $\mathcal{N}$ = 4 Super Yang-Mills theory at finite temperature and zero chemical potential at infinite ’t Hooft coupling, $p/p_{\rm free}$ = 3/4 (and 1 for non interacting theory).

Therefore, quark matter, whether weakly or strongly coupled, is expected to exhibit an $N_{\rm eff}$ of the order of one, with only slight variations across the density range. In this analysis, this quantity serves as a key indicator, supporting the conclusion that the conformal matter inside neutron stars exhibits deconfined behavior, distinguishing it from any other nearly conformal phases.

\subsection{Quark matter posterior}
\label{subsec:QMposterior}

\begin{figure}[ht!]
    \centering
\includegraphics[width=0.7\textwidth]{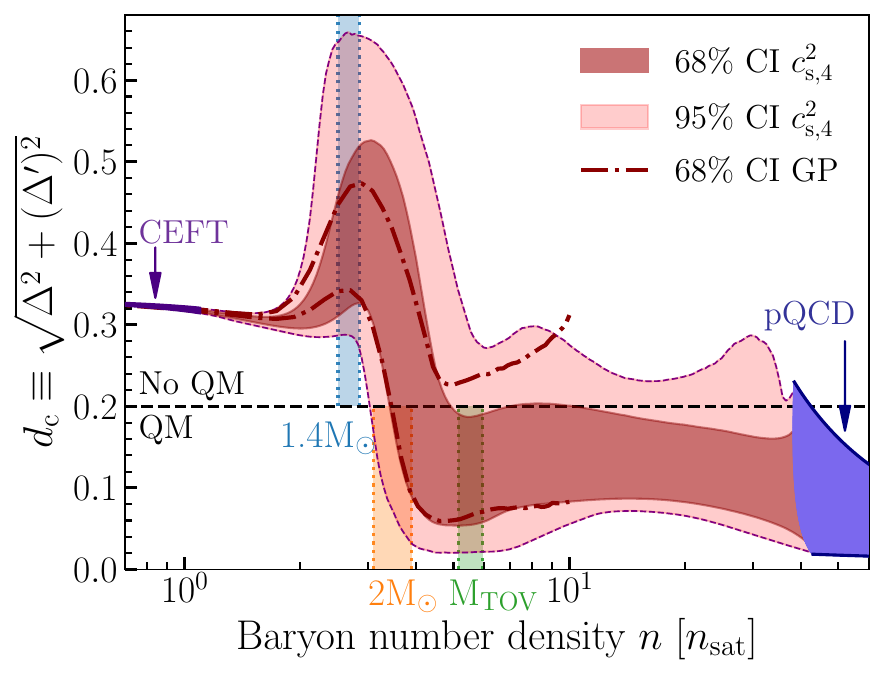}
    \caption{The conformal parameter $d_c$, defined in \cref{eq:criteria}, with a value of 0.2 shown as a black dashed line, plotted as a function of number density. The dark and light bands represent the 68\% and 95\% credible intervals (CIs) obtained using a four-segment sound speed interpolation, denoted as $c^2_{s,4}$. In addition, the 68\% CI obtained from the GP ensemble is shown with a red dash-dotted line. The colored bands correspond to the 68\% CI for the central densities of different masses.}
    \label{fig:dc}
\end{figure}

Turning to the Bayesian inference result, the $c^2_{s,4}$ and the GP posteriors for the parameter $d_{\rm c}$ as a function of number density are shown in \cref{fig:dc}. As evident from the figure, the behavior of this quantity has a clear separation of the two phases. Specifically, the first phase is characterized by an increase in $d_{\rm c}$, followed by a rapid drop to the second phase with a lower value of the parameter that remains approximately constant up to the pQCD limit. Both the GP and the $c^2_{s,4}$ priors agree well within the GP density range. The posterior probability for conformal matter inside NSs differs significantly between light and heavy stars. For sound-speed interpolation, the probabilities are 0\% and 11\% for 1.4$M_\odot$ and 2$M_\odot$ stars, respectively, but rise dramatically for the heaviest stars, with an 88\% probability of conformal matter cores for $M_{\rm TOV}$. The GP prior predicts a 75\% probability of conformalization of neutron-star matter for TOV stars.

Similar separation of phases is observed in various neutron-star-matter properties, as illustrated by the CIs for $\cs$, $\gamma$, and $\Delta$ in \cref{fig:QM_cs_gamma_delta}. Notably, for any quantity the behavior of the EoS near the TOV density closely resembles that of the pQCD EoS at higher densities and stays nearly unchanged across the density range between the TOV and pQCD. As evident from the middle panel, the previously employed criterion from \cite{Annala:2019puf}, $\gamma < 1.75$, results in a significantly higher posterior compared to $d_{\rm c} < 0.2$, specifically 99.8\% and 97.8\% for $c^2_{s,4}$ and the GP prior, respectively. The right column of the figure in \cref{fig:QM_cs_gamma_delta} presents these quantities as functions of $M/M_{\rm TOV}$. The phase change of EoS toward conformality is particularly clear for the most massive neutron stars.

\begin{figure}[ht!]
    \centering
\includegraphics[width=0.86\textwidth]{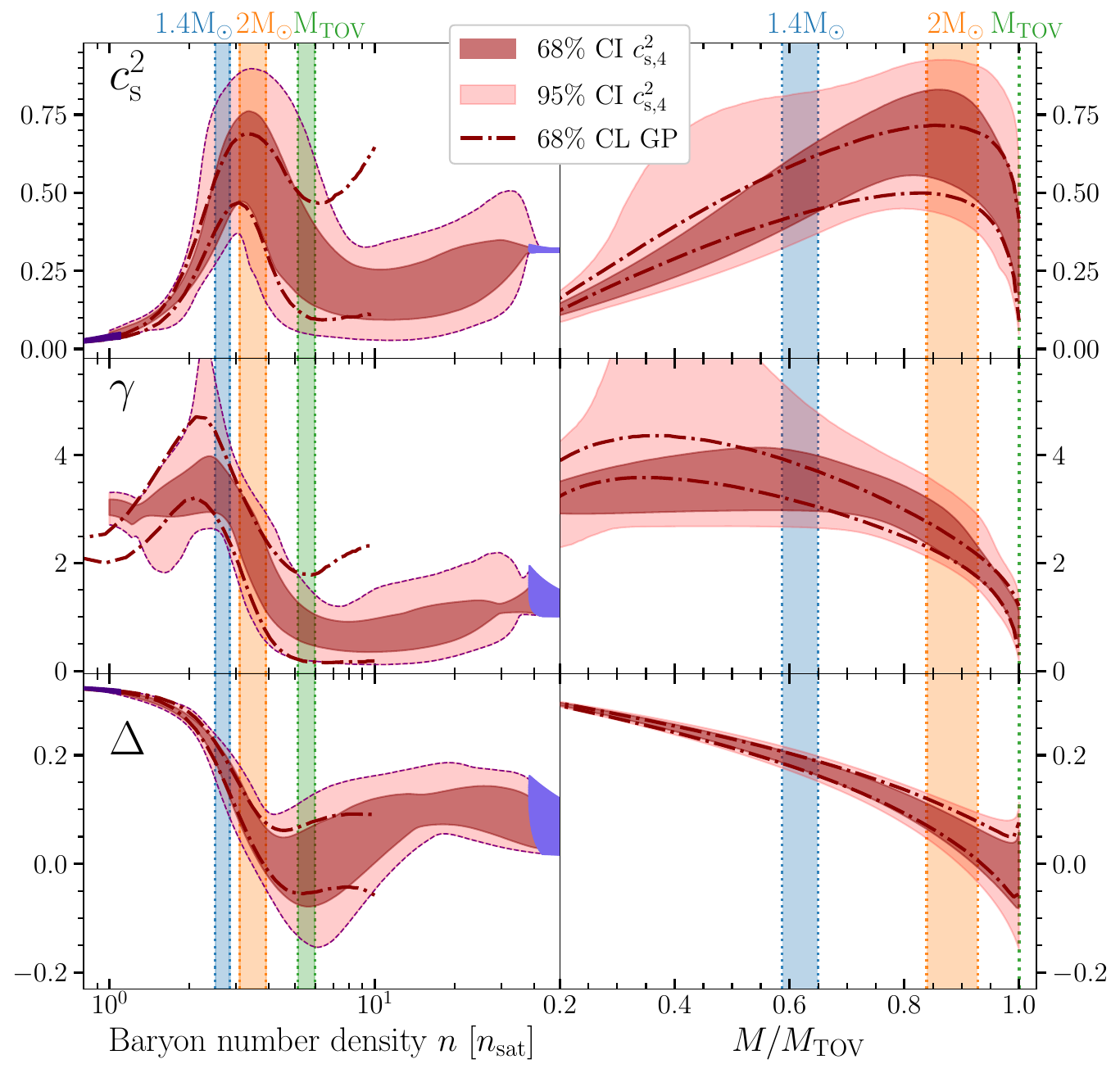}
    \caption{The sound speed $c^2_s$, polytropic index $\gamma$, and normalized trace anomaly $\Delta$ are shown as functions of number density and the mass ratio $M/M_{\rm TOV}$. The dark and light bands represent the 68\% and 95\% CIs obtained using $c^2_{s,4}$. Additionally, the 68\% CI obtained from the GP ensemble is shown with a red dash-dotted line.}
    \label{fig:QM_cs_gamma_delta}
\end{figure}

\begin{figure}[ht!]
    \centering
\includegraphics[width=0.67\textwidth]{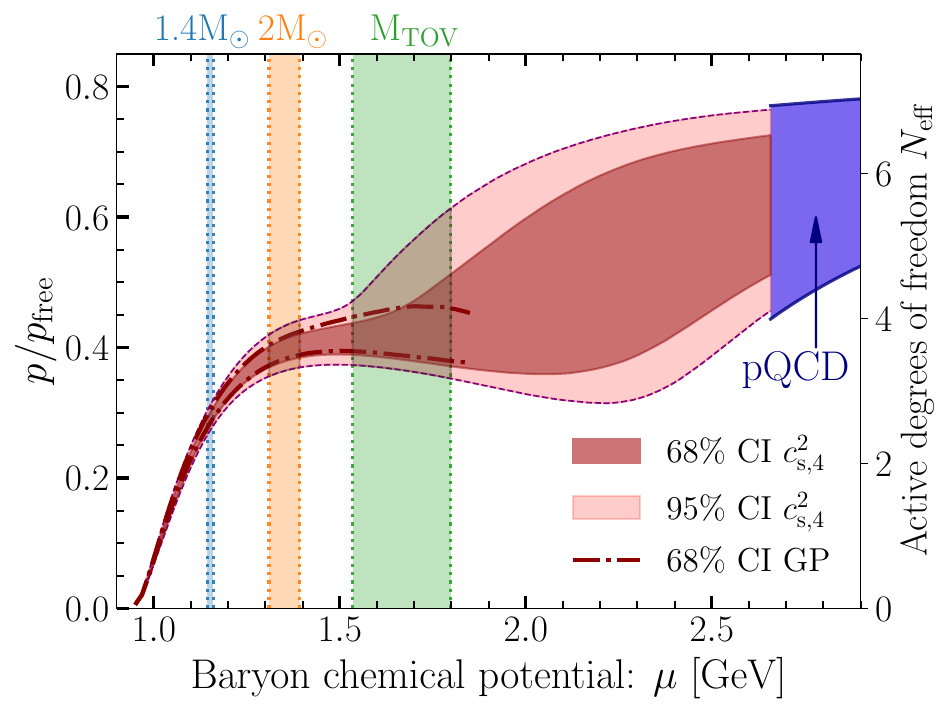}
    \caption{Pressure normalized to that of a free Fermi gas of quarks is shown as a function of chemical potential. The dark and light bands represent the 68\% and 95\% CIs obtained using $c^2_{s,4}$. Additionally, the 68\% CI obtained from the GP ensemble is shown with a red dash-dotted line. The colored bands correspond to the 68\% CI for the central densities of different masses.}
    \label{fig:ppfree}
\end{figure}

As explained in \cref{subsec:conformal}, the normalized pressure, which is proportional to the number of active degrees of freedom, $N_{\rm eff}$, can be used to differentiate deconfined quark matter from other types of near-conformal behavior. The CIs for the normalized pressure as a function of the chemical potential are shown in \cref{fig:ppfree}. Starting from around $2M_\odot$, the CIs for the normalized pressure flatten out at approximately $p/p_{\rm free} = 0.4 \pm 0.03$, which is roughly two-thirds of the pQCD value. This value is consistent with that of weakly interacting quark matter. Altogether, this provides evidence for the presence of quark matter cores in the most massive neutron stars.

The final point to consider is how this behavior contrasts with hadronic models. In the appendix, \cref{fig:models} displays $\cs$, $\gamma$, and $\Delta$ for hadronic models at $T=0$ from the CompOSE database, which are used to derive the values in \cref{table:conformal}. The red bar highlights the density range at $n=3\ns$, where the models generally agree. However, even beyond this density, the models exhibit strongly non-conformal behavior, characterized by a rapidly decreasing trace anomaly, a steeply increasing speed of sound, and conformal parameter $d_c$.

Additionally, \cref{fig:ppfree_models} illustrates the behavior of the normalized pressure across these hadronic models. In most cases, the normalized pressure shows a gradual decline near the TOV densities, diverging from the posterior for interpolated EoSs. While a few specific nuclear models obtain $d_{\rm c} < 0.2$ and $p/p_{\rm free}\approx0.4$ values around the TOV density, these models have negligible posterior weight in the analysis as well as fail to represent the general behavior of dense NM models.

\mybox{Summary of \cref{sec:qm}}{blue!20}{white!10}{
\begin{itemize}
    \item The matter in the cores of the most massive NSs exhibits near-conformal behavior, as indicated by the analysis of $\cs$, $\gamma$, $\Delta$, and the newly introduced conformality parameter $d_{\rm c}$.
    \vspace{0.1cm}
    \item The effective number of active degrees of freedom flattens out around the TOV density, varying only slightly around the value consistent with weakly coupled quark matter.  
\end{itemize}
}

\clearpage

\section{First-order phase transitions}
\label{sec:fopt}

A first-order phase transition (FOPT) is characterized by a discontinuous jump in energy and number density, with $\cs = 0$, representing the most extreme form of softening possible in stable matter. While some EoSs in the GP ensemble exhibit segments with low sound speed, effectively mimicking FOPT-like behavior, these cases are exponentially suppressed in the prior with a non-zero mean. The goal of this section is to modify the GP prior to include an explicit FOPT.

Although one might argue that the explicit inclusion of FOPT does not yield a fully model-agnostic approach, it is worth noting that the suppression of such cases can be considered in a similar manner. The posterior probability of a crossover to quark matter can change significantly when FOPTs are explicitly included in the prior. This section investigates these effects.

The procedure for generating FOPTs using the GP prior is detailed in \cref{subsec:fopt_constr}, including the identification of the most compelling scenarios associated with FOPTs. The results of Bayesian inference with the updated prior are presented in \cref{subsec:fopt_results}, along with the posterior distributions of FOPT parameters, such as its location and strength. \Cref{subsec:bayes_fopt} presents a comparison of various scenarios based on Bayes factors to evaluate the preferences indicated by the current data. Finally, \cref{subsec:pt} combines the analysis from the previous section with the updated prior, including FOPTs, to evaluate the likelihood of phase changes in the cores of NSs, providing a detailed comparison between crossovers to QM and FOPTs.

The astrophysical input in this section is identical to that of the previous section, with pulsar measurements summarized in \cref{table:astro} and the binary TD and BH hypotheses introduced in \cref{sec:bayesian}. The theoretical input, however, differs. Specifically, additional information from the low-density limit within cEFT is incorporated. While the GP remains conditioned on the same cEFT data below 1.1$\ns$, a cEFT likelihood function is employed in the range 1.1$\ns$ to 2$\ns$, with further details provided in the appendix \cref{fig:ceft_likelihood}.

Each generated EoS is utilized only up to the TOV density, where two QCD inputs are considered. The first, a conservative QCD input, utilizes the simple check from \cref{sec:bayesian}(d) with $\nterm = \ntov$. The second utilizes the marginalized QCD likelihood function, introduced in \cref{sec:termination}(c), which marginalizes over the GP prior in the range $[\ntov, 40\ns]$ and is conditioned on the pQCD sound speed within [25, 40]$\ns$.

\subsection{FOPT construction}
\label{subsec:fopt_constr}

The EoS with an explicit FOPT is constructed using two uncorrelated samples of GP, separated by a segment in $n$ with zero $\cs$. Both GP samples share the same kernel and hyperparameters, as they are drawn from the same GP conditioned on cEFT data up to $1.1\ns$.\footnote{The division of the cEFT input into two parts has no specific justification: first, the GP is conditioned on cEFT data up to $1.1\ns$, and then the cEFT likelihood function is used within the range $[1.1, 2]\ns$.}

The prior for the FOPT is chosen to be uniform in $n_{\rm PT} - \Delta n$, where $n_{\rm PT}$ represents the starting density of the PT, and $\Delta n$ denotes the strength of the PT. The prior is constructed as follows:
\begin{enumerate}
\item The location of the FOPT is sampled from a uniform distribution: $n_{\rm PT} \sim \mathcal{U}(1.1\ns, 10\ns)$.
\item The strength of the FOPT is sampled from a uniform distribution: $\Delta n \sim \mathcal{U}(0\ns, 8\ns)$.
\item Two independent GP samples are drawn over the ranges $[1.1\ns, n_{\rm PT}]$ and $[n_{\rm PT} + \Delta n, 10\ns]$.
\item A segment with $\cs = 0$ is inserted in the range $[n_{\rm PT}, n_{\rm PT} + \Delta n]$.
\end{enumerate}

An example of the resulting EoS is illustrated in \cref{fig:help_fig_fopt}. As shown in the figure, the sound speed values before and after the PT are uncorrelated, which is an important property of the FOPT. 

\begin{figure}[ht!]
    \centering
\includegraphics[width=0.65\textwidth]{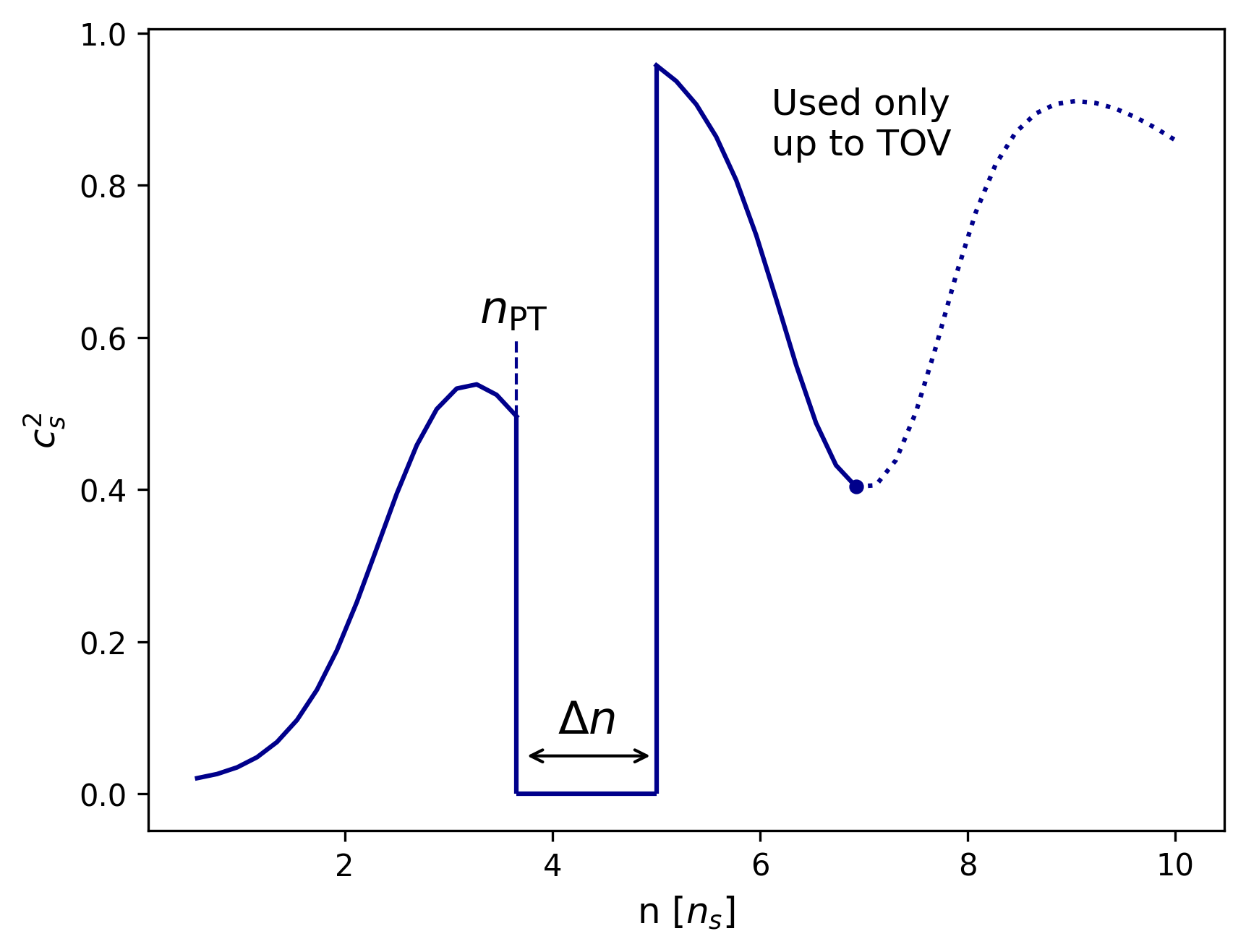}
    \caption{An example of an EoS generated using two segments of GP and an explicit FOPT, with $c^2_s=0$ in between.}
    \label{fig:help_fig_fopt}
\end{figure}

Each EoS from the ensemble can be categorized into distinct sets based on the location of the FOPT relative to the end of the stable branch of NSs, i.e., the TOV density. The differences between these sets are summarized as follows:

\begin{itemize}
    \item \textbf{NO FOPT:} There is no FOPT within the stable branch of NSs. This occurs if the FOPT happens above the TOV density. In this case, only one segment of the GP is relevant, as densities above the TOV density are not used. Alternatively, this can also occur if $\Delta n \lesssim 0.1\ns$, corresponding to the grid spacing of the GP.

    \item \textbf{FOPT inside NS:} The FOPT occurs within the stable branch of NSs, and the star remains stable afterward, i.e., $n_{\rm PT} + \Delta n < \ntov$.

    \item \textbf{Destabilizing FOPT:} The FOPT occurs within the stable branch of NSs, but the star becomes unstable and collapses into a BH afterward. The first grid point above the FOPT lies within the unstable branch, and the TOV density is identified as $\ntov = n_{\rm PT} + \Delta n$.

    \item \textbf{Twin Stars:} A second stable branch appears in the mass-radius curve, regardless of the FOPT's location. For twin stars, astrophysical likelihoods are marginalized over both stable branches, and QCD constraints are imposed at the maximal density of the second branch.
\end{itemize}
The ensemble consists of a total of 300k EoSs, categorized as follows: 66k identified as \texttt{no FOPT}, 37k as \texttt{FOPT inside NS}, 121k as \texttt{destabilizing FOPT}, and 76k characterized as \texttt{twin stars}.

\subsection{FOPT posterior}
\label{subsec:fopt_results}

The main inference results are presented in \cref{fig:cs2_fopt,fig:ndn_fopt,fig:bayes_fopt}. The CIs for the speed of sound are displayed in \cref{fig:cs2_fopt} (upper left), while the other panels of the figure represent representative samples of EoSs categorized into one of the following set: \texttt{no FOPT}, \texttt{FOPT inside NS}, and \texttt{destabilizing FOPT}. The figure does not include \texttt{twins} set due to its negligible evidence, as will be clarified in the next section.

The CIs of the sound speed exhibit very similar shapes for different sets, comparable to those in \cref{fig:QM_cs_gamma_delta,fig:posterior_CIs}. Note that the CIs in \cref{fig:cs2_fopt}, unlike those in other figures, represent the conditional probability $P(c^2_s \,|\, n, n < \ntov)$, as the EoS contributes only up to the TOV density. While this figure cannot be directly compared to others, the \texttt{no FOPT} set reproduces the same behavior as inferred in previous sections and can be compared to the prior that includes FOPT.

As observed throughout this thesis, the peak in the sound speed around 2–3$\ns$ remains stable when using different inputs and QCD likelihood functions. The previously inferred behavior — strong stiffening due to mass constraints, followed by rapid softening caused by the QCD input — remains unaffected by the inclusion of FOPT in the prior. \Cref{fig:cs2_convers_fopt} in the appendix presents the CIs of $\cs$ for a conservative QCD input with $\nterm = \ntov$, where no significant softening is observed. This is attributed to the reasons outlined in \cref{sec:termination}. 

\begin{figure}[ht!]
    \centering
\includegraphics[width=0.95\textwidth]{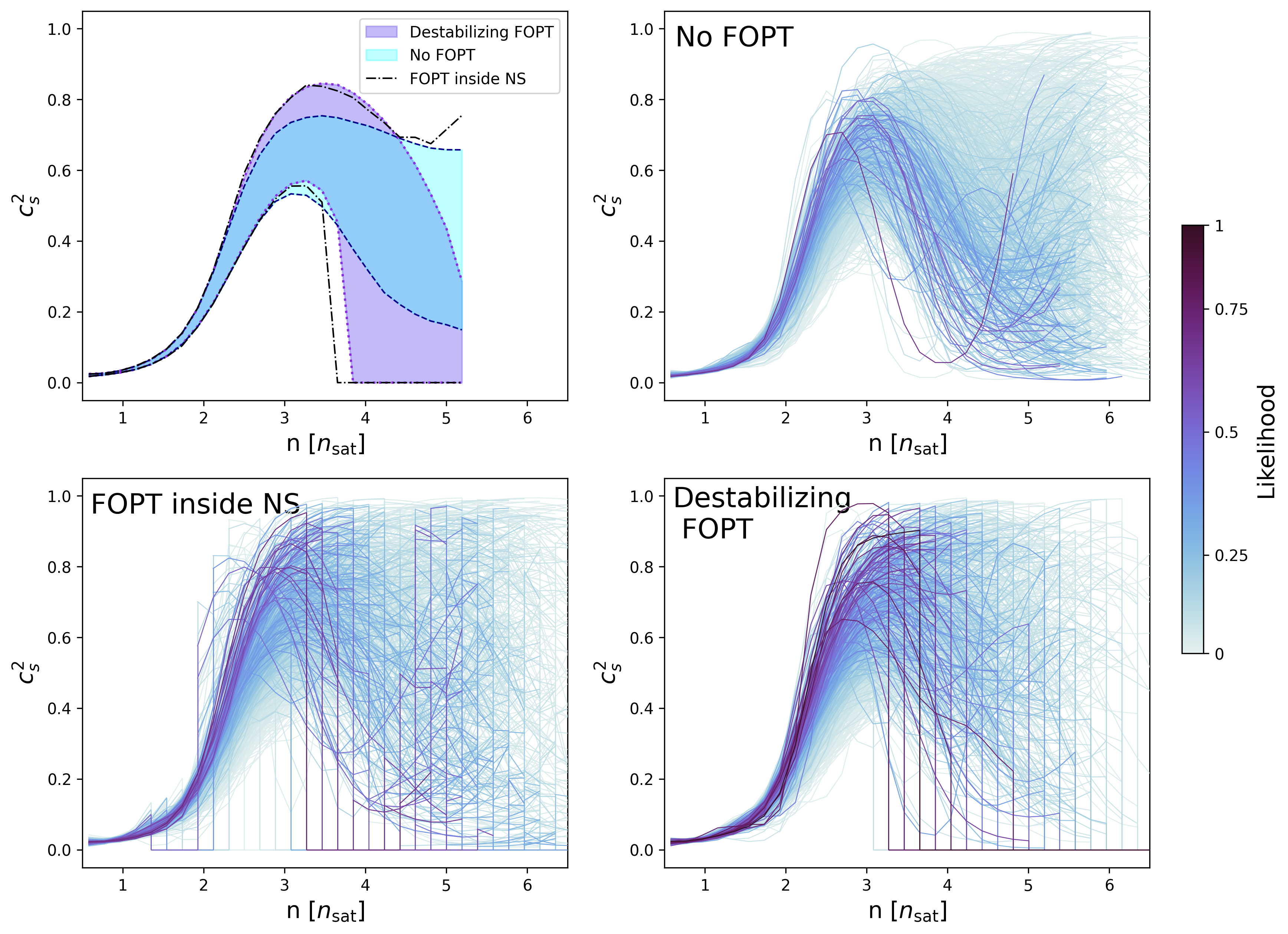}
    \caption{(Upper left) The 68\% CI for the sound speed for three different sets, categorized based on the location of the FOPT relative to the end of the stable branch of NSs. (Other panels) A representative sample of EoSs, with color-coded likelihood obtained by incorporating all astrophysical inputs considered in this chapter, along with the cEFT likelihood function and the marginalized QCD likelihood function. The likelihood is normalized to the maximum likelihood within the ensemble}
    \label{fig:cs2_fopt}
\end{figure}

An important consequence of including FOPT is that the peak in $c^2_s$ tends to be higher than in scenarios without any FOPT. This can be understood as follows: as shown in \cref{fig:extension}, stiff EoSs with a high tension index require significant softening. Consequently, EoSs with FOPT incorporated by construction can achieve higher values of $\cs$ before being penalized by the QCD input. EoSs with destabilizing FOPT exhibits the highest likelihood within the ensemble, with $\cs$ values exceeding 0.8 — a behavior not observed in the \texttt{no FOPT} set.

Turning to the specific characteristic quantities of the FOPT, the posterior distribution of $n_{\rm PT}-\Delta n$ is shown in \cref{fig:ndn_fopt}. The prior distribution is uniform for these quantities, with any values above $n_{\rm PT} + \Delta n > 10\ns$ discarded, as indicated by the prior cut in the figure.

Starting with \texttt{FOPT inside NS}, shown in black in the figure, first-order phase transitions are largely ruled out in the intermediate density range of $n_{\rm PT} \sim 2$–$3\ns$. In this region, the matter must remain stiff to satisfy astrophysical constraints, particularly the 2-solar-mass constraint. This effectively excludes FOPT inside NSs in the mass range of approximately $[0.5, 1.9]M_\odot$. Small FOPTs, of the order of the grid spacing, are allowed by current data in the density region just above the cEFT limit below 2$\ns$. To the best of my knowledge, there are no models that suggest such behavior.

For an EoS to be classified within the \texttt{FOPT inside NS} set, it must remain stable after the phase transition, imposing additional prior constraints on the strength of the PT. The intersection between the \texttt{FOPT inside NS} and \texttt{destabilizing FOPT} posteriors occurs around $\Delta n\approx1.2\ns$, where any larger FOPTs lead to the collapse of the NS.

In principle, if a star is destabilized by an FOPT of strength $\Delta n$, any larger FOPT would produce the same outcome, resulting in a uniform distribution of $\Delta n$ above a certain threshold. However, larger FOPTs may be penalized by the marginalized QCD likelihood function, as they might be inconsistent with upper integral constraints from \cref{fig:mu_n}. This is evident from the slight reduction in posterior weight for larger $\Delta n$ within the \texttt{destabilizing FOPT} set. The twins’ posterior shows a slight extension below $3\ns$, with a pronounced peak in the range of $3-4\ns$. Twin-star solutions are mostly produced by FOPTs with strengths around $1-2\ns$.

\begin{figure}[ht!]
    \centering
\includegraphics[width=0.8\textwidth]{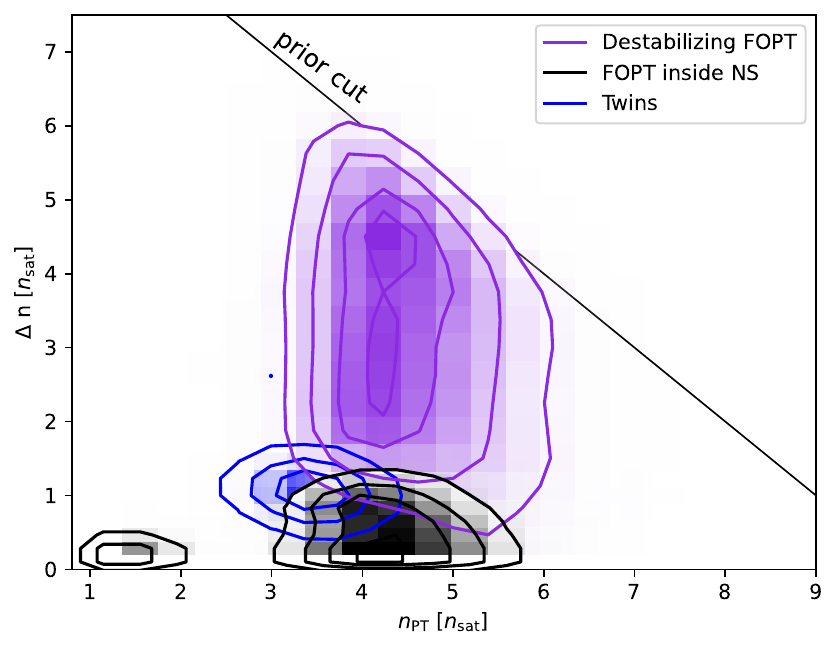}
    \caption{The posterior distribution for the location of the FOPT, $n_{\rm PT}$, and its strength, $\Delta$, for different sets. The prior distribution is uniform in the $n_{\rm PT}$–$\Delta n$ plane by construction. A Gaussian filter is applied to smooth the data.}
    \label{fig:ndn_fopt}
\end{figure}

\subsection{Bayes factors}
\label{subsec:bayes_fopt}

To compare different sets and scenarios, the Bayes factor is employed to quantify the preference of the data for one set over another. Each set is treated as a competing statistical model, with the Bayes factor representing the ratio of evidence between the two, which can be expressed as:
\begin{equation}
\label{bayes_factor}
B^{\rm set_1}_{\rm set_2} = \frac{P(  {\rm set_1} \,|\, \rm data \,)}{P(  {\rm set_2} \,|\, \rm data \,)}\frac{P({\rm set_2})}{P({\rm set_1})},
\end{equation}
Here, $P({\rm set_i} ,|, \rm data)$ represents the posterior probability of set $i$ while $P({\rm set_i})$ denotes the prior probability of set $i$. Assuming that all sets are equally probable \textit{a priori}, $P({\rm set_i})$ simplifies to being proportional to the number of EoSs within each set.

The Bayes factors resulting from the inference are summarized in \cref{table:Bayes} in the form of ${B^{\rm set}_{\mathrm{noFOPT}}}$, where each number in the table represents a comparison between the given set in the column and the \texttt{no FOPT} set. Large values, of the order of 10 or more, indicate a strong preference for the given set over the \texttt{no FOPT} set. Both conservative and aggressive inputs are used to assess Bayes factor variation. For all entries, the ensemble is conditioned on radio measurements and binary TD constraints, while varying QCD inputs and incorporating either all X-ray measurements from \cref{table:astro} or a single NICER measurement of PSR J040 + 6620.

\begin{table}[h]
\centering
\label{table:Bayes}

\begin{NiceTabular}{c|c|c|c}[code-before = \rowcolor{blue!15}{2,4}]
\toprule
$\mathbf{B^{\rm set}_{\mathrm{noFOPT}}}$ & \textbf{Destab. FOPT} & \textbf{FOPT inside NS} & \textbf{Twins} \\
\addlinespace[0.3em]

\makecell{Marginalized QCD\\ $\ \ \ \ \ \ \ $+ X-rays} & \change{1.5} & \change{0.7} & 0.001\\ 
\addlinespace[0.3em]
\makecell{Marginalized QCD\\ $\ \ \ \ \ \ \ $+ PSR J0740}& \change{1.5} & \change{1.0} & 0.001\\ 
\addlinespace[0.3em]

\makecell{Conservative QCD\\ $\ \ \ \ \ \ \ $+ X-rays} & \change{0.8} & \change{0.5} & 0.001 \\
\addlinespace[0.3em]

\makecell{Conservative QCD\\ $\ \ \ \ \ \ \ $+ PSR J0740}& \change{0.8} & \change{0.7} & 0.001\\

\bottomrule
\end{NiceTabular}

\caption{A summary of the Bayes factors for various sets compared to the \texttt{no FOPT} set. The evidence is calculated for ensembles conditioned on radio measurements, the cEFT likelihood function, and GW data, in addition to the two likelihoods specified in the first column.}
\end{table}

Most of the Bayes factors in the table are indecisive, except for the twin-star solutions, which are disfavored by the current data, particularly by the mass and mass-radius measurements. The remaining Bayes factors indicate the insensitivity of the data to the different types of phase changes inside NS, primarily because the astrophysical inputs (except for the BH hypothesis) do not propagate beyond $2M_\odot$, and QCD inputs are consistent with both smooth softening in the \texttt{no FOPT} set and a destabilizing FOPT. The \texttt{FOPT inside NS} set closely resembles the \texttt{no FOPT} set, as the constraints allow only small phase transitions that do not significantly alter the EoS, as illustrated in \cref{fig:cs2_fopt,fig:ndn_fopt}.

The marginalized QCD input slightly favors scenarios involving FOPT, as higher values of sound speed, close to the speed of light, are allowed before the drastic softening of the EoS.
Such EoSs are mostly excluded in the \texttt{no FOPT} set, as they require more extreme behavior above TOV density that is not present in the GP ensemble from \cref{sec:termination}(c), over which the marginalization is constructed. Note that the same results can be achieved by imposing the QCD input at a slightly higher density than TOV, without utilizing the marginalized QCD likelihood function.

\begin{figure}[ht!]
    \centering
\includegraphics[width=0.8\textwidth]{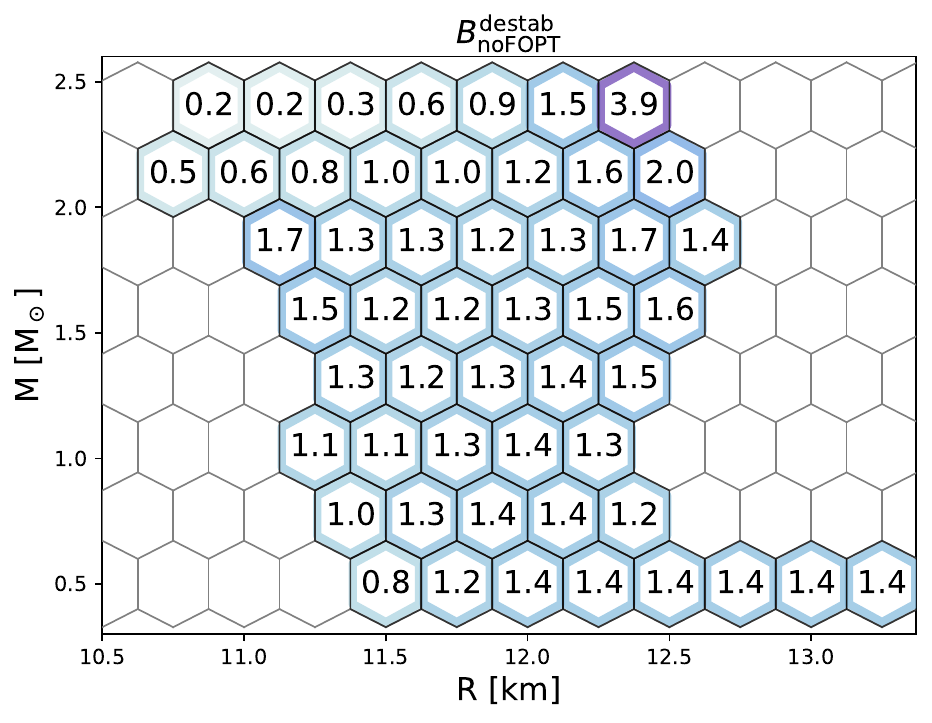}
    \caption{A summary of the Bayes factors for a potential future mass-radius observation, comparing sets with a destabilizing FOPT to those without any FOPT. Each hexagon represents a single measurement, with a likelihood of 1 inside and 0 otherwise.}
    \label{fig:bayes_fopt}
\end{figure}

The current data cannot decisively differentiate between scenarios with and without FOPT. However, it is possible to explore how future mass-radius measurements could affect the Bayes factor, indicating the preference of the data for one of the scenarios. The results of such an analysis are shown in \cref{fig:bayes_fopt}, where each hexagon\footnote{To address a potential question, there is no particular reason for the hexagonal shape of the measurements.} represents a possible future mass-radius measurement. The likelihood for such future measurement is set to one if the EoS passes through the hexagon and zero otherwise. To prevent numerical issues, at least 100 EoSs must pass through a hexagon for the Bayes factor to be included in the analysis. The Bayes factors in the figure, denoted as $\mathbf{B}^{\rm destab}_{\rm noFOPT}$, compares the \texttt{destabilizing FOPT} set with the \texttt{no FOPT} set. As evident from the figure, no single mass-radius measurement decisively skews the results toward a specific scenario. However, observations of large mass and radius show a slight preference for the destabilizing FOPT.

\subsection{Phase transitions in the core}
\label{subsec:pt}

To conclude this chapter, the analysis of quark matter cores in \cref{sec:qm} can be combined with the study of FOPT. The criteria defined in \cref{eq:criteria} for a crossover to QM can be applied to the new ensemble. For the \texttt{no FOPT} set, using the marginalized QCD likelihood function at TOV instead of the conservative QCD input at $10\ns$ \change{decreases the posterior probability for a crossover to QM, which is now 64\% (cf. 75\% in \cref{sec:qm}), for the reasons outlined in \cref{sec:termination}.} The inclusion of FOPT reduces the probability to \change{50\%} for the \textit{FOPT inside NS} set and \change{30\%} for the \textit{destabilizing FOPT} set. Examples of EoSs from these sets that contribute to the QM posterior include early FOPT with a crossover at higher densities or a crossover to QM occurring right before a destabilizing FOPT.

Taking a subset of crossover to QM from the \texttt{no FOPT} set allows for an explicit comparison between the destabilizing FOPT scenario and a crossover. This results in
\begin{equation}
\label{eq:Bayes_QM_destab}
    \mathbf{B^{\rm destab}_{\rm QM}} \approx \change{0.85},
\end{equation}
indicating even less significance and further emphasizing that the current data are insufficient to distinguish between different behaviors in the cores of NSs.

What is more intriguing is that the posterior probability of some phase change inside an NS can be computed. In \cite{Komoltsev:2024lcr}, I reported a \change{91\%} probability for the occurrence of a non-trivial phase transition inside NS cores. This value was obtained by comparing the evidence of EoSs featuring a phase transition — either a first-order PT within stable branch of NS or a crossover to quark matter at the TOV density — against the entire ensemble. However, this probability is significantly affected by the manual inclusion of an FOPT into the prior set. To address this and further quantify the evidence in favor of phase transitions, one can compute the Bayes factor under the assumption that the two competing models, “with phase transition” and “without phase transition”, have equal prior probability. This leads to
\begin{equation}
    \mathbf{B^{\rm PT}_{\mathrm{noPT}}}\approx\change{2.5}
\end{equation}
Any attempt to identify EoSs that exhibit FOPT-like behavior (but are not classified within the FOPT sets or as crossovers to QM) would increase this factor. \change{Additionally, removing the discontinuity in $c_s^2$ — which, as explained in \cref{sec:termination}, arises at the TOV point where the marginalized QCD likelihood is imposed — necessitates further softening of the EoS, effectively increasing the Bayes factor. Nevertheless, with current data and methods, the preference for a non-trivial phase transition in NS cores remains only marginal.}

\begin{figure}[ht!]
    \centering
\includegraphics[width=1\textwidth]{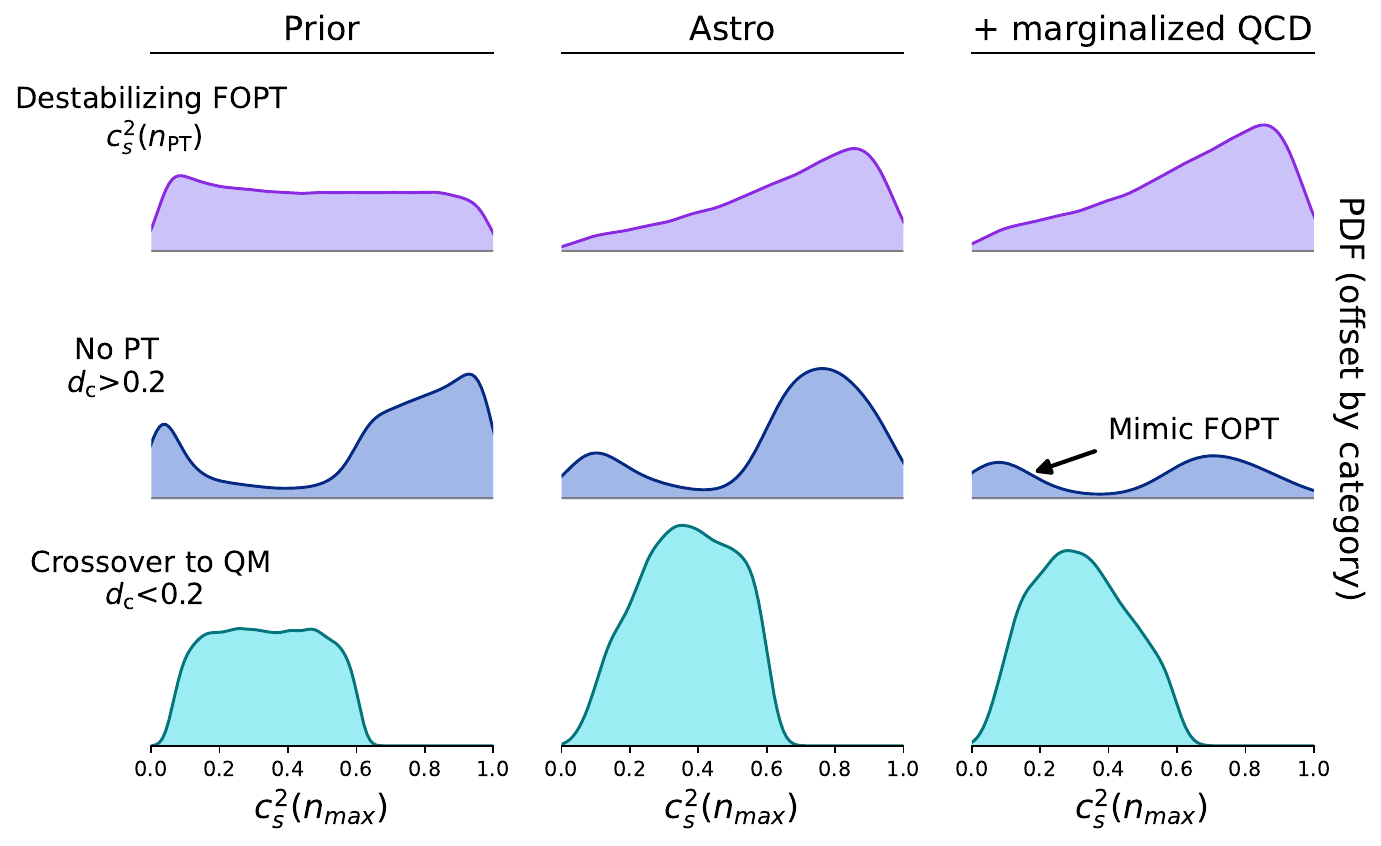}
    \caption{The distribution of the sound speed offset across three scenarios: destabilizing FOPT, no PT, and crossover to QM. Each column represents different inputs, with the first corresponding to the prior. The second column shows the posterior density when all astrophysical data is imposed, while the last column illustrates the effect of the marginalized QCD input, imposed on top of the astrophysical likelihoods.}
    \label{fig:cs2nmax}
\end{figure}

This analysis can be presented in a more visual form, as shown in \cref{fig:cs2nmax}, which depicts the distribution of the sound speed at the maximum central density. The distributions are displayed for three distinct scenarios: (1) crossover to QM with $d_{\rm c}(n_{\rm max}) < 0.2$, (2) no PT, corresponding to EoSs without any FOPT and with $d_{\rm c}(n_{\rm max}) > 0.2$, and (3) destabilizing FOPT, where the distribution corresponds to $c^2_s(n_{\rm PT})$ at the last grid point before the phase transition. Each column represents a different input: the prior, astrophysical data only, and astrophysical data with the marginalized QCD input.

Each distribution is normalized so that the area is proportional to the evidence. Consequently, the ratio of the areas provides a direct visual interpretation of the Bayes factor. For instance, the ratio of the areas between the first and last rows in the final column corresponds to the Bayes factor from \cref{eq:Bayes_QM_destab}. This clearly demonstrates how the QCD input disfavors stiff EoSs with $d_{\rm c} > 0.2$ in the no PT row.

\mybox{Summary of \cref{sec:qm}}{blue!20}{white!10}{
\begin{itemize}
    \item Non-trivial phase transitions, such as a crossover to QM or a FOPT, can be explored in NS cores, with current data showing a \change{slight} preference for such scenarios, yielding a Bayes factor of $\mathbf{B^{\rm PT}_{\mathrm{noPT}}} \approx 2.5$
    \vspace{0.1cm}
    \item The Bayes factor comparing models with and without FOPTs is of the order of one, indicating that both scenarios are equally consistent with current astrophysical data and theoretical inputs. However, twin-star solutions are largely ruled out.
    \vspace{0.1cm}
    \item Scenarios involving first-order phase transitions can either correspond to destabilizing FOPTs with $\Delta n \lesssim 1.2\ns$ starting around $3$–$4\ns$, or FOPTs inside neutron stars with $\Delta n \gtrsim 1.2\ns$ within the same density range. FOPTs occuring in the mass range [0.5, 1.9]$M_\odot$ are inconsistent with astrophysical observations.
    \vspace{0.1cm}
   \item Any single future mass-radius measurement would be insufficient to distinguish between a smooth crossover and a destabilizing FOPT.
    
\end{itemize}
}
}


%
\mainfont{\chapter{Conclusion}

The original focus of this thesis was to test the impact of pQCD calculations on the inference of the EoS of neutron-star matter, which is explored in great detail in \cref{chpt:QCD_and_NS}. However, the research naturally expanded toward a more fundamental question: the nature of the phase transition between hadronic and quark matter. This is where the novel QCD input plays a crucial role in determining the physics of NS cores, where such a phase transition could potentially occur.

Let me now summarize the results in order. Thermodynamical requirements on the EoS, such as stability, causality, and consistency, impose global constraints on the behavior of dense cold matter between the cEFT and pQCD limits. This framework explicitly demonstrates how pQCD input can propagate constraints from around $40\ns$ down to lower densities, such as those found in neutron stars.

The novel constraints on the EoS can then be incorporated into Bayesian inference to assess the impact of the QCD input. The results show that the QCD input provides significant and nontrivial constraints on the neutron-star EoS, extending beyond the current astrophysical observations. The crucial insight into the physics of the cores of NSs seems to lie in the interplay between astrophysical observations and pQCD calculations. This interplay between astrophysical data — particularly mass constraints — and the QCD input introduces a peak structure in the sound speed of neutron-star matter. Above the peak, the QCD input forces EoS to soften, driving it toward conformality. 

The novel pQCD constraints has been widely used in various studies \cite{Cai:2025nxn, Brandes:2024wpq, Cuceu:2024llk, Roy:2024sjx, Gholami:2024ety, Bai:2024amm, Alarcon:2024hlj, Biswas:2024hja, Jimenez:2024hib, Marquez:2024bzj, Albino:2024ymc, Tang:2024jvs, Malik:2024qjw, Guerrini:2024gzu, Kurkela:2024xfh, Brandes:2023bob, Malik:2024nva, Fujimoto:2023unl, Roy:2023gzi, Yao:2023yda, Tang:2023owf, Mroczek:2023zxo, Fan:2023spm, Bartolini:2023wis, Cao:2023rgh, Pang:2023dqj, Grams:2023sml, Cai:2023gol, Brandes:2023hma, Lope-Oter:2023urz, Providencia:2023rxc, Takatsy:2023xzf, Brodie:2023pjw, Mroczek:2023eff, Gorda:2022lsk, Malik:2023mnx, Han:2022rug, Somasundaram:2022ztm, Lugones:2021bkm,Koehn:2024set}, establishing a new community standard, supported by several publicly available codes to facilitate an easy integration of the QCD inputs into other frameworks \cite{komoltsev_oleg_2023_7781233,komoltsev_2025_15407795}.

While the constraining power of the QCD input is not sensitive to perturbative uncertainties, it strongly depends on the termination density of the EoS. Sensitivity emerges from EoSs requiring drastic softening beyond termination density, followed by a high sound-speed segment to match the high-density limit. Methods that penalize such an extreme behavior above the TOV density introduce additional model dependence but ultimately yield a similar softening of the EoSs.

The observed EoS softening can be interpreted as a signature of a phase transition, with one possible explanation being that matter in the cores of the most massive NSs exhibits near-conformal behavior, consistent with weakly coupled quark matter. An alternative scenario involves a first-order phase transition - the strongest form of softening - that destabilizes the star. While current astrophysical and theoretical constraints cannot distinguish between these two possibilities, the findings of this thesis provide a \change{slight evidence} for nontrivial phase transitions of some kind occurring in the cores of the most massive neutron stars.

This leaves the fundamental question about the nature of the softening of the neutron matter EoS open, making it an exciting area of research. Further advancements in our understanding can be achieved by improving both theoretical and experimental side. On the theory side, the global constraints on EoS arise from low- and high-density limits, making it essential to advance both cEFT and pQCD calculations in the future. This includes computing the next order in the expansion, improving uncertainty estimation of theoretical calculations, and exploring other methods \cite{Gorda:2021znl,Karkkainen:2025nkz,Navarrete:2024zgz,Gorda:2021kme,Gorda:2018gpy,PhysRevC.103.025803,Drischler:2024ebw,PhysRevD.105.014025,Tews:2018kmu,Moore:2023glb}.

General-relativistic simulations of BNS mergers provide a rich environment for exploring various phenomenological aspects of NS physics. To facilitate meaningful comparisons between theory and experiment, a significant number of BNS simulations are required. The next generation of GW detectors are expected to detect many more BNS events, leading to tighter constraints on tidal deformability. With extensive observational data, precise mapping of the EoS will become possible. 

However, mapping the region near the TOV limit may be challenging, as it requires observations of stars near their maximum mass to constrain this area effectively. For the physics of the cores of NSs the post-merger GW signal may provide important information. Different types of phase transitions can leave distinct signatures, as they can significantly impact the dynamics of the remnant. The next generation of GW detectors may have the sensitivity needed to observe the post-merger signal of BNS mergers \cite{2021arXiv210909882E, Punturo_2010, Baiotti_2017}. Additionally, on the experimental side, low-energy nuclear experiments and heavy-ion collision experiments can provide tighter constraints on the low-density EoS.

As our understanding of matter under extreme conditions grows, a multidisciplinary approach becomes increasingly essential. While all current inputs to the NS EoS remain mutually consistent, future observations and improved theoretical calculations may reveal discrepancies between different inputs. Identifying such discrepancies could provide evidence of new physics beyond the Standard Model and general relativity, allowing NSs to be used as powerful laboratories for probing fundamental physics.}


%
\mainfont{\chapter{Appendix}
\label{chpt:appendix}

\subsection{The Love number and tidal deformability}

Tidal deformability $\lambda$ is defined as the ratio of the induced mass quadrupole moment $Q_{ij}$ and the tidal field $\mathcal{E}_{ij}$  and is related to the second dimensionless tidal Love number $k_2$ \cite{Hinderer_2008,Han:2018mtj}. Here, $\Lambda$ represents the dimensionless TD, while $M$ and $R$ denote the mass and radius of the star, respectively:
\begin{align}
    Q_{ij} &= -\lambda \mathcal{E}_{ij},\\
    \lambda &= \frac{2}{3} k_2 R^5, \\
    \Lambda &= \frac{\lambda}{M^5},
\end{align}

Tidal Love number $k_2$ can be calculated using the compactness parameter $\beta=M/R$ and an auxiliary variable $y$ \cref{eq:Love}:
\begin{align}
\label{eq:Love}
k_2 &= \frac{8\beta^2}{5} (1 - 2\beta)^2 \left[ 2 + 2\beta (y - 1) - y \right] \notag \\
&\quad \times \bigg\{ 2\beta \left[ 6 - 3y + 3\beta (5y - 8) \right] + 4\beta^3 \left[ 13 - 11y + \beta (3y - 2) \right] \notag \\
&\quad + 2\beta^2 (1 + y) + 3(1 - 2\beta)^2 [2 - y] + 2\beta (y - 1) \ln(1 - 2\beta) \bigg\}^{-1}.
\end{align}
The variable $y(r)$ is determined by solving a first-order differential
equation, which is solved simultaneously with the TOV equation, using
the boundary condition $y(0)=2$:
\begin{align}
\label{eq:Love2}
r y'(r) &+ y(r)^2 + y(r) e^{\lambda(r)} \bigg\{ 1 + 4\pi r^2 [p(r) - \epsilon(r)] \bigg\} \notag + r^2 Q(r) = 0, \\
Q(r) &= 4\pi e^{\lambda(r)} \bigg[ 5\epsilon(r) + 9p(r) + \frac{\epsilon(r) + p(r)}{dp/d\epsilon} \bigg] - \frac{6 e^{\lambda(r)}}{r^2} - \left( \frac{\nu'(r)}{r} \right)^2.
\end{align}
Additionally, the metric coefficients required for these computations are given by:
\begin{align}
\label{eq:metric_coef}
e^{\lambda(r)} &= \left[ 1 - \frac{2m(r)}{r} \right]^{-1},\\
\frac{d\nu}{dr} &= \frac{2}{r} \left[ \frac{m(r) + 4\pi p(r)r^3}{r - 2m(r)} \right].
\end{align}

\subsection{cEFT likelihood function}

\begin{figure}[ht!]
    \centering
\includegraphics[width=0.6\textwidth]{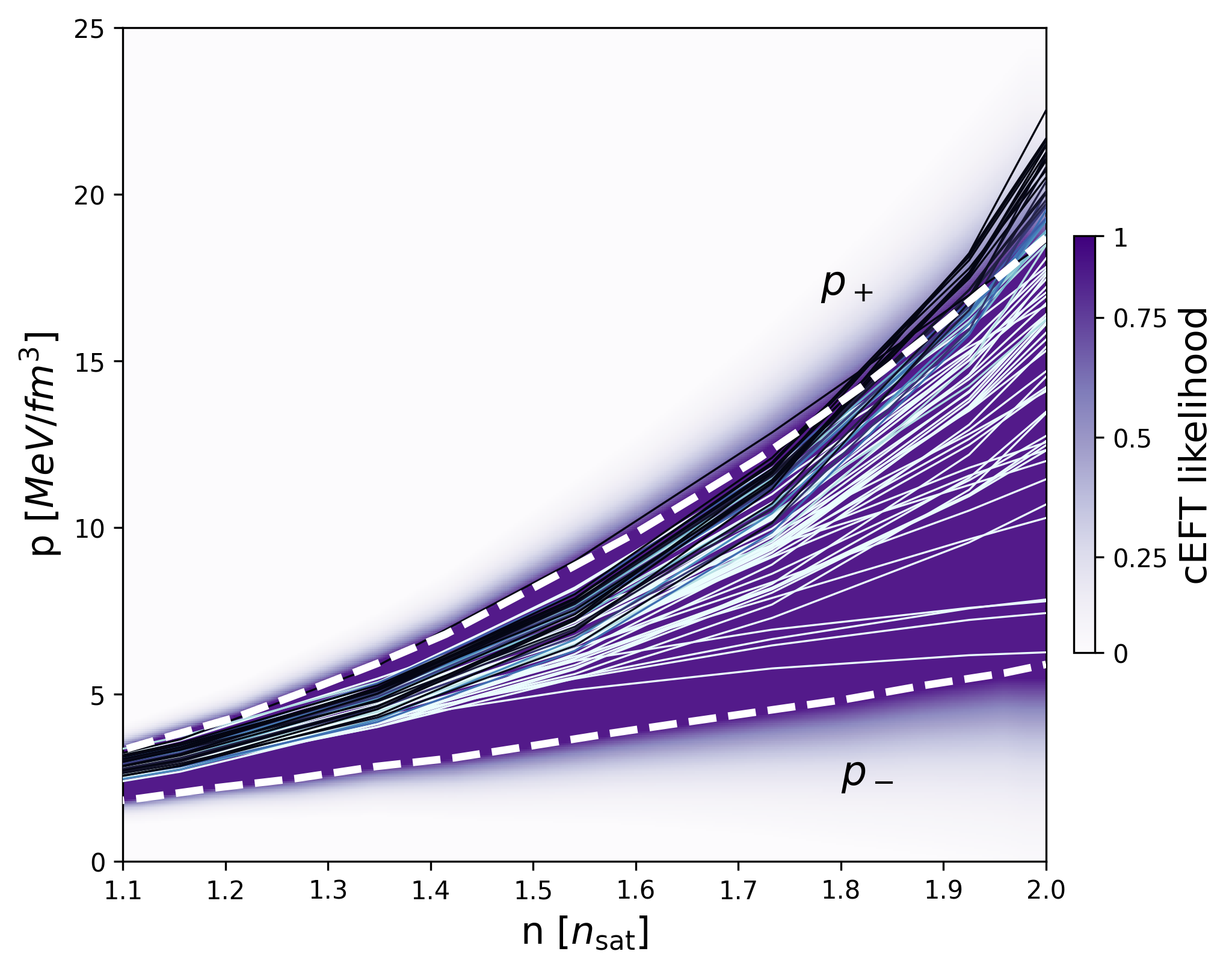}
    \caption{The cEFT likelihood function in the range $n \in [1,2]\ns$ is constructed according to \cref{eq:score_function}. The color bar represents the likelihood values, where darker shades of purple correspond to higher likelihood. The sample of EoSs, weighted according to the cEFT likelihood function, is displayed using a contrasting color scheme—white indicates high likelihood, while black denotes excluded EoSs.}
    \label{fig:ceft_likelihood}
\end{figure}

The cEFT likelihood function is constructed according to \cite{Tews:2018kmu,Koehn:2024set} and is defined as:
\begin{align}
    f(p, n) = \begin{cases} \exp\left(-\beta \frac{p-p_{+}}{p_{+} - p_{-}}\right) \qquad \text{if $p > p_{+}$\,,}\\
         \exp\left(-\beta \frac{p_{-}-p}{p_{+} - p_{-}}\right) \qquad \text{if $p < p_{-}$\,,}\\
        1 \qquad \qquad \qquad \qquad  \text{otherwise\,.}
    \end{cases}
\label{eq:score_function}
\end{align}
The cEFT likelihood function, $P(\mathrm{cEFT} \,|\, \mathrm{EoS}) \propto \prod_i f(p(\mathrm{EoS}, n_i), n_i)$, along with $p_+$ and $p_-$, is illustrated in \cref{fig:ceft_likelihood}, together with a sample of EoSs weighted according to the cEFT likelihood function.

\subsection{Additional plots}

\begin{figure}[ht!]
    \centering
\includegraphics[width=1\textwidth]{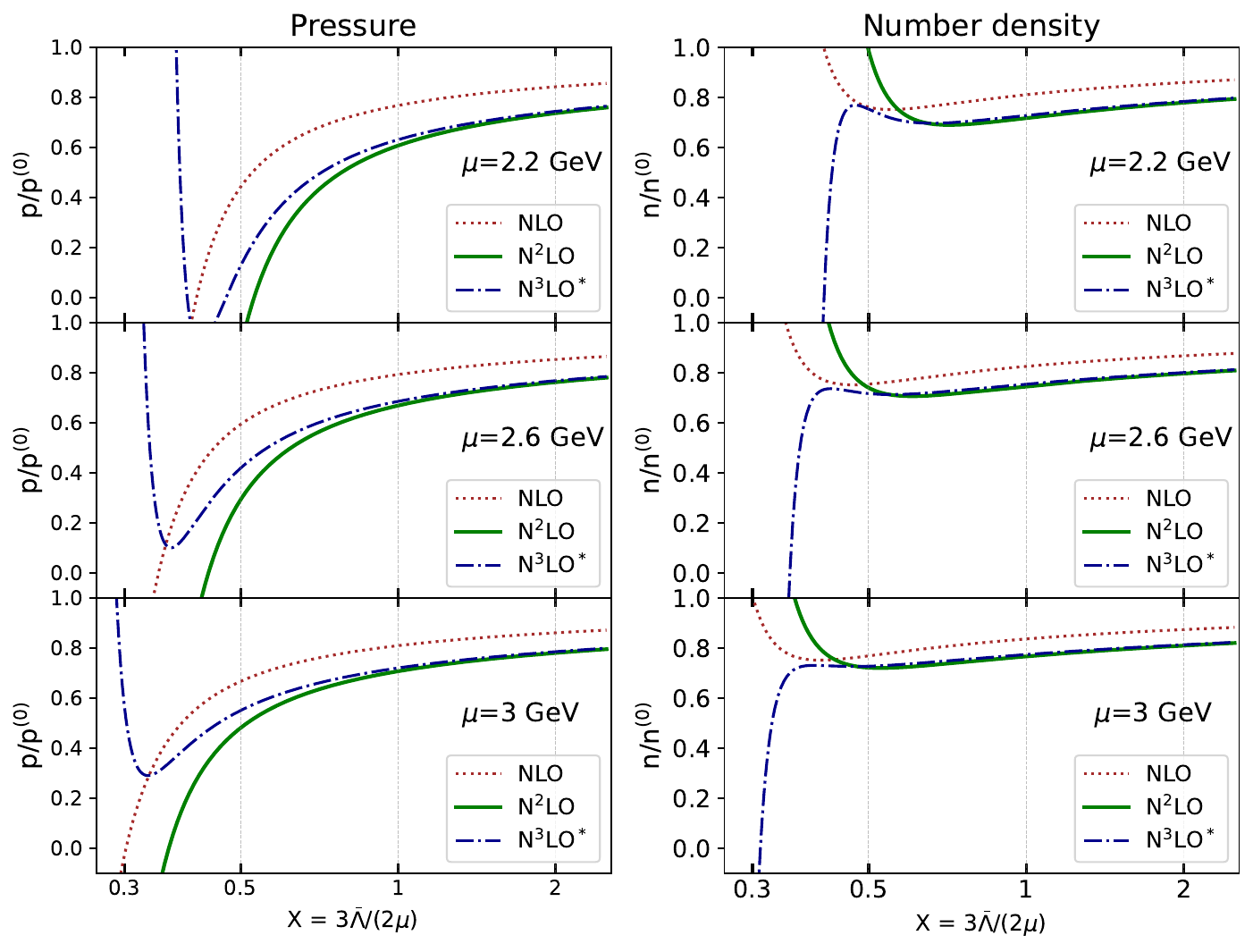}
    \caption{Fully computed NLO, N$^2$LO, and partially computed N$^3$LO results in perturbative QCD are shown for (left) the normalized pressure and (right) the normalized density as functions of the renormalization scale parameter $X$. Each row represents a fixed chemical potential, $\mu_\h = {2.2, 2.6, 3.0}$ GeV, which approximately corresponds to densities of $n \approx {23, 40, 63} , \ns$, respectively.}
    \label{fig:pQCD_p_n}
\end{figure}

\begin{figure}[ht!]
    \centering
\includegraphics[width=0.9\textwidth]{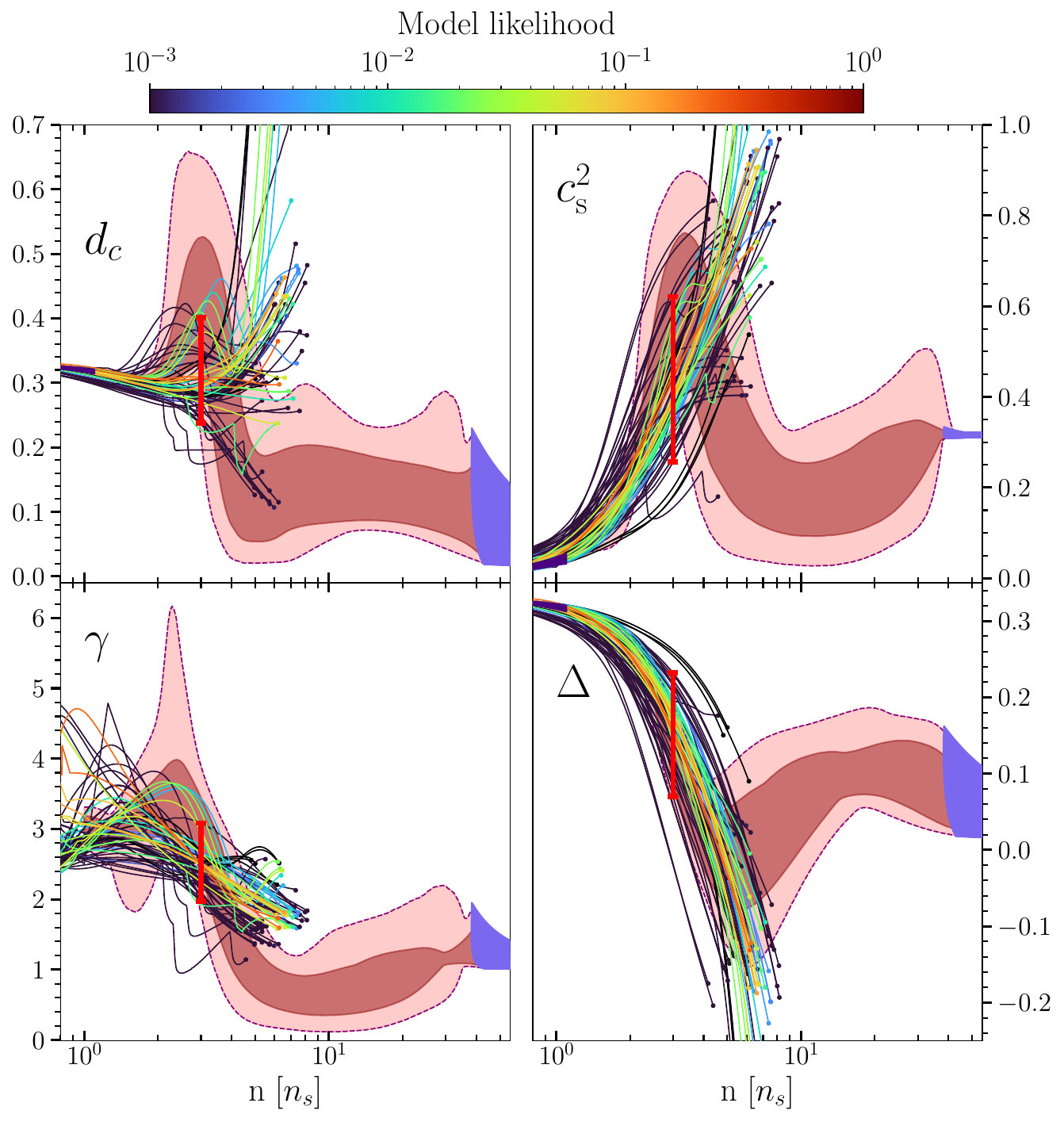}
    \caption{Comparison of the thermodynamic quantities from \cref{fig:dc,fig:QM_cs_gamma_delta}, obtained using the $c^2_{s,4}$ interpolation, with nuclear matter models from the CompOSE database at $T=0$ in $\beta$-equilibrium \cite{Typel:2013rza}. The coloring of each model corresponds to the likelihood function used in \cref{sec:qm}, normalized to the maximum likelihood in the GP ensemble. All models are terminated at the TOV density, where QCD inputs are imposed. The red solid bars represent the densities at which the values for dense NM in \cref{table:conformal} are chosen.}
    \label{fig:models}
\end{figure}

\begin{figure}[ht!]
    \centering
\includegraphics[width=0.8\textwidth, trim=0 0 0 40, clip]{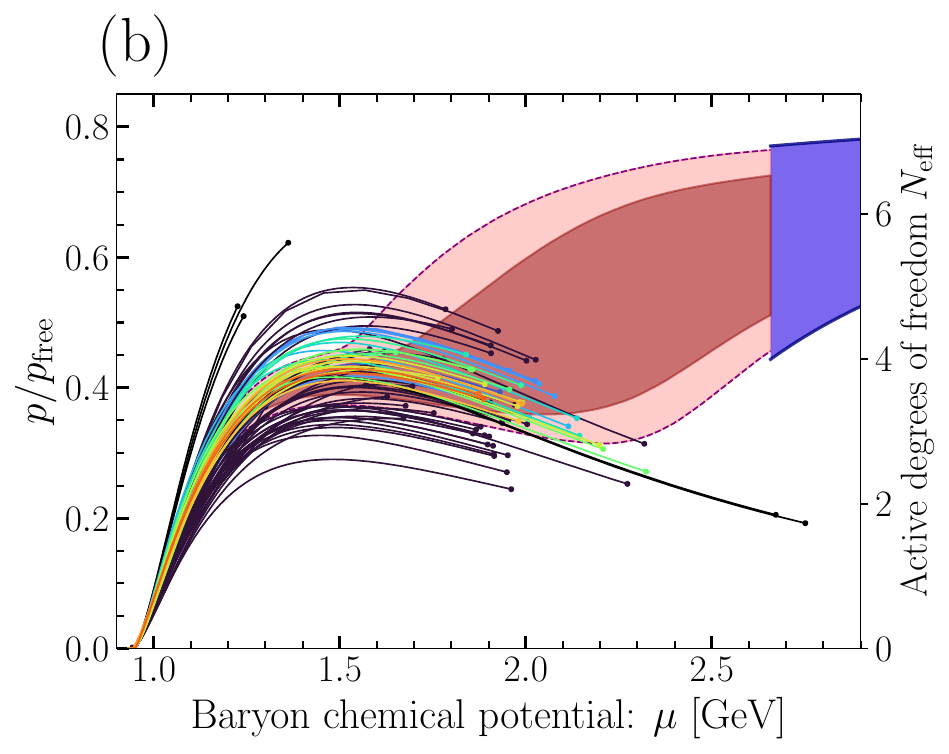}
    \caption{Comparison of the normalized pressure $p/p_{\rm free}$ from \cref{fig:ppfree}, obtained using the $c^2_{s,4}$ interpolation, with nuclear matter models from the CompOSE database at $T=0$ in $\beta$-equilibrium \cite{Typel:2013rza}. The coloring of each model corresponds to the likelihood function used in \cref{sec:qm}, normalized to the maximum likelihood in the GP ensemble. All models are terminated at the TOV density, where QCD inputs are imposed.}
    \label{fig:ppfree_models}
\end{figure}

\begin{figure}[ht!]
    \centering
\includegraphics[width=1\textwidth]{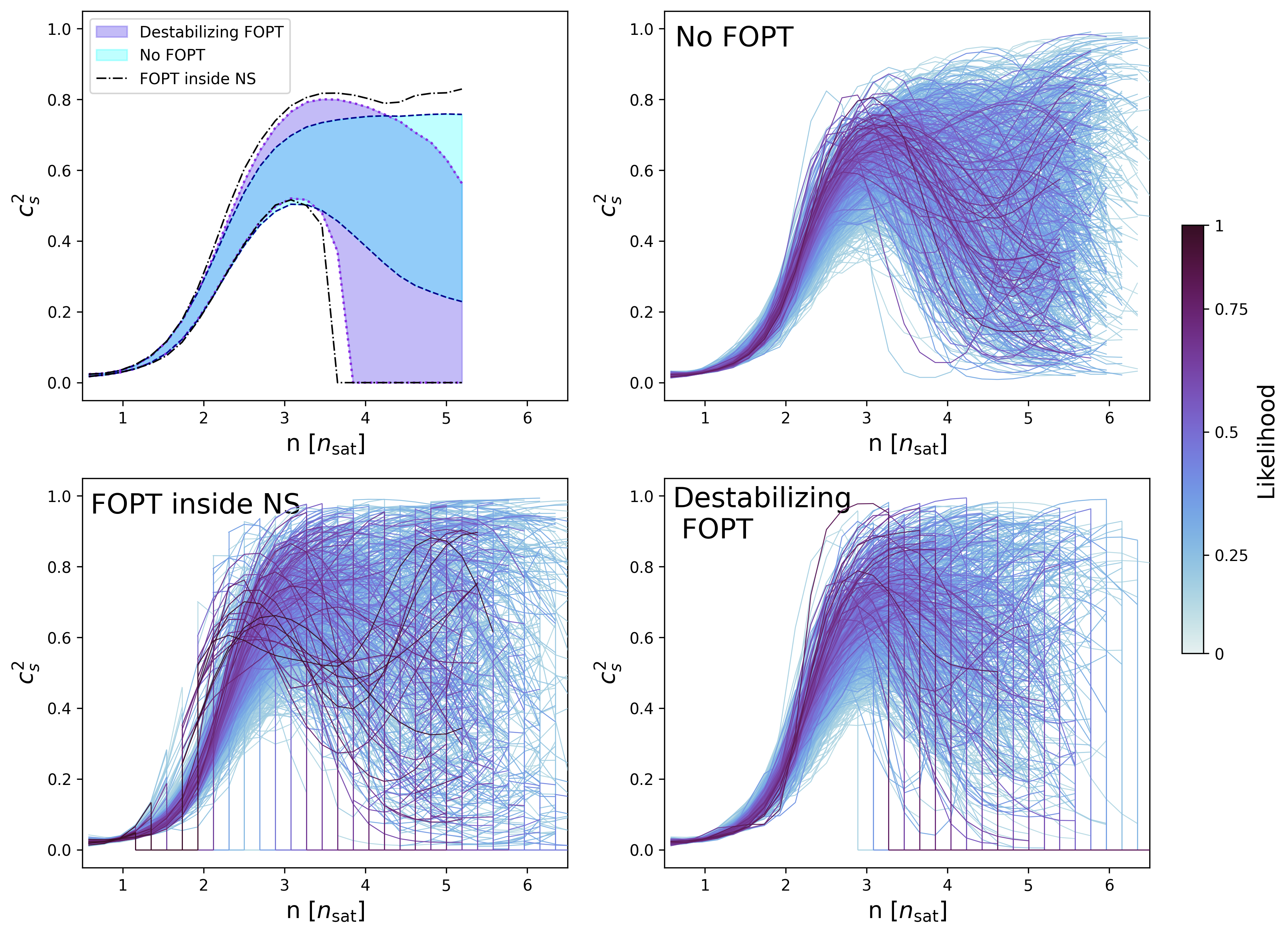}
    \caption{A modified version of \cref{fig:cs2_fopt} with less aggressive inputs: cEFT up to 1.1$\ns$, conservative QCD input, NICER PSR J0740+6620, radio measurements of PSR J0348+0432, and TD constraints from GW170818 data. (Upper left) The 68\% CI for the speed of sound for three different sets. (Other panels) A representative sample of EoSs, color-coded by likelihood based on conservative inputs. The likelihood is normalized to the maximum value within the ensemble.}
    \label{fig:cs2_convers_fopt}
\end{figure}
}

\displayReferences{Thesis_bib}
\printbibliography
\end{document}